\documentclass[manuscript]{aastex}
\usepackage{emulateapj5}
\usepackage{graphicx}
\usepackage{amsmath}

\begin{document}

\title{Evidence of thermonuclear flame spreading on neutron stars from burst rise oscillations}
\author{Manoneeta Chakraborty\altaffilmark{1} and Sudip Bhattacharyya\altaffilmark{1}}
\altaffiltext{1}{Department of Astronomy and Astrophysics, Tata Institute of Fundamental Research,
1 Homi Bhabha Road, Mumbai 400005, India; manoneeta@tifr.res.in; sudip@tifr.res.in}

\begin{abstract}\label{Abstract}
Burst oscillations during the rising phases of thermonuclear X-ray bursts are usually 
believed to originate from flame spreading on the neutron star surface. 
However, the decrease of fractional oscillation amplitude with rise time,
which provides a main observational support for the flame spreading model, 
have so far been reported from only a few bursts.
Moreover, the non-detection and intermittent detections of rise oscillations from many bursts
are not yet understood considering the flame spreading scenario.
Here, we report the decreasing trend of fractional oscillation amplitude 
from an extensive analysis of a
large sample of {\it Rossi X-ray Timing Explorer} Proportional Counter Array bursts 
from ten neutron star low-mass X-ray binaries. This trend is 99.99\% significant for
the best case, which provides, to the best of our knowledge, 
by far the strongest evidence of such trend. 
Moreover, it is important to note that an opposite
trend is not found from any of the bursts. The concave shape of the fractional
amplitude profiles for all the bursts suggests latitude-dependent flame speeds,
possibly due to the effects of the Coriolis force.
We also systematically study the roles of low fractional amplitude and low count rate for
non-detection and intermittent detections of rise oscillations, and attempt to understand
them within the flame spreading scenario.
Our results support a weak turbulent viscosity for flame spreading, and imply that
burst rise oscillations originate from an expanding hot spot, thus making these oscillations
a more reliable tool to constrain the neutron star equations of state.
\end{abstract}


\section{Introduction}\label{Introduction}

Thermonuclear or type I X-ray bursts are eruptions in X-rays, observed from many neutron star low-mass X-ray binary (LMXB) systems 
\citep{Strohmayer2006, Galloway2008}. Intermittent unstable burning of the accreted matter accumulated on the neutron star surface causes
these bursts \citep{Lamb1978, Swank1977}. During such a burst, the  observed X-ray intensity increases by a factor of $\sim 10$ in $\approx 0.5-5$ s, and then decays in $\approx 10-100$ s as the stellar surface cools down. These bursts may belong to various burning regimes based on the chemical composition of the fuel and the accretion rate per unit stellar surface area. For example, a burst may be a mixed hydrogen and helium burst triggered by hydrogen ignition, or a pure helium burst, or a mixed hydrogen and helium burst triggered by helium ignition (\citet{Strohmayer2006} and references therein). The spectral and timing properties of these bursts can be a useful tool to measure neutron star parameters, and hence to constrain the equation of state (EoS) models of super-dense degenerate matter of stellar cores (\citet{Bhattacharyya2010} and references therein).

Some bursts from several neutron star LMXBs show X-ray intensity fluctuations, termed as burst oscillations 
(\citet{Watts2012} and references therein). They originate from the azimuthally asymmetric brightness pattern on the surface of a spinning neutron star \citep{Chakrabarty2003, Strohmayer2003}. This feature can appear during the rise and/or the decay of the burst. The frequency of these oscillations is close (better than $\approx 1$\%) to the stellar spin frequency, and can evolve by a value between a fraction of a Hz to a few Hz during a burst (\citet{Watts2012} and references therein).
Burst oscillations are used to measure neutron star spin rates (e.g., \citet{Bhattacharyya2007:2, Chakraborty2012, Watts2012}).
Moreover, the phase-folded light curves of burst oscillations can be fitted with 
appropriate relativistic models to constrain other stellar parameters, 
such as mass and radius \citep{Bhattacharyya2010, Miller1998, Nath2002, Bhattacharyya2005:1, 
Lo2013}. However, in order to use this method reliably, it will be helpful to 
understand the origin and nature of the asymmetric brightness pattern on the stellar surface. 

In this paper, we focus on the oscillations during burst rise. Such rise oscillations
could originate from an expanding hotspot or burning region \citep{Watts2012}. This is because ignition is expected to happen at a certain point on the neutron star surface, and then the thermonuclear flames should spread to ignite all the fuel \citep{Fryxell1982, Cumming2000}.

Thermonuclear flames should spread by deflagration on neutron stars \citep{Spitkovsky2002}. These authors studied flame spreading on rapidly spinning neutron stars considering the effects of the Coriolis force and the lift-up of the burning ocean. Their calculations show that the flame spreads rapidly by geostrophic flow initially after ignition. A typical speed of this geostrohic 
flow can be $\vartheta_{geostrophic}\thickapprox(gh)^{1/2} \thicksim 4.5\times10^3$ 
km s$^{-1}$, where $g$ is the gravitational acceleration and $h$ is the scale height of the burning region \citep{Spitkovsky2002}. But when the burning region becomes sufficiently large making the Rossby number less than $1$, the Coriolis force comes into play, 
and slows down the burning front. The flame then spreads by much slower 
ageostrophic speed. For a weak turbulent viscosity, this ageostrophic flame 
speed is $\vartheta_{\rm flame} \sim (gh)^{1/2}/ft_{\rm n} \sim 5-20$ km s$^{-1}$.
Here, $t_{\rm n}$ is the time scale of nuclear burning, and the Coriolis parameter $f = 2\Omega\sin\theta$, where $\Omega$ is the stellar angular speed and $\theta$ is the latitude 
of the burning front. For a dynamically important turbulent viscosity, 
the ageostrophic flame speed is maximum: $\vartheta_{\rm flame} \sim (gh/ft_{\rm n})^{1/2} \sim 100-300$ km s$^{-1}$. Therefore, an observational estimate of the time-scale of flame spreading can be useful to probe the viscosity and $g$, $h$ and $t_{\rm n}$ (note that $\Omega$ is known from burst oscillations).
After \citet{Spitkovsky2002}, a recent work \citep{Cavecchi2013} has reported vertically resolved hydrodynamic simulations of flame spreading via deflagration in the thin helium ocean of a spinning star. But this paper assumes a constant Coriolis parameter, and hence does not study the latitude dependence of flame speed.

Observational indications of thermonuclear flame spreading have been found from decreasing fractional amplitude of burst rise oscillations \citep{Strohmayer1997, Strohmayer1998, Straatenetal2001, Nath2002, Bhattacharyya2005:2, Bhattacharyya2006:3}. This is because, as the burning region expands on a spinning neutron star, an increasingly larger fraction of this region usually remains visible throughout the spin period. 
However, to the best of our knowledge, the decreasing trend has not been 
quantified by fitting with empirical models and by doing F-tests. 

The evolution of fractional amplitude during burst rise can not only be useful to 
probe thermonuclear flame spreading, but also to detect the effects of the 
Coriolis force on such spreading. This is because, as 
\citet{Bhattacharyya2007:3} showed by numerical calculations, a convex-shaped
fractional amplitude profile implies isotropic flame speeds and a concave-shaped
fractional amplitude profile implies  latitude-dependent flame speeds, possibly
influenced by the Coriolis force. \citet{Bhattacharyya2005:2, Bhattacharyya2006:3, 
Bhattacharyya2007:3} showed two examples of apparently concave-shaped profiles for 
two bursts, but these authors or any other authors did not quantify the shapes of
fractional amplitude profiles of burst rise oscillations in an attempt to
detect latitude-dependent flame speeds.
Apart from fractional amplitude profiles, indications of flame spreading were also 
found from weak double-peaked bursts \citep{Bhattacharyya2006:1, Bhattacharyya2006:2}, 
as well as from an unusual precursor burst with oscillations \citep{Bhattacharyya2007:1}. 
Finally, \citet{Maurer2008} divided the bursts into three groups based on their peak fluxes 
and `rise light curve morphology', and connected these groups to various 
burning regimes, ignition latitudes and flame spreading parameters.

In spite of some observational indications of thermonuclear flame spreading during burst rise, several issues exist. The first one is why oscillations are not detected during the rise of most bursts. Is it because (1) the count rate is too low; or (2) flames spread too fast, so that the spreading is over at the beginning of the rise when the intensity is low; or (3) the ignition happens at a latitude which is not visible due to the viewing angle of the observer, and then, when the burning region can be seen, it becomes azimuthally symmetric; or (4) the visible portion of the burning region is azimuthally symmetric almost from the beginning? One needs to answer this question to know whether burst rise oscillations originate from flame spreading. Therefore,
a more systematic and extensive study is required to check if the available burst 
rise data, including the upper limits, are consistent with the expanding burning 
region model \citep{Watts2012}.

In this paper, we systematically analyze the {\it Rossi X-ray Timing Explorer} 
({\it RXTE}) Proportional Counter Array (PCA) data of 51 thermonuclear bursts from 
ten neutron star LMXBs. We primarily focus on the bursts from the prolific burster 
4U 1636--536, harboring a rapidly spinning (frequency $\approx 582$ Hz;
\citet{Strohmayer2002}) neutron star. We find that the non-detection of oscillations 
during the rise of most bursts from 4U 1636--536 is not solely due to the low count
rates. We extensively examine the fractional amplitude evolution during the rising 
phase of all the 51 bursts, and show that such evolution is consistent with rise 
oscillations originating from flame spreading, possibly influenced by the Coriolis force.
We explain our methods in \S~\ref{DataAnalysis}, describe the results with 
figures in \S~\ref{Results}, discuss the implications of 
the results in \S~\ref{Discussions}, and summarize the key points of the paper
in \S~\ref{Summary}.

\section{Data Analysis}\label{DataAnalysis}

\subsection{Data description}\label{Data}

We choose the {\it RXTE} PCA bursts for our analysis in the following way.
We use only the data from the \citet{Galloway2008} catalogue,
because (1) {\it RXTE} usually
operated with only one or two proportional counter units (PCUs) on in the
recent years, and hence the later data generally had poorer
statistics, and (2) \citet{Galloway2008} give a good representative sample of bursts.
In this catalogue, 12 sources with burst rise oscillations
are mentioned, out of which two are pulsars. We exclude pulsars from our
analysis, because the periodic pulsations and the burst oscillations are difficult
to distinguish. The rest ten sources are 4U 1636--536, 4U 1608--52, MXB 1659--298, 
4U 1702--429, 4U 1728--34, KS 1731--260, 1A 1744--361, SAX J1750.8--2900, 4U 1916--053 
and Aql X--1. Neutron stars in all these sources are rapidly spinning (see 
Tables~\ref{Log} and \ref{Log2}).
Galloway et al. (2008) reported rise oscillations from 32 bursts out
of 172 PCA bursts from 4U 1636--536, and rise oscillations from 41 bursts out
of 315 PCA bursts from the other nine sources. We use the bursts from
all the ten sources to study the evolution of fractional amplitudes of
rise oscillations (\S~\ref{Rms} and \ref{othersources}). 
In order to probe the cause of non-detection and intermittent detection
of burst rise oscillations from many bursts 
(\S~\ref{Detection}), as well as to study the burst properties in
various bursts groups (\S~\ref{morphology}), we use the bursts only
from 4U 1636--536. This is because this source showed by far the highest number
of bursts with rise oscillations (see Tables~\ref{Log} and \ref{Log2}).
We consider 161 out of 172 bursts for 4U 1636--536, because the rising phases
of the rest of the bursts were at most partially observed.
Finally, we note that the analysis of event mode data (time resolution 
$\approx 122$ $\mu$s) from {\it RXTE} PCA is reported in this paper \citep{Jahodaetal2006}.

\subsection{Fractional amplitude calculation}\label{amplitude_calculation}

A primary aim of this work is to study the fractional amplitude evolution of
burst rise oscillations observed from neutron star LMXBs. To begin with, we define
the rise time of each burst as the time in which the pre-burst level subtracted
PCA count rate increases from the 5\% of the peak to the peak.
Then we divide the rise time of every burst into 0.25 s bins for the sake of
a systematic and uniform analysis. Such a small time bin is necessary to 
track the burst oscillation amplitude evolution, especially just after the 
burst onset, when the amplitude possibly evolves fast \citep{Bhattacharyya2007:3}.
Moreover, equal time bins are required to use a uniform detection criterion
and to study the non-detection and intermittent detections of burst rise 
oscillations from many bursts, as discussed later.
After dividing the rise time into bins, we check whether
burst oscillation is detected in a time bin, and estimate the fractional amplitude
for that bin.

\begin{figure*}
\centering
\includegraphics[width=0.35\textheight]{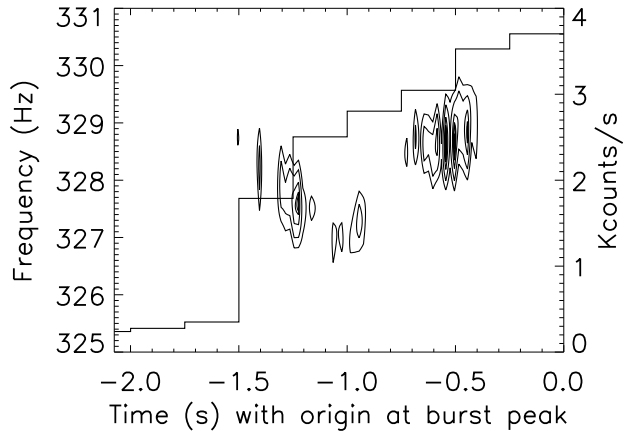} 
\caption{Power contours of oscillations in the frequency-time space during the 
rising phase of burst 6 (Table~\ref{Log2}) from 4U 1702-429. 
The power contours show that the frequency of oscillations changes
apparently erratically, and oscillations are present for short intervals
during the rise. The histogram exhibits the burst intensity with time. This figure
explains why phase-timing analysis technique cannot be used to estimate the 
frequency$-$time profile for the rising phase of many bursts
(see \S~\ref{amplitude_calculation}). \label{freqevol}}
\end{figure*}

A time bin size of 0.25 s implies a very coarse frequency resolution of 4 Hz,
when the light curve is Fourier transformed. Consequently, a conventional method of 
static power spectrum cannot be used either to estimate the fractional amplitude,
or to estimate the frequency of oscillations for a time bin. Therefore, a phase-timing 
analysis is usually used to estimate the frequency evolution
\citep{Muno2000, Strohmayer2002, Bhattacharyya2006:3}. 
Then the resulting best-fit frequency versus time profile is used to calculate
the fractional amplitude for each time bin using the phase-folded light curve (see below). 
The results of this phase-timing analysis
very well match with those of an alternative technique of Z$^2$ power maximization
for the entire rise interval \citep{Strohmayer2002, Bhattacharyya2006:3}.
However, the frequency model should have a small number of free
parameters, and the oscillation should be present in several rise time bins,
for the phase-timing analysis technique to work well. But frequency appears to
change apparently erratically, and oscillation is present only for short intervals
during the rise of many bursts (Fig.~\ref{freqevol} shows an example of such a burst). 
Therefore, we cannot use the phase-timing analysis technique for many bursts.
Consequently, we use an alternative method to estimate the frequency$-$time 
profile, as we aim for a uniform analysis for all the bursts.

Here we describe this method.
We adopt the following procedure to estimate the oscillation frequency. 
We assume that this frequency remains constant in a time bin, which may usually be 
justified for a small bin size of 0.25 s. Then we estimate this frequency by
maximizing the fractional amplitude using the phase-folded light curves for 
frequency values in the range ($\nu_1$, $\nu_2$) with a step size of $\Delta\nu$ 
(e.g., \citet{Strohmayer1998}).  Here, $\nu_1 < \nu_{\rm star} < \nu_2$, 
where $\nu_{\rm star}$ is the known spin frequency of the neutron star.
Moreover, we consider $\nu_2-\nu_1 \sim 5$ Hz, since the burst oscillation frequency
of a given source does not change by more than 5 Hz.
This fractional amplitude maximization technique is similar to the  Z$^2$ 
power maximization technique mentioned above.
In the next paragraph, we describe how the fractional amplitude is estimated for
a given frequency value.

For an oscillation frequency $\nu_{\rm osc}$ for a given time bin, 
we extract the phase-folded light curve from 
the event mode data within the 0.25 s bin, after subtracting the persistent intensity level 
\citep{Bhattacharyya2005:2}. The persistent level is calculated averaging 20 s of
pre-burst emission, or emission well after the bursts if the pre-burst data
are not available. The number of phase bins used is 16 and the error of a bin of the 
phase-folded light curve is computed by propagating the 
errors of total emission and persistent emission. Following the standard technique, 
a phase-folded light curve 
is fitted with a constant plus a sinusoid: $A+B\sin(2 \pi \nu t +\delta)$, where 
$A$ and $B$ are free parameters and $\nu = \nu_{\rm osc}$ 
\citep{Bhattacharyya2005:2, Muno2000}.
Any additional harmonic content, if present at all, is too weak to detect.
We find that a `constant+sinusoid' model with harmonic is {\it not} better 
than a simple `constant+sinusoid' model with more than 95.45\% (i.e, $2 \sigma$)
significance. This is supported by the fact that the only report of harmonic component 
in burst rise oscillations from any source was by \citet{Bhattacharyya2005:2}, 
who combined the first 1/3rd of the
rise time intervals of nine bursts from 4U 1636--536 to detect the 
harmonic with $3\sigma$.
If the oscillation is detected (see the next paragraph) for the chosen time bin,
then the fractional root-mean-squared (rms) amplitude for that bin
is $r = B/(\sqrt{2}A)$, where the best-fit values of $A$ and $B$,
for the frequency which maximizes the fractional amplitude, are used. The error ($r^{\rm err}$) of $r$
is estimated by propagating the errors of $A$ and $B$.

Now we describe our procedure to check if the oscillation is detected 
for the chosen time bin. For this, we fit the phase-folded light curve
also with a constant $A'$, and perform an F-test between the constant model and the
`constant+sinusoid' model. If the latter model is at least $3\sigma$ more 
significant than the former one, i.e., if the probability that the 
`constant+sinusoid' model is better that the constant model only by random chances
is 0.0027 or less, we consider that the burst oscillation is
detected. If the oscillation is not detected,
we estimate the upper limit of $r$, which is the maximum value of $r+r^{\rm err}$ for
all the values of $\nu$ in the range ($\nu_1$, $\nu_2$) with the step size of $\Delta\nu$
(see before in this subsection for the meanings of $\nu_1$, $\nu_2$ and $\Delta\nu$).
\citet{Galloway2008} reported the detection of burst rise oscillations for 32 PCA bursts
from 4U 1636--536 and for 41 PCA bursts from other nine sources (Table~\ref{Log2}; 
\S~\ref{Data}). With our above mentioned criterion, we detect rise oscillation 
for at least one 0.25 s time bin for 27 bursts from 4U 1636--536 (Table~\ref{Log}) 
and for 24 bursts from other nine sources (Table~\ref{Log2}). This is because
our detection criterion is quite stringent
considering the small numbers of total counts in short 0.25 s time bins.
Such stringent criterion contributes to the reliability of our conclusions.

\begin{figure*}[h!tba]
\centering
\includegraphics[width=0.35\textheight]{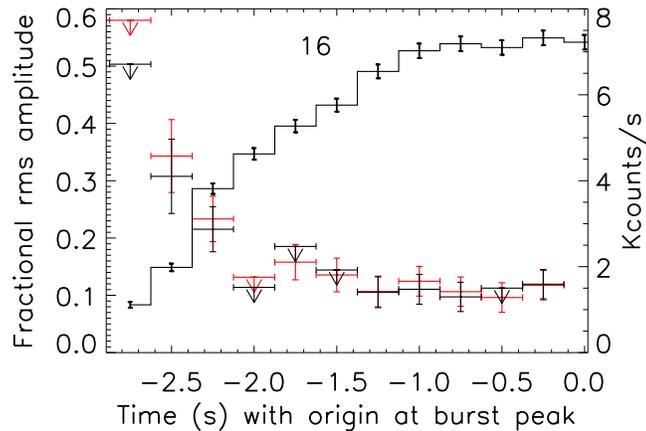}
\caption{Fractional rms amplitude evolution (red and black points with error bars) of
burst rise oscillations, and burst rise light curve (histogram) for the {\it RXTE} PCA burst 16 
from 4U 1636--536 (Table~\ref{Log}). For the red points, the oscillation frequencies are
estimated using our method (\S~\ref{amplitude_calculation}), while for the black points,
the oscillation frequencies are estimated using the phase-timing analysis technique
\citep{Muno2000}. This figure shows that fractional rms amplitude versus time
curves from the two techniques are consistent with each other (\S~\ref{amplitude_calculation}).
\label{rmsampevolcompare}}
\end{figure*}

Since we do not use the standard phase-timing analysis technique to estimate
the oscillation frequencies, we need to do a sanity check for our results.
Therefore, we estimate the fractional rms amplitude profiles using our method
and the phase-timing analysis technique for a few bursts. We find that for a 
given burst, such profiles from two methods are consistent with each other.
We show this with an example burst in Fig.~\ref{rmsampevolcompare}. Here the fractional
rms amplitude values from the two methods differ typically by 4.5\% 
($\approx 0.26$ times the 1 $\sigma$ error from our analysis).

We furthermore note that the non-constant intensity within a 0.25 s burst 
time bin does not tangibly affect our fractional rms amplitude value,
as we find from detailed simulations. In our simulations, we assume a constant 
fractional amplitude in a time bin with the steepest count rate slope available
from the data. We find that our estimated fractional rms amplitude value 
is systematically less than the correct value which includes the effect of
non-constant intensity, but this systematic error contributes to the total 
error by less than 0.4\% for our typical data count rates.

Now an important question to ask is what determines the detection of burst oscillations.
This will be useful to understand why oscillations are not detected during the rise
of most bursts (see \S~\ref{Introduction}). The detection of burst oscillation in a
time bin depends on the signal to noise ratio (S/N). To represent the
S/N, we use the figure of merit quantity
\begin{eqnarray}
\frac{\Delta N}{N^{\rm err}} &=& \frac{\Delta N}{N_{\rm T}-N_{\rm Per}}\frac{N_{\rm T}-N_{\rm Per}}{N^{\rm err}}\nonumber \\
&=&\sqrt{2} r \frac{N_{\rm T}-N_{\rm Per}}{N^{\rm err}},
\label{eqncrp}
\end{eqnarray}
where, $\Delta N$ is the absolute peak amplitude (in count rate) of oscillation, $N_{\rm T}$ is the 
total count rate averaged over a time bin, $N_{\rm Per}$ is the average persistent count rate,
$N^{\rm err}$ is the estimated error on $N_{\rm T}-N_{\rm Per}$, 
and $r$ is the fractional rms amplitude. Note that $\Delta N$ and $N_{\rm T}-N_{\rm Per}$
are essentially $B$ and $A$ respectively, where the latter two are free parameters in the
previously mentioned `constant+sinusoid' model $A+B\sin(2 \pi \nu t +\delta)$ of phase-folded
oscillation light curves. Eqn.~\ref{eqncrp} shows that
for a given strength (i.e., $r$) of the burst rise oscillation, the detection of oscillation depends
on the count rate parameter $\frac{N_{\rm T}-N_{\rm Per}}{N^{\rm err}}$. Therefore, we calculate
this parameter for all the 0.25 s rise time bins of all the 161 bursts (\S~\ref{Data}) from 
4U 1636--536.

\subsection{Burst rise morphology characterization}\label{morphology_characterization}

Following \citet{Maurer2008} (see also \S~\ref{Introduction}), 
we estimate the shapes of the rise light curves of 27 bursts 
with oscillations detected during rise (see Table~\ref{Log}). For this morphology analysis,
we consider the burst rise as the time 
interval between 5\% and 90\% of the peak PCA count rate, after correcting for the persistent
emission. In order to make an uniform comparison among different bursts, we normalize
both the rise intensity and the rise time of each burst so that they increase from 0 to 10 in
dimensionless units. The shape of a burst rise light curve is expressed with a quantity $\mathcal{C}$,
which gives a measure of the degree of light curve convexity as
\begin{eqnarray}
\mathcal{C} &=& \displaystyle\sum\limits_{i=0}^M (R_i-x_i) \Delta t,
\label{morphology1}
\end{eqnarray}
where $R_i$ and $x_i$ are respectively the re-normalized count rate and the re-normalized
rise time in each bin, $\Delta t$ is the re-normalized time bin size, and  $M$ is the number of 
such re-normalized time bins. In the chosen units, $\mathcal{C}$ lies between $-50$ and $50$, and 
it is positive for an overall convex shape and negative for an overall concave shape.

\subsection{Estimation of persistent emission change}\label{fa}

The recent work of \citet{Worpel2013} suggests a change of persistent emission from
its pre-burst value during thermonuclear X-ray bursts.
These authors multiplied the best-fit pre-burst emission spectrum with a factor $f_a$ to
represent the persistent emission spectrum during a burst. They generally found $f_a \ne 1$,
which indicated a change of persistent emission during a burst.
Such change would introduce a systematic error in our calculated fractional 
rms amplitudes, as described in \S~\ref{amplitude_calculation}.
Hence we estimate this persistent emission change (following \citet{Worpel2013}) 
for each of 27 bursts from 4U 1636--536 with detected rise oscillations 
(see Table~\ref{Log}), and examine 
if this change has any significant effect on fractional amplitude evolution.

For spectral analysis we use the same 0.25 s time bins during burst rise 
(see \S~\ref{amplitude_calculation}), and extract the energy spectrum from each bin.
Each spectrum is fitted within $3-15$ keV in XSPEC using the model {\tt wabs*bbodyrad+constant*$P(E)$},
where {\tt wabs} is an absorption model, {\tt bbodyrad} is a blackbody model,
the constant is the multiplicative factor $f_a$, and $P(E)$ is the fixed persistent 
emission spectral model {\tt wabs*(bbodyrad+powerlaw)} with best-fit parameters. These 
best-fit parameters of persistent emission model are estimated from 
spectral fitting of a 100 s time segment either prior to the burst or well after the burst.
The background for this fitting is calculated using the FTOOLS command `PCABACKEST'.
The neutral hydrogen column density ($N_{\rm H}$) parameter in the {\tt wabs} component 
is frozen at $0.25\times10^{22}$ cm$^{-2}$ \citep{Asai2000}, 
and a 0.5\% systematic error is assumed. For the fitting we have used Churazov weighting to take care 
of the low count statistics. From the time resolved spectroscopy we obtain the 
value of $f_a$ and its error for each time bin. 

Then the fractional rms amplitude and its error are recalculated,
or the upper limit of the amplitude is re-estimated (in case of non-detection)
for each 0.25 s time bin using this $f_a$ and its error for each bin. However, given the available
quality of data, there are the following difficulties in reliably estimating the 
fractional amplitude evolution including the effects of $f_a$.
(1) The usually large statistical error of $f_a$ makes the errors on the count rates
of the phase-folded light curves larger. This reduces the number of 0.25 s time bins
with detection of oscillations, increases the error on fractional amplitudes, and 
makes the trend of the amplitude evolution less clear. 
(2) The fitting cannot be performed for some time bins as the statistics is very poor. 
$f_a$ and its error cannot be calculated for these bins and consequently the 
modified rms amplitude cannot be determined. This 
results in some missing data points in the rms amplitude evolution making the trend weaker.
For these reasons, we report the fractional amplitudes primarily without considering the change of
persistent emission during a burst, and check if the resulting conclusions are consistent 
with those from the `$f_a$ method' mentioned in this subsection.
 
\section{Results}\label{Results}

\subsection{Detection of oscillations for 4U 1636--536}\label{Detection}

\begin{figure*}
\centering
\hspace{-1cm}
\includegraphics[width=0.66\textheight]{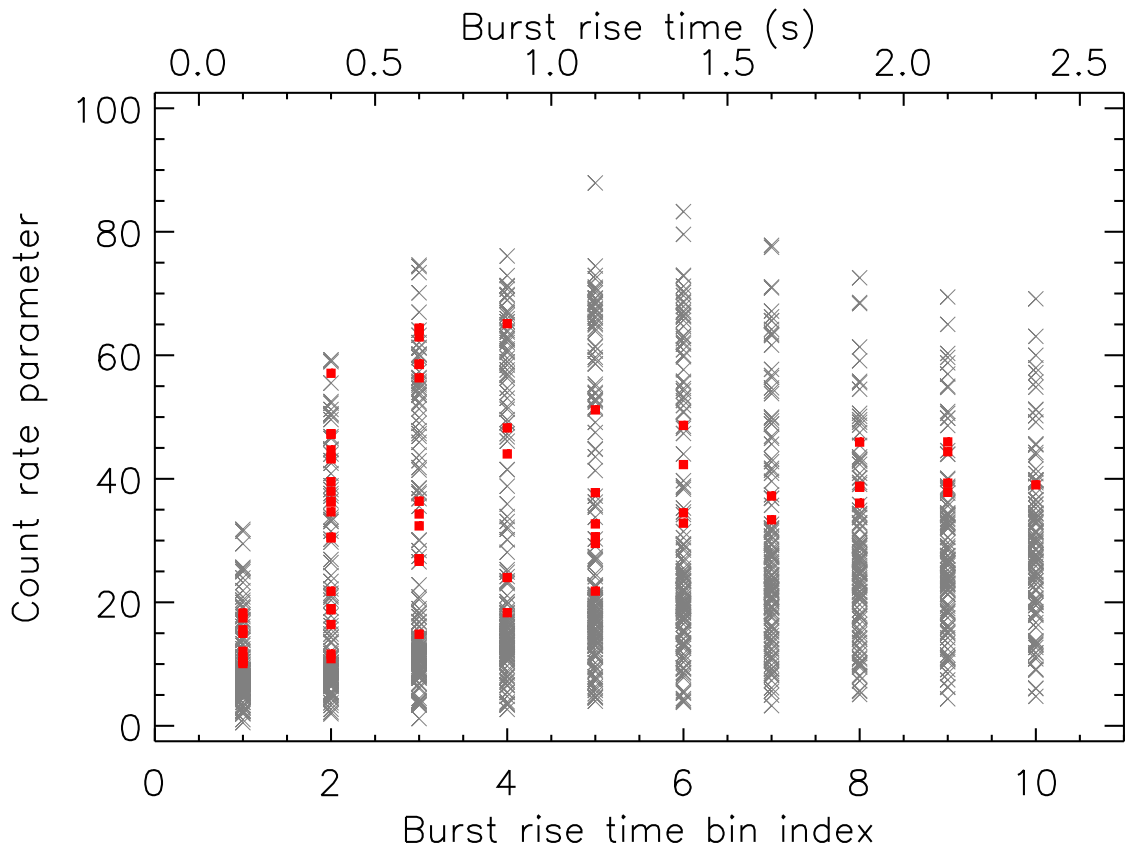} 
\caption{Count rate parameter [$= (N_{\rm T}-N_{\rm Per})/N^{\rm err}$] of thermonuclear bursts versus
time bin (each of 0.25 s) index during burst rise (see \S~\ref{amplitude_calculation}).
Here  $N_{\rm T}$ is the total count rate, $N_{\rm Per}$ is the average background 
count rate and $N^{\rm err}$is the $1\sigma$ error.
A total number of 161 bursts from 4U 1636--536 observed with {\it RXTE} PCA are
used in this plot (see \S~\ref{DataAnalysis}). A grey cross corresponds to a time bin of a burst.
The small red squares correspond to the cases where burst oscillation is detected.
The distribution of red squares on the grey crosses clearly shows that the non-detections
of burst oscillations are not solely due to low observed count rates
(see \S~\ref{Detection}). 
\label{countbindetection}}
\end{figure*}

\begin{figure*}
\centering
\hspace{-1cm}
\includegraphics[width=0.66\textheight]{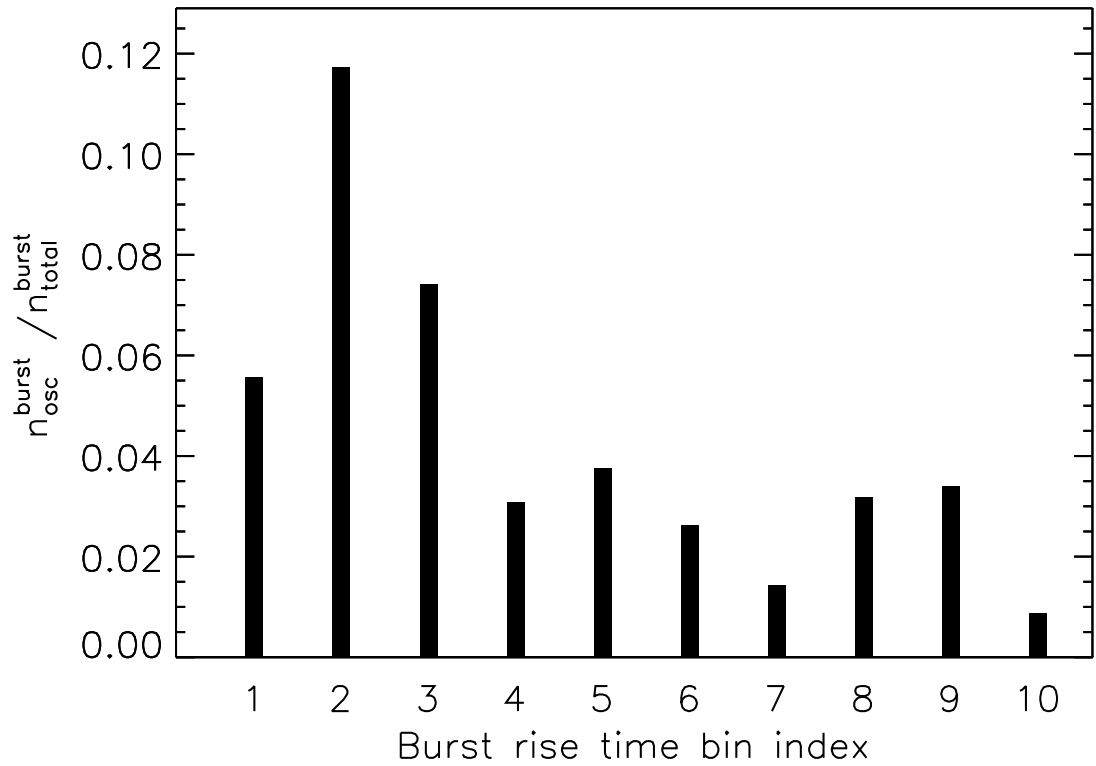} 
\caption{The fraction of thermonuclear bursts with detected burst rise oscillations versus
time bin (each of 0.25 s) index during burst rise (see \S~\ref{DataAnalysis}).
A total number of 161 bursts from 4U 1636--536 observed with {\it RXTE} PCA are
used in this plot. For a given time bin, $n^{\rm burst}_{\rm osc}$ is the number of  bursts
with detected oscillations, and $n^{\rm burst}_{\rm total}$ is the total number of  bursts.
This figure shows that the fraction of bursts with detected oscillations overall
decreases with time, although the burst intensity is expected to increase with time
during burst rise (see \S~\ref{Detection}). 
\label{detectionfractionbin}}
\end{figure*}

In order to understand why burst rise oscillations are not detected from most bursts
from 4U 1636--536, we plot the count rate parameter versus 0.25 s time bin index for all 161
bursts in Fig.~\ref{countbindetection} (see \S~\ref{amplitude_calculation} for some details). This figure
clearly shows that, for each bin, some bursts with oscillations can have much lower count rate parameter 
values than some of those without oscillations.
Oscillations during rise are detected in 1 bin for 9 bursts, in 2 bins for 9 bursts, in 3 bins for 5 bursts and in 4, 6, 7 and 9 bins for 1 burst each.
In order to further check whether oscillations from more bursts are detected as the intensity
(and also the count rate parameter)
increases during the rise, we show, for each bin, the ratio of the number of bursts with oscillations 
to the total number of bursts in Fig.~\ref{detectionfractionbin}. 
This figure shows that the detected fraction has an overall decreasing trend with time.
Therefore, Figs.~\ref{countbindetection} and \ref{detectionfractionbin} imply
that the detection of oscillation depends on the fractional amplitude value.
Note that, in both Figs.~\ref{countbindetection}
and \ref{detectionfractionbin}, we consider the first ten 0.25 s time bins (or all the bins if the number is
not more than ten for a burst), because the later time bins do not usually have any detected oscillation.

\begin{figure*}
\centering
\hspace{-1cm}
\includegraphics[width=0.66\textheight]{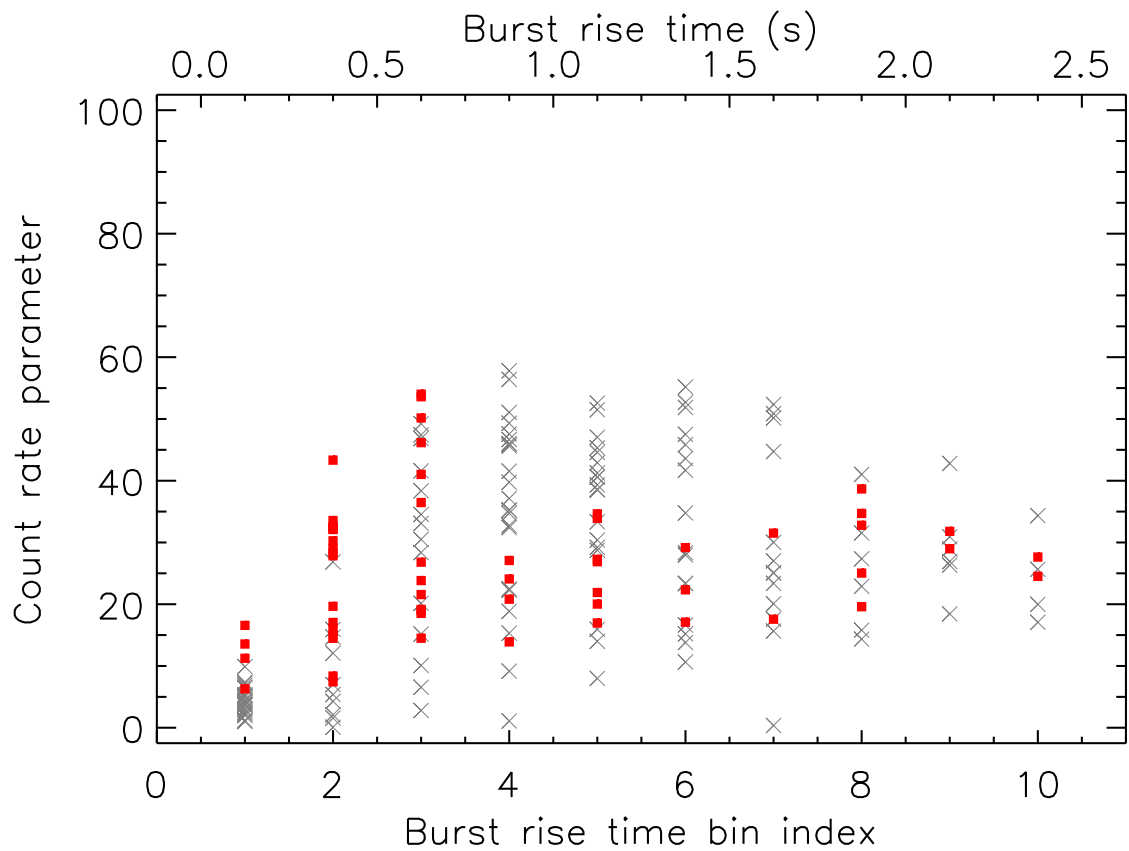} 
\caption{Similar to Fig.~\ref{countbindetection}, but (a) including the effects of the changing 
persistent emission, i.e., considering $f_a$ as described in \S~\ref{fa}; and
(b) for only the 27 bursts with oscillations detected in at least one time bin (see \S~\ref{DataAnalysis}
and \S~\ref{Detection}). This figure shows that the non-detections
of burst oscillations are not solely due to low observed count rates, even when
the effect of the changing persistent emission is considered.
\label{countbindetectionfa}}
\end{figure*}

In order to find out if the evolution of persistent emission during a burst changes
the above finding, we calculate the count rate parameter values including $f_a$ 
(see \S~\ref{fa}). With these modified count rate parameter values, 
we make a figure (Fig.~\ref{countbindetectionfa})
similar to Fig.~\ref{countbindetection} for the 27 bursts mentioned in Table~\ref{Log}.
We find that, even for the change of persistent emission, the non-detections of
burst rise oscillations are not solely due to the low count rates.

\begin{figure*}
\centering
\hspace{-1cm}
\includegraphics[width=0.66\textheight]{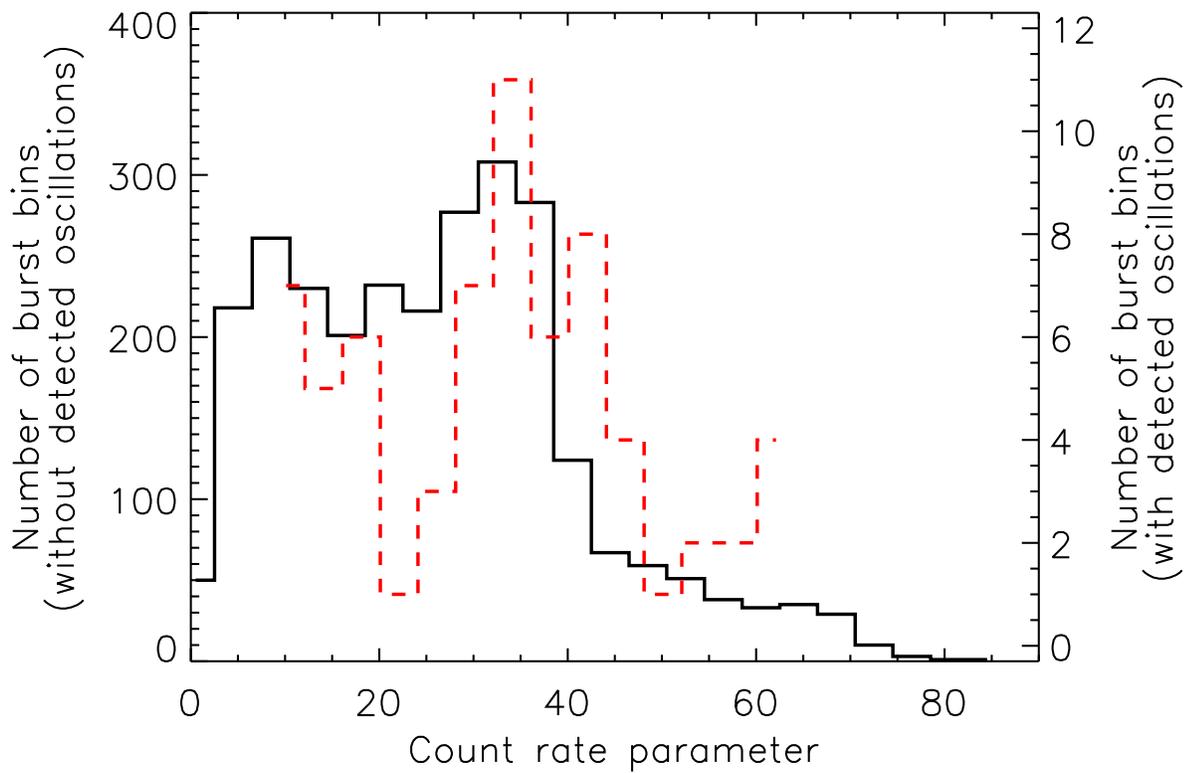}
\caption{The distributions of 0.25 s rise time bins
of all the 161 {\it RXTE} PCA bursts from 4U 1636--536 over the count rate
parameter values in the following two cases: (1) bins without detected 
oscillations (solid black histogram; left y-axis),
and (2) bins with detected oscillations  (broken red histogram; right y-axis)
(see \S~\ref{Detection}). From this figure, it appears that the second distribution
is somewhat shifted towards the higher values of the count rate parameter
relative to the first distribution, and hence the detection of oscillations 
somewhat depends on the burst intensity (\S~\ref{Detection}).
\label{KS}}
\end{figure*}

However, the burst intensity may also affect the detection of oscillations.
In order to check this, we plot the distributions of 0.25 s rise time bins 
of all the 161 bursts from 4U 1636--536 over the count rate
parameter values in the following two cases: (1) bins without detected oscillations,
and (2) bins with detected oscillations; both without considering $f_a$
(see Fig.~\ref{KS}). From this figure, it appears that the second distribution
is somewhat shifted towards the higher values of the count rate parameter
relative to the first distribution. If true, this would imply that the burst intensity
affects the detection of oscillations.  In order to quantify the
difference between the two distributions, we perform a Kolmogorov-Smirnov (K-S) test 
between them. We obtain a maximum deviation of 
0.243 between the two distributions with a significance level of $6.9\times10^{-4}$, 
i.e, the probability that the two distributions are same is small ($6.9\times10^{-4}$).
This implies that the detection of oscillations depends, not only
on the fractional amplitude value as shown earlier, but also 
on the count rate parameter, and hence on the burst intensity.

\subsection{Fractional amplitude evolution for 4U 1636--536}\label{Rms}

We compute the fractional rms amplitude evolution of burst rise 
oscillations of all the 27 {\it RXTE} PCA bursts from 4U 1636--536 (mentioned in Table~\ref{Log}).
The fractional amplitudes are plotted with time 
in Fig.~\ref{rmsampevol}. This figure indicates that, in spite of the 
oscillations being usually intermittent and short-lived, the fractional amplitude
appears to decrease with time. Moreover, the fractional amplitude evolution seems 
to have a concave shape, i.e., the amplitude initially decreases fast and then does not
change much. In order to verify these, we fit the fractional amplitude curves
with an empirical model of the form 
\begin{eqnarray}
\phi = a-bc(1-e^{-t/c}),
\label{model}
\end{eqnarray}
with the parameters $a>0$, $b>0$ and $c$. This is a convenient function because
constraints on $a$ and $b$ ensure that it monotonically decreases with time $t$.
A positive value of $c$ implies a concave fractional rms amplitude versus time curve,
while a negative value implies a convex curve.
 
For two bursts (16 and 18) the rise oscillations are detected
for relatively long durations, i.e., in minimum five time bins, including at least 
one of the first two bins (see Fig.~\ref{rmsampevol}).
We fit the time evolution of fractional rms amplitude of detected points (excluding 
the upper limits) with a constant model and the $\phi$-model (Eqn.~\ref{model}) 
for these bursts. The $\chi^2/{\rm d.o.f.}$ values obtained for the constant 
and the $\phi$-model are $22.5/8$ and $1.0/6$ respectively
for the burst 16, and $19.4/6$ and $4.0/4$ respectively for the burst 18.
Then we perform F-tests between the two models for these two bursts, and 
find that the $\phi$-model is better than the constant model with a significance
of $\approx 4\sigma$ and $\approx 2\sigma$ for burst 16 and 18 respectively.
Therefore, it may be inferred that the
fractional amplitude decreases with time with a significance of $\approx 99.99$\% 
and $\approx 95.45$\% for these two bursts.
The best-fit values of the $c$-parameter are $0.46\pm0.21$ and $0.42\pm0.19$
for burst numbers 16 and 18 respectively, which imply concave-shaped
fractional amplitude versus time curves (see Fig~\ref{convexpar}).

In order to check if the $c$-parameter is positive when the upper limits are considered
for fitting, we fit the fractional amplitude versus time curves, including
both the upper limits and the detected points, with the $\phi$-model.
The best-fit parameter values are obtained by maximizing the likelihood function modified for censored
data (i.e, data comprising of upper limits; \citet{Feigelson1985, Isobe1986,
Wolynetz1979a, Wolynetz1979b}). This method can be effectively employed if the number
of detected points is not much lesser than the number of censored points
\citep{Isobe1986}. With this constraint, we can apply this technique
for modelling the fractional amplitude evolution of seven bursts, and estimate
the best-fit parameter values. Examples of the best-fit model curves
are shown in panels 5 and 16 of Fig.~\ref{rmsampevol} (green dash-dot curves).
The best-fit $c$-parameter values are plotted in the left panel of Fig.~\ref{convexpar}.
This panel shows that all the seven best-fit $c$ values
are positive. Moreover, the best-fit values of $c$ from fits with and without upper limits
are consistent with each other for bursts 16 and 18 (left panel of Fig.~\ref{convexpar}).

Since the above mentioned likelihood maximization technique cannot be applied 
for all the 27 bursts (Table~\ref{Log}), we fit all the fractional amplitude versus time curves
(including upper limits) with the $\phi$-model using the weighted least
square (a generalized form of $\chi^2$; \citet{Feigelson2013}) minimization method.
The weighted least square $L^2$ is defined as $\sum \frac{1}{\sigma_i^2} (R_i-R_{mi})^2$,
where $R_i$ is the observed fractional rms amplitude, $R_{mi}$ is the model fractional rms
amplitude and $\sigma_i$ is either an assumed error for an upper limit or the
measured error for a detected point. For an upper limit, we consider an asymmetric error,
with the upper error weighted with a very small factor (to ensure that the model curve usually remains 
below the upper limit) and the lower error weighted with a sufficiently large factor.
Since such an error is not Gaussian, we minimize a more general $L^2$ instead of a $\chi^2$.
The panels 5 and 16 of Fig.~\ref{rmsampevol} show the examples of the best-fit model curves
(red dashed curves). These curves are similar (typically within observational
$1\sigma$ error bars) to the best-fit model curves from the
likelihood maximization method mentioned above. The best-fit $c$-parameter values are plotted 
in the right panel of Fig.~\ref{convexpar}. This figure shows that all the best-fit $c$ values
are positive, and are consistent with the values from the likelihood maximization method.
Moreover, the best-fit values of $c$ from fits with and without upper limits
are consistent with each other for bursts 16 and 18 (right panel of Fig.~\ref{convexpar}). 

Finally, we check if the change of persistent emission during a burst 
can qualitatively affect the fractional amplitude evolution. We calculate
the amplitudes for all 0.25 s time bins of all 27 bursts (Table~\ref{Log}),
as described in \S~\ref{fa}. We find that, although the absolute values of
fractional amplitudes are somewhat affected by $f_a$ (the amplitude values 
typically change by 0.9 times the data error calculated 
considering $f_a$), their relative values for a burst
are not largely affected. This means that the shapes of the fractional rms amplitude
versus time curves do not qualitatively change when the effects of persistent emission
evolution is considered. We give examples for two bursts in Fig.~\ref{rmsampevolfa}.

\subsection{Fractional amplitude evolution for sources other than 4U 1636--536}\label{othersources}

We perform a similar analysis, as mentioned in \S~\ref{Rms}, for the 24 bursts with rise
oscillations from nine neutron star LMXBs listed in Table~\ref{Log2}.
Oscillations during the rise were detected in 1 bin for 13 bursts, in 2 bins for 7 bursts and in 3 bins for 2 bursts and 4 bins for 2 bursts. 
Fig.~\ref{rmsampevol2} shows the fractional oscillation amplitude versus time curves for
the rising phases of these bursts. These curves suggest that the amplitude decreases
with time. Moreover the amplitude profiles appear to have concave shapes. To verify these,
we fit the these curves with the $\phi$-function of Eqn.~\ref{model}. 
Since for each of these
24 bursts the number of time bins with detected rise oscillations is small,
we can use only the weighted Least-square minimization method (see \S~\ref{Rms}) to find 
the best-fit parameter values. The best-fit $c$-parameter values cluster on the right of the
$c=0$ line (Fig. \ref{convexpar2}), implying that the fractional amplitude profiles are of 
concave shapes. Thus the results for these nine sources are similar to those for 4U 1636--536
(\S~\ref{Rms}).

\subsection{Burst groups for 4U 1636--536}\label{morphology}

As mentioned in \S~\ref{Introduction}, \citet{Maurer2008} reported that the
burst rise morphology is affected by burning regime, ignition latitude and
thermonuclear flame spreading. They divided the bursts from 4U 1636--536 into 
three groups based on their peak fluxes and rising light curves,
and discussed in detail how burning regime and ignition latitude, and hence
the flame spreading parameters, are different in different burst groups.
If this interpretation is true and if flame spreading causes the evolution of 
burst rise oscillations, then some properties of such evolution should also
be different in different burst groups. We plan to test this.

To begin with, we need to identify which of the 27 bursts with rise oscillations
belongs to which group. Since \citet{Maurer2008} did not mention which 
group a given burst belongs to, we calculate the peak flux and the burst rise 
morphology (i.e., $\mathcal{C}$-parameter; see \S~\ref{morphology_characterization}) 
for each of 27 bursts mentioned in Table~\ref{Log}.
Following the definition of \citet{Maurer2008}, these bursts are
categorized into three groups: Group 1 (burst peak flux
$<50\times10^{-9}$ erg s$^{-1}$ cm$^{-2}$; $\mathcal{C}<0$), Group 2 (burst peak flux
$<50\times10^{-9}$ erg s$^{-1}$ cm$^{-2}$; $\mathcal{C}>0$) and Group 3 (burst peak flux
$>50\times10^{-9}$ erg s$^{-1}$ cm$^{-2}$). We find 4, 6 and 17 bursts belonging to Groups
1, 2 and 3 respectively. 
The mean rise times of bursts from Groups 1, 2 and 3 are 3.66 s, 2.75 s and 1.76 s respectively.
We divide the bursts from only 4U 1636--536 in groups, because sufficient
number of bursts with rise oscillations is not available for any of the other nine sources
(Table~\ref{Log2}).

Now we try to find out if the distribution of burst rise time bins with
detected oscillations are different in different burst groups. 
In the upper panel of Fig.~\ref{detectiowithgroup}, 
for each burst group, we plot the fraction ($X_{\rm n,osc}$) of burst rise time bins (each 0.25 s) 
with detected oscillations versus time bin index. 
This fraction for a given bin index indicates the probability (among bursts with
rise oscillations) of having detected oscillations, or the likely strength of oscillations
for that bin index for a particular burst group. 
So overall this panel gives an idea about the typical and maximum durations of oscillations
during burst rise for each group.
More specifically, this panel indicates that oscillations can exist for at most
$\sim 2.25$ s for Group 1 bursts, $\sim 2.75$ s for Group 2 bursts,
and $\sim 1.25$ s for Group 3 bursts, measured from the burst onset.
The fraction ($X_{\rm n,osc}$) fluctuates substantially for Group 1, may be partially
due to a small (4) number of available bursts (see the lower panel of Fig.~\ref{detectiowithgroup}). 
However, $X_{\rm n,osc}$ shows a weak but overall increasing trend, which indicates
that the oscillation becomes stronger in the later part of burst rise for this group.
For the Group 2, $X_{\rm n,osc}$ is zero in the first time bin (may be because of the lowest
count rates among all bins of all groups), then sharply goes to a high value, and finally
gradually decreases, indicating a gradual decrease of the oscillation strength.
For the Group 3, $X_{\rm n,osc}$ slightly increases first from a non-zero value,
and then relatively quickly decreases, indicating a quick decrease of the oscillation strength.
The lower panel of Fig.~\ref{detectiowithgroup} shows the available total number of burst rise
time bins versus the time bin index. This panel essentially reveals typical and maximum durations 
of burst rise for each burst group.

The above mentioned maximum duration of oscillations for each burst group 
(i.e., $\sim 2.25$ s, $\sim 2.75$ s and $\sim 1.25$ s; see Fig.~\ref{detectiowithgroup}) depends on
(1) the burst rise durations of that group, and (2) the fraction of rise time the 
oscillations persist for the bursts of that group. In order to disentangle these two causes,
and to find out what fraction of rise time the oscillations persist for, we 
normalize the rise time of each burst to 1, and then replot the upper 
panel of Fig.~\ref{detectiowithgroup} in Fig.~\ref{detectiowithgroupscaled}
in a way mentioned in the caption of Fig.~\ref{detectiowithgroupscaled}.
This figure shows that for Group 3, the oscillations persist at most 
60\% of the rise time, while for the other two groups, the oscillations
can survive for the entire durations of burst rise for some bursts.
Finally, we note that Figs.~\ref{detectiowithgroup} and \ref{detectiowithgroupscaled}
show that the evolution of burst rise oscillations is different in different
burst groups, as expected (see the first paragraph of this subsection).

\section{Discussion}\label{Discussions}

In this paper, we investigate if the thermonuclear flame spreading gives rise
to oscillations during the rising phases of bursts from several neutron star LMXBs, 
and if so, then what the implications of these oscillations are.
In \S~\ref{non-detection}, we discuss the roles of low fractional amplitude and 
low count rate for non-detection and intermittent detections of burst rise oscillations.
In \S~\ref{evolution}, we argue that the observed fractional amplitude evolution
during burst rise is consistent with thermonuclear flame spreading, and the flame
speed is possibly latitude-dependent.
In \S~\ref{groups}, we mention three remaining puzzles regarding the flame spreading origin 
of burst rise oscillations, and discuss them in the context of three burst 
groups proposed by \citet{Maurer2008}.
Finally in \S~\ref{implications}, we detail the implications of our results.

\subsection{Intermittent- and non-detection of burst oscillations}\label{non-detection}

One of the most puzzling aspects of burst rise oscillations is that it is not 
detected from most bursts, and even when it is found from a burst, it is 
usually detected intermittently and for a fraction of the rise time. A simple
reason for this could be the low count rates or intensities (see \S~\ref{Introduction}).
In \S~\ref{amplitude_calculation}, we show that the detection of the burst rise oscillations depends on 
two factors: the fractional rms amplitude and the count rate parameter.
Fig.~\ref{countbindetection} clearly shows
that the red-square detection points spread over a large range of count rate parameter values 
(see \S~\ref{Detection}) for the individual 0.25 s time bins. Moreover, these detection points
do not cluster at the higher values of the count rate parameter.
These mean that the low count rates cannot entirely explain 
the non-detections of oscillations, and low fractional amplitudes must also be 
a reason for such non-detections. Note that this conclusion
remains valid, even when the change of persistent emission during bursts is considered
(see Fig.~\ref{countbindetectionfa}). This is also supported by Fig.~\ref{detectionfractionbin},
which shows that the fraction of bursts with detection of oscillations overall decreases with
the time bin index, even though the intensity (and hence the count rate parameter)
is expected to increase with the bin index. However, as expected, intensity also
affects the detection. This can be inferred from the increase
and decrease of the ratio of Fig.~\ref{detectionfractionbin} in a shorter time scale, 
as well as from Fig.~\ref{KS} and the corresponding K-S test described in \S~\ref{Detection}.

\subsection{Fractional amplitude evolution and flame spreading}\label{evolution}

Now the question is, if low fractional amplitudes are an important reason
for non-detection and intermittent detections 
of oscillations, can burst rise oscillations originate
from thermonuclear flame spreading? Before addressing this question (see
\S~\ref{groups}), we attempt to find out if the fractional amplitude evolution
for 27 bursts from 4U 1636--536 (Table~\ref{Log}) and 24 bursts from the 
other nine sources (Table~\ref{Log2}) support the flame spreading scenario.
A visual examination of Fig.~\ref{rmsampevol} and Fig.~\ref{rmsampevol2} 
clearly shows that the fractional amplitude evolution for each burst is consistent with
decreasing with time. As mentioned in \S~\ref{Rms},
a quantitative measurement for the best case (burst 16 of 4U 1636--536) reveals that the 
amplitude decreases with time with a significance of $\approx 99.99$\%. 
Furthermore, the minimum count rate parameter value at which the oscillation is 
detected increases with the time bin index (Fig.~\ref{countbindetection}; see also
Fig.~\ref{countbindetectionfa} for a weaker trend). From Eqn.~\ref{eqncrp},
this implies that the fractional rms amplitude decreases with time.
As discussed in \S~\ref{Introduction}, such a decrease of fractional amplitude
strongly suggests that the thermonuclear flame spreading gives rise
to burst rise oscillations in these sources.

Figs.~\ref{convexpar} and \ref{convexpar2} 
show the best-fit $c$-parameter values of the empirical model given in
Eqn.~\ref{model}. All these values for 4U 1636--536 from three different methods (see \S~\ref{Rms})
clearly cluster on the positive side of the $c = 0$ line (Fig~\ref{convexpar}).
Similarly, all the best-fit $c$ values for nine other sources also 
cluster on the positive side of the $c = 0$ line (Fig~\ref{convexpar2}).
This strongly suggests that the fractional rms amplitude versus time curves are
concave, i.e., at first the fractional amplitude decreases fast, and then it 
does not change much. Note that the same suggestion was made by \citet{Bhattacharyya2007:3},
but for only two bursts, that too without any quantification. Therefore, the current paper
reports a significant progress, because it confirms the previous result with a much larger sample
size (51 bursts from ten sources), and a quantification with an empirical model. 
\citet{Bhattacharyya2007:3}
numerically showed that a convex shape of a fractional amplitude versus time curve
implies an isotropic flame spreading, while a concave shape could be a result of the 
latitude-dependent flame speed, likely due to the 
Coriolis force (see also \S~\ref{Introduction}). Therefore, our detailed analysis
tentatively reveals the possible effects of the Coriolis force on flame spreading, as expected 
for rapidly spinning neutron stars, such as the ten sources 
considered in this paper \citep{Spitkovsky2002}.
Finally, we note that the change of persistent emission during 
a burst does not affect our conclusions (\S~\ref{Rms}). 

\subsection{Flame spreading in three burst groups}\label{groups}

Although the fractional amplitude evolution reported in \S~\ref{Rms} 
and \S~\ref{othersources} supports the 
flame spreading origin of burst rise oscillations, the following puzzles remain.
(1) Why is the oscillation intermittently detected for a given burst? 
In the flame spreading scenario, if the fractional amplitude decreases monotonically,
how can the oscillation be not detected in a former time bin, but be detected in a later time bin?
(2) Why is the oscillation not detected, i.e., fractional amplitudes are quite low for most bursts?
What kind of flame spreading process can explain this?
(3) What determines the duration of detected oscillations during the rise of a given burst?

One may attempt to understand the puzzle 1 in the following ways.
(a) The fractional amplitude may not always decrease monotonically. The increase of
the burning area causes the decrease of amplitude, but the change of the polar angle of the 
burning region center (as the flames spread) may sometimes increase the fractional amplitude due to
the viewing geometry (see more discussion later regarding the Group 1 bursts). So if the 
fractional amplitude increases by the net effect, the oscillation may be detected 
in a time bin after a non-detection in the previous bin.
(b) Eqn.~\ref{eqncrp} implies that the detection of oscillations depends on 
fractional amplitude and burst intensity. Hence, if the intensity does not increase
sufficiently as the amplitude decreases in a time bin, the oscillation may not be 
detected in that bin, but may be detected in a later bin with sufficiently higher
intensity. 

In order to address the puzzles 2 and 3, we need to discuss the burst oscillations
together with various burst groups. As mentioned in \S~\ref{morphology},
the maximum duration of oscillations for each burst group depends on
(1) the durations of burst rise in that group, and (2) the fractions of burst rise time 
with oscillations in that group. The former may be primarily
determined by the time scale of flame spreading and fuel burning, while the latter may be
determined by the fraction of rise time in which an azimuthal asymmetry caused
by the burning region persists. After disentangling the above two causes (\S~\ref{morphology}),
we find that, for Group 3, the visible portion of the burning region becomes
azimuthally symmetric much before the flame spreading is over.
But for Groups 1 and 2, the burning region can remain somewhat azimuthally 
asymmetric till the end of the flame spreading.
We now examine if these findings are consistent with the conclusions
of \citet{Maurer2008}, and attempt to address the puzzles 2 and 3 mentioned above.

\citet{Maurer2008} concluded that the Group 3 bursts are He bursts with 
equatorial ignition. The He burst interpretation is consistent with the shorter rise times of the 
Group 3 bursts (see the lower panel of Fig.~\ref{detectiowithgroup}; also
mentioned by \citet{Maurer2008}). According to \citet{Spitkovsky2002},
an exactly equatorial ignition will make the burning region azimuthally
symmetric very quickly, and hence no burst rise oscillation is 
expected. This may explain why rise oscillations are not detected for more than half 
(64\%) of the bursts of Group 3 \citep{Maurer2008}. If the ignition of some of the Group 3 bursts
are off (but near) equatorial, then we expect initial oscillations, which
may disappear soon. This is because, since the flame
speeds at lower latitudes are more than those at higher latitudes \citep{Spitkovsky2002},
the azimuthal asymmetry of the burning region may disappear soon (much before the flame spreading
is over) by spreading in the direction of latitudes. This is consistent 
with the above mentioned observation (Fig.~\ref{detectiowithgroupscaled}) 
that the oscillations exist during a fraction ($\le 60$\%) of the rise time.
Note that a slightly off-equatorial ignition could happen because of an off-equatorial
higher temperature (e.g., by random fluctuation or due to the effects of previous bursts).

According to \citet{Maurer2008}, the bursts of Groups 1 and 2 are mixed H/He bursts
with off-equatorial H ignition. These authors further concluded that the Group 1
bursts are ignited near the north pole while the Group 2 bursts are ignited near the
south pole, considering that the observer's line of sight passes through the northern 
hemisphere. The longer rise times of the bursts from Groups 1 and 2 (see the lower panel 
of Fig.~\ref{detectiowithgroup}; also mentioned by \citet{Maurer2008}) are consistent
with the mixed H/He burst interpretation. 

But are our results consistent with the Group 2 bursts being ignited near the south pole?
In case of such ignitions of
the Group 2 bursts, we expect the azimuthal asymmetry of burning region (near the south 
pole) to survive for quite some time, because of the low flame speed near the
pole \citep{Spitkovsky2002}. During this time, the northern burning front should 
propagate fast towards the equator, and become azimuthally symmetric. So the
asymmetry at the southern front should be the origin of the oscillations. 
These expectations are supported by the observed
rise oscillations with larger durations for Group 2 bursts (Fig.~\ref{detectiowithgroupscaled}).
However we note that for a near-south-pole ignition, 
the entire southern hemisphere should be ignited and the southern 
azimuthal asymmetry should disappear much before the
northern burning front reaches the north pole. 
But we find that the burning region can remain somewhat azimuthally
asymmetric till the end of the flame spreading for Group 2 bursts (see above; also
Fig.~\ref{detectiowithgroupscaled}). 
One possible explanation to this is that the northern burning front
stalls at the equator, as suggested by \citet{Cavecchi2013}.
Another question is why rise oscillations are not detected for most (77\%) of the
bursts of Group 2 \citep{Maurer2008}. We note that,
if the ignition happens too close to the south pole, the
southern azimuthal asymmetry (if formed) may be out of sight, and hence no
oscillations will be observed. 

Now we examine if our results are consistent with the 
Group 1 bursts being ignited near the north-pole.
In case of such ignitions of the Group 1 bursts, the fractional amplitude
is expected to be initially small because of the near-face-on viewing geometry, unless
the the observer's line of sight is far from the stellar spin axis. As the
burning region expands towards the equator, the fractional amplitude may decrease
because of the increasing burning area, but may increase due to the
changed viewing geometry (see above). The latter is because, as the center of the 
burning region moves towards the equator, a larger fraction of this region 
periodically goes out of the view, increasing the fractional amplitude.
A competition between the above two effects might partially cause the fluctuations of the 
ratios $X_{\rm n,osc}$ and $X'_{\rm n,osc}$, as we find (see Figs.~\ref{detectiowithgroup}
and \ref{detectiowithgroupscaled}).
Another question is why rise oscillations are not detected for most (93\%) 
of the bursts of Group 1 \citep{Maurer2008}. We note that,
if the ignition happens too close to the north pole,
a significant azimuthal asymmetry may not be formed (depending on the viewing geometry), 
and hence no oscillations will be observed.

Therefore, our findings based on the evolution of burst rise
oscillation amplitude are broadly consistent with the conclusions of 
\citet{Maurer2008} based on an independent method of burst rise morphology, if we consider the 
latitude-dependent flame speeds suggested by \citet{Spitkovsky2002}. Such comparison
provides a required sanity check and gives confidence to the flame spreading
model of burst rise oscillations. Moreover, our study of the evolution of burst rise
oscillations for various burst groups (\S~\ref{morphology}) helps to probe
the reasons of the absence of rise oscillations in most bursts (puzzle 2 mentioned above)
and what determines the durations of burst rise oscillations (puzzle 3 mentioned above).

\subsection{Implications of results}\label{implications}

Now we briefly discuss the implications of our finding, that the burst rise oscillations
are likely to be caused by the thermonuclear flame spreading possibly 
under the influence of the Coriolis force. As we discuss below, our finding 
will be very important (1) to probe the physics of thermonuclear bursts and 
flame spreading, and (2) to make the burst rise oscillation method 
\citep{Bhattacharyya2010} to measure the neutron star parameters more reliable.

(1) Even without going into the details, our finding implies that flames cover the 
entire surface of the neutron star in $\sim 2.5$ s. Note that, if the burst rise oscillation amplitude evolutions
were not due to the flame spreading, then that might imply the flames covering the
neutron star surface within the first time bin, i.e., in 0.25 s (see \S~\ref{Introduction}).
This may have important implications on the flame spreading conditions like viscosity.
For example, for a near-polar ignition on a typical neutron star of 10 km radius, if the
flame spreads within 0.25 s to cover the stellar surface, a rough estimate gives
a flame speed of $\sim 120$ km s$^{-1}$. On the other hand, a rough estimate of
flame speed from a spreading time scale of 2.5 s is $\sim 12$ km s$^{-1}$.
Now, as mentioned in \S~\ref{Introduction} and \citet{Spitkovsky2002}, for a typical
neutron star of mass 1.4 M$_{\odot}$, radius 10 km, spin frequency 582 Hz, and
considering $h$ and $t_{\rm n}$ to be 10 m and 0.1 s respectively, the values of
the flame speed are $\sim 6$ km s$^{-1}$ and $\sim 160$ km s$^{-1}$ assuming 
a weak and a maximum turbulent viscosity respectively. Therefore, our results
support a weak turbulent viscosity for flame spreading. More burst parameters
can be constrained by future observations with the LAXPC instrument of {\it Astrosat} \citep{Agrawal2006}, 
and more effectively with a next generation X-ray mission like {\it LOFT} \citep{DelMonte2012, Mignani2012}.

(2) If burst oscillations originate from one or two hot spots on the neutron star surface,
then such oscillations can be used as a tool to measure mass and radius of neutron stars,
and hence to constrain the stellar equation of state models (\citet{Bhattacharyya2010, Lo2013};
also \S~\ref{Introduction}). Therefore, it is very important to verify if burst oscillations 
originate from hot spots, because constraining equation of state models is 
a fundamental problem of physics and an important science goal 
for future X-ray missions (e.g., {\it NICER}, {\it LOFT}; see \citet{Lo2013}).
Our finding suggests that burst rise oscillations originate from a hot spot (albeit
expanding due to flame spreading), and hence is important to make burst oscillations 
a more reliable tool.

\section{Summary}\label{Summary}

Here we summarize the key points of this paper.\\
(1) We study the evolution of fractional amplitude of burst rise oscillations for 51 bursts from
ten neutron star LMXBs. Previously, this was reported only for a few bursts
(see \S~\ref{Introduction}).\\
(2) Our detection criterion of burst rise oscillations in 0.25 s time bins is quite
stringent ($3\sigma$; \S~\ref{amplitude_calculation}), considering the small
numbers of total counts in such short time bins.
This contributes to the reliability of our conclusions.\\
(3) With detailed simulations, we find that the contribution of the 
systematic error due to the non-constant intensity within a 0.25 s burst
time bin to the total fractional amplitude error is less than 0.4\% 
for our typical data count rates (see \S~\ref{amplitude_calculation}).\\
(4) We study the roles of low fractional amplitude and low count rate for 
non-detection and intermittent detections of rise oscillations in many bursts.\\
(5) We find a decreasing trend of burst rise oscillation amplitude with time. This
is consistent with thermonuclear flame spreading. To the best of our knowledge,
we, for the first time, quantify this trend by fitting with empirical models and by
doing F-tests (\S~\ref{Rms} and \S~\ref{othersources}).
We find that the decrease of amplitude is $4\sigma$ significant for the 
best case (\S~\ref{Rms}). Moreover, it is important to note that an opposite 
trend is not found from any of the 51 bursts.\\
(6) From our fits of the burst rise oscillation amplitude versus time, we find not only 
decreasing trends, but also concave-shaped profiles for all the 51 bursts (\S~\ref{Rms} and
\S~\ref{othersources}; Figs.~\ref{convexpar} and \ref{convexpar2}). 
This implies latitude-dependent flame speeds, possibly
due to the effects of the Coriolis force (see \S~\ref{Introduction}).\\
(7) We, for the first time, explore and find that the shapes of oscillation 
amplitude profiles with and without persistent emission variation during
burst rise are consistent with each other (see \S~\ref{Rms}).\\
(8) \citet{Maurer2008} reported that the burst rise morphology can be used to
probe burning regime, ignition latitude and thermonuclear flame spreading
(see \S~\ref{Introduction} and \S~\ref{morphology}). Our findings based on
an independent method of the evolution of burst rise oscillation amplitude
are broadly consistent with their interpretation (\S~\ref{Discussions}). This
comparison provides a sanity check for our conclusions.\\
(9) Our findings suggest a weak turbulent viscosity scenario for flame spreading
(\S~\ref{Discussions}).\\
(10) Our findings imply that burst rise oscillations originate from an
expanding hot spot, thus making burst oscillations a more reliable tool
to constrain the neutron star equation of state models (\S~\ref{Discussions}).

\acknowledgments

We thank an anonymous referee for constructive comments which improved the paper.

{}

\clearpage
\begin{table*}
\small
\centering
\caption{Log table of 27 thermonuclear X-ray bursts with detected burst rise oscillations from 
the neutron star LMXB 4U 1636--536 (stellar spin frequency $\approx$ 582 Hz) 
observed with \textit{RXTE} PCA (see \S~\ref{Data}). \label{Log}}
\begin{tabular}{ccccc}
\hline
Burst ID\footnotemark[1] & Burst ID\footnotemark[2] & Date & Time\footnotemark[3] & ObsID \\
\hline
1  &  1  &  1996-Dec-28  &  22:39:24  &  10088-01-07-02 \\
2  &  2  &  1996-Dec-28  &  23:54:04  &  10088-01-07-02 \\
3  &  6  &  1998-Aug-19  &  11:44:39  &  30053-02-02-02 \\
4  &  8  &  1998-Aug-20  &  05:14:12  &  30053-02-02-00 \\
5  &  9  &  1999-Feb-27  &  08:47:29  &  40028-01-02-00 \\
6  &  13  &  1999-Jun-18  &  23:43:04  &  40028-01-08-00 \\
7  &  15  &  1999-Jun-21  &  19:05:53  &  40031-01-01-06 \\
8  &  16  &  1999-Sep-25  &  20:40:49  &  40028-01-10-00 \\
9  &  24  &  2000-Oct-03  &  23:32:48  &  40028-01-20-00 \\
10  &  25  &  2000-Nov-05  &  04:21:59  &  50030-02-01-00 \\
11  &  26  &  2000-Nov-12  &  18:02:28  &  50030-02-02-00 \\
12  &  31  &  2001-Apr-30  &  05:28:34  &  50030-02-10-00 \\
13  &  37  &  2001-Aug-23  &  00:50:33  &  60032-01-04-04 \\
14  &  45  &  2001-Sep-30  &  14:47:17  &  60032-01-12-00 \\
15  &  61  &  2002-Jan-09  &  00:26:38  &  60032-01-20-00 \\
16  &  75  &  2002-Jan-12  &  21:35:34  &  60032-05-03-00 \\
17  &  77  &  2002-Jan-13  &  01:29:03  &  60032-05-03-00 \\
18  &  84  &  2002-Jan-14  &  01:22:36  &  60032-05-05-00 \\
19  &  102  &  2002-Jan-22  &  07:07:20  &  60032-05-10-00 \\
20  &  109  &  2002-Jan-30  &  23:06:55  &  60032-05-12-00 \\
21  &  110  &  2002-Feb-05  &  22:21:51  &  60032-05-13-00 \\
22  &  111  &  2002-Feb-11  &  17:35:07  &  60032-05-14-00 \\
23  &  115  &  2002-Apr-26  &  05:07:18  &  60032-05-18-00 \\
24  &  127  &  2005-Mar-23  &  05:27:58  &  91024-01-10-00 \\
25  &  138  &  2005-Jun-11  &  02:42:04  &  91024-01-50-01 \\
26  &  148  &  2005-Aug-10  &  05:36:36  &  91024-01-80-00 \\
27  &  150  &  2005-Aug-16  &  01:45:36  &  91024-01-83-00 \\
\hline 
\end{tabular}
\begin{flushleft}
\begin{footnotesize}
$^1$ Burst ID used in this paper.\\
$^2$ Corresponding burst indices from \citet{Galloway2008}.\\
$^3$ Start time of burst in UT.
\end{footnotesize}
\end{flushleft}
\end{table*}

\clearpage
\begin{table*}
\small
\centering
\caption{Log table of 24 thermonuclear X-ray bursts with detected burst rise oscillations 
from nine neutron star LMXBs observed with \textit{RXTE} PCA (see \S~\ref{Data}). \label{Log2}}
\begin{tabular}{ccccccc}
\hline
Source name & Frequency (Hz)\footnotemark[1] & Burst ID\footnotemark[2] & Burst ID\footnotemark[3] & Date & Time\footnotemark[4] & ObsID \\
\hline
4U 1608--52 & 620 
&	1	&	8	&	1998 Apr 11 &	06:35:31	&	30062-01-02-05 \\
&&	2	&	10	&	2000 Mar 11 &	01:42:36	&	50052-01-04-00 \\	
&&	3	&	21	&	2002 Sep 7 &	02:26:15	&	70059-01-20-00 \\
&&	4	&	22	&	2002 Sep 9 &	03:50:29	&	70059-01-21-00 \\
\hline
MXB 1659--298 & 567 
& 	1	&	2	&	1999 Apr 9 &	14:47:34	&	40050-04-01-00 \\
\hline
4U 1702--429 & 329 
&	1	&	4	&	1997 Jul 26 &	14:04:18 	& 	20084-02-02-00 \\ 
&&	2	&	8	&	1999 Feb 22 &	04:56:05 	& 	40025-04-01-01 \\ 
&&	3	&	9	&	2000 Jun 22 &	11:57:46 	&	50030-01-01-04 \\ 
&&	4	&	15	&	2001 Apr 1 &	15:47:17 	& 	50030-01-11-00 \\ 
&&	5	&	24	&	2004 Feb 29 &	 06:32:16 	& 	80033-01-04-02 \\ 
&&	6	&	26	&	2004 Mar 1 &	23:26:40 	&	80033-01-05-01 \\ 
\hline 
4U 1728--34 & 363 
&	1	&	14	&	1997 Sep 19 &	12:32:58& 	20083-01-01-01 \\
&&	2	&	15	&	1997 Sep 20 &	10:08:52	&	20083-01-01-02 \\
&&	3	&	16	&	1997 Sep 21 &	15:45:31	&	20083-01-02-01 \\
&&	4	&	17	&	1997 Sep 21 &	18:11:07	&	20083-01-02-01 \\
&&	5	&	64	&	1999 Aug 19 &	09:33:48	&	40019-03-02-00 \\
&&	6	&	68	&	1999 Aug 20 &	05:54:45	&	40019-03-03-00 \\
&&	7	&	96	&	2001 Oct 27 &	23:53:44	&	50030-03-09-01 \\
\hline 
KS 1731--260 & 524 
& 	1	&	7	&	1999 Feb 23 &	03:09:01	&	40409-01-01-00 \\
&&  2	&	8	&	1999 Feb 26 &	17:13:09	&	30061-01-04-00 \\
\hline
1A 1744--361 & 530 
&	1	&	1	&	2005 Jul 16 &	22:39:56	&	91050-05-01-00 \\
\hline
SAX J1750.8--2900 & 601 
&	1	&	2	&	2001 Apr 12 &	14:20:31	&	60035-01-02-02\\
\hline
4U 1916--053 & 270 
&	1	&	9	&	1998 Aug 1 &	18:23:49 	& 	30066-01-03-03 \\
\hline
Aql X--1 & 549 
&	1	&	24	&	2001 Jul 1 &	14:18:37	&	60054-02-02-01 \\
\hline
\end{tabular}
\begin{flushleft}
\begin{footnotesize}
$^1$ Spin frequency of the sources in Hz. \\
$^2$ Burst ID used in this paper.\\
$^3$ Corresponding indices from \citet{Galloway2008}.\\
$^4$ Start time of burst in UT.
\end{footnotesize}
\end{flushleft}
\end{table*}

\clearpage
\begin{figure*}
\centering
\begin{tabular}{lr}
\includegraphics[width=0.32\textheight]{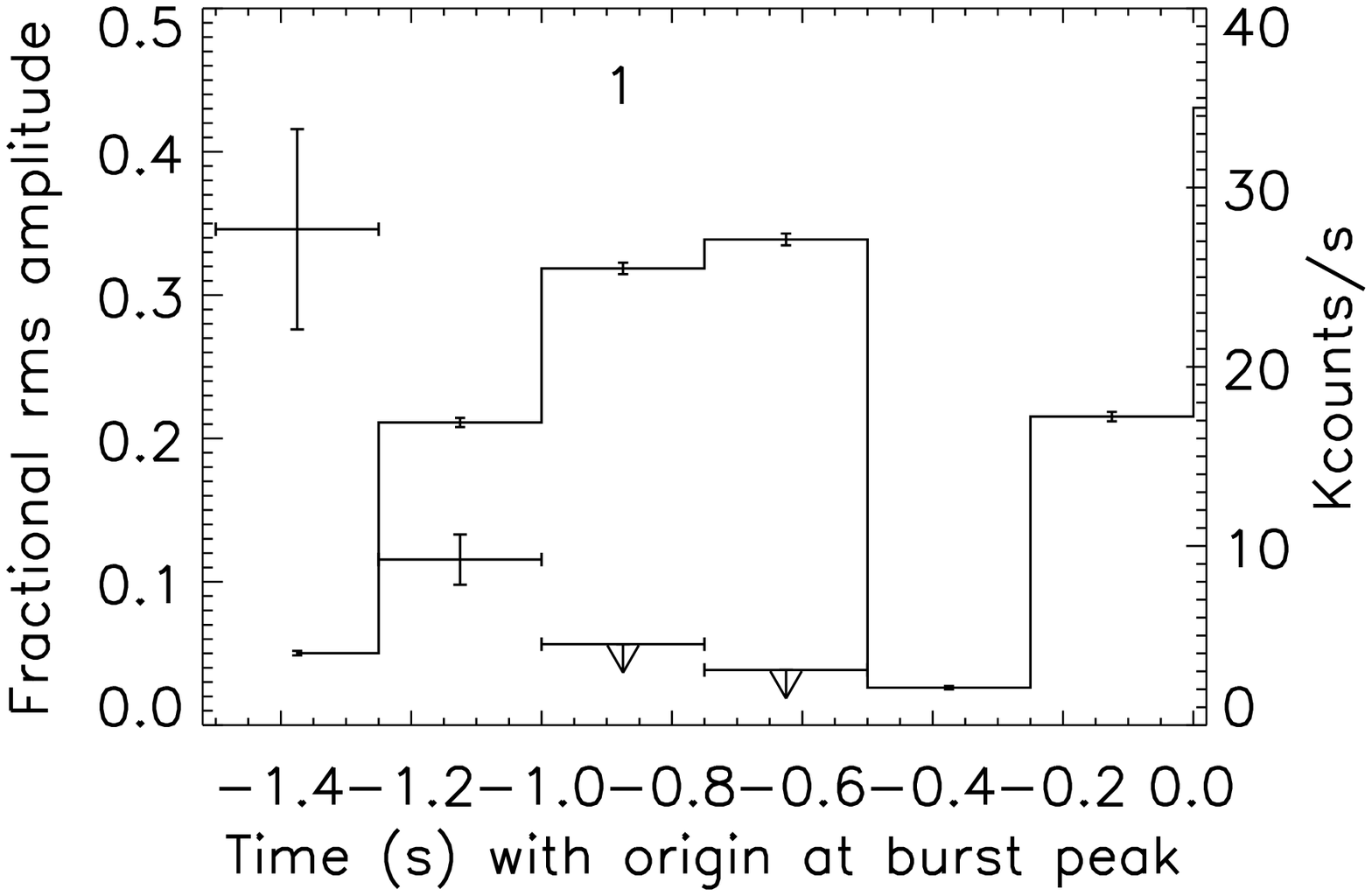} &
\includegraphics[width=0.32\textheight]{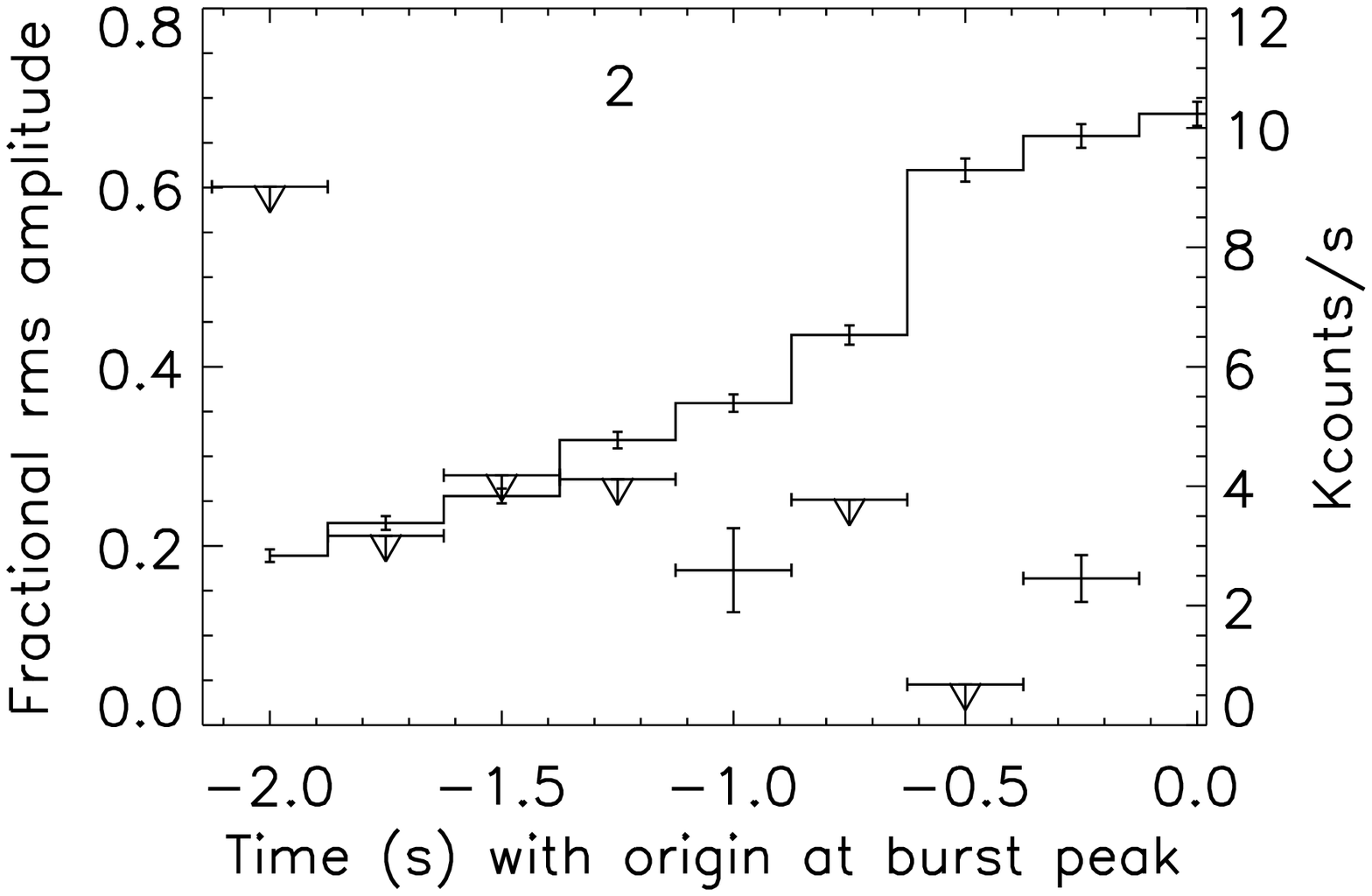} \\ 
\includegraphics[width=0.32\textheight]{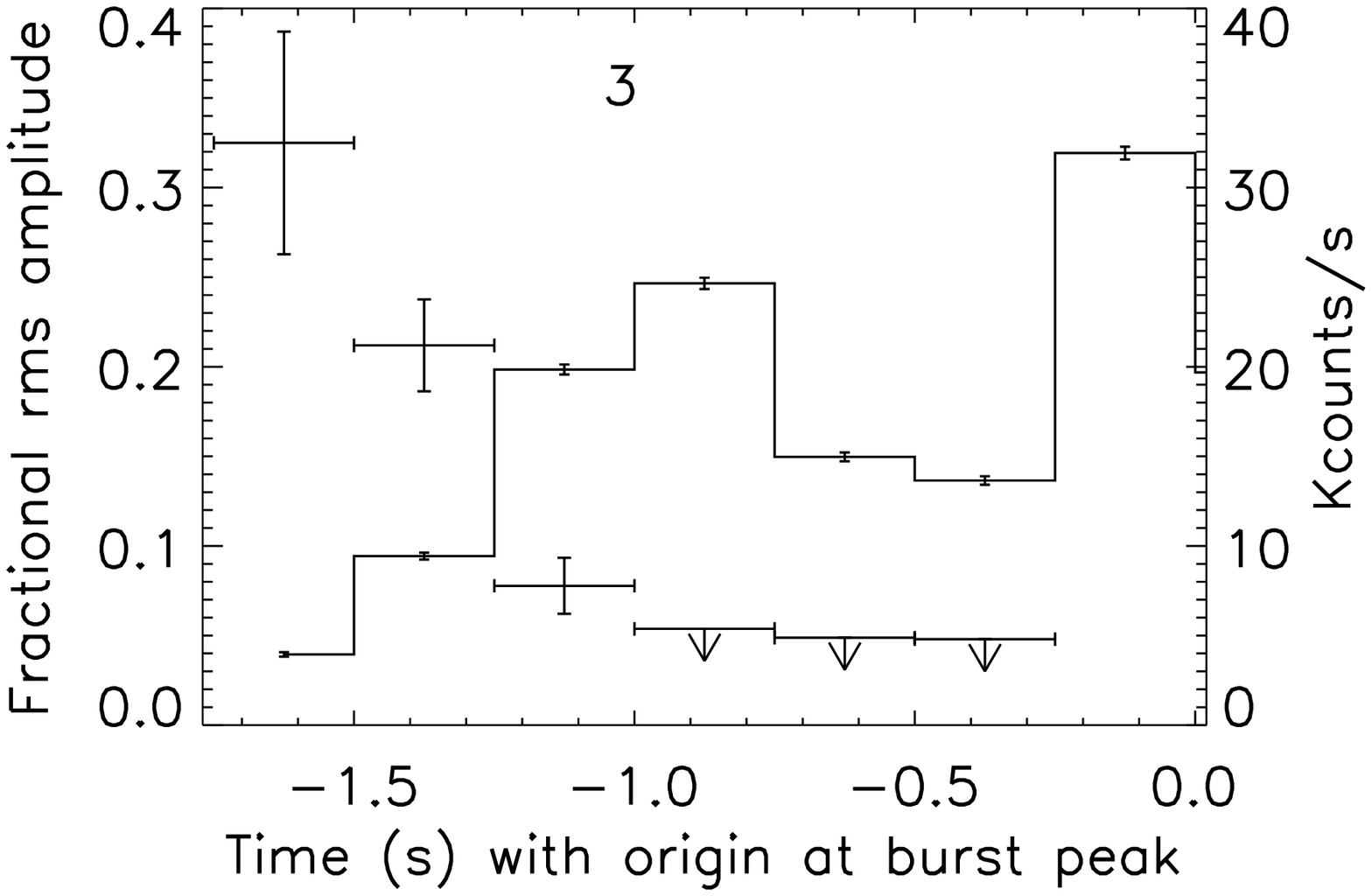} &
\includegraphics[width=0.32\textheight]{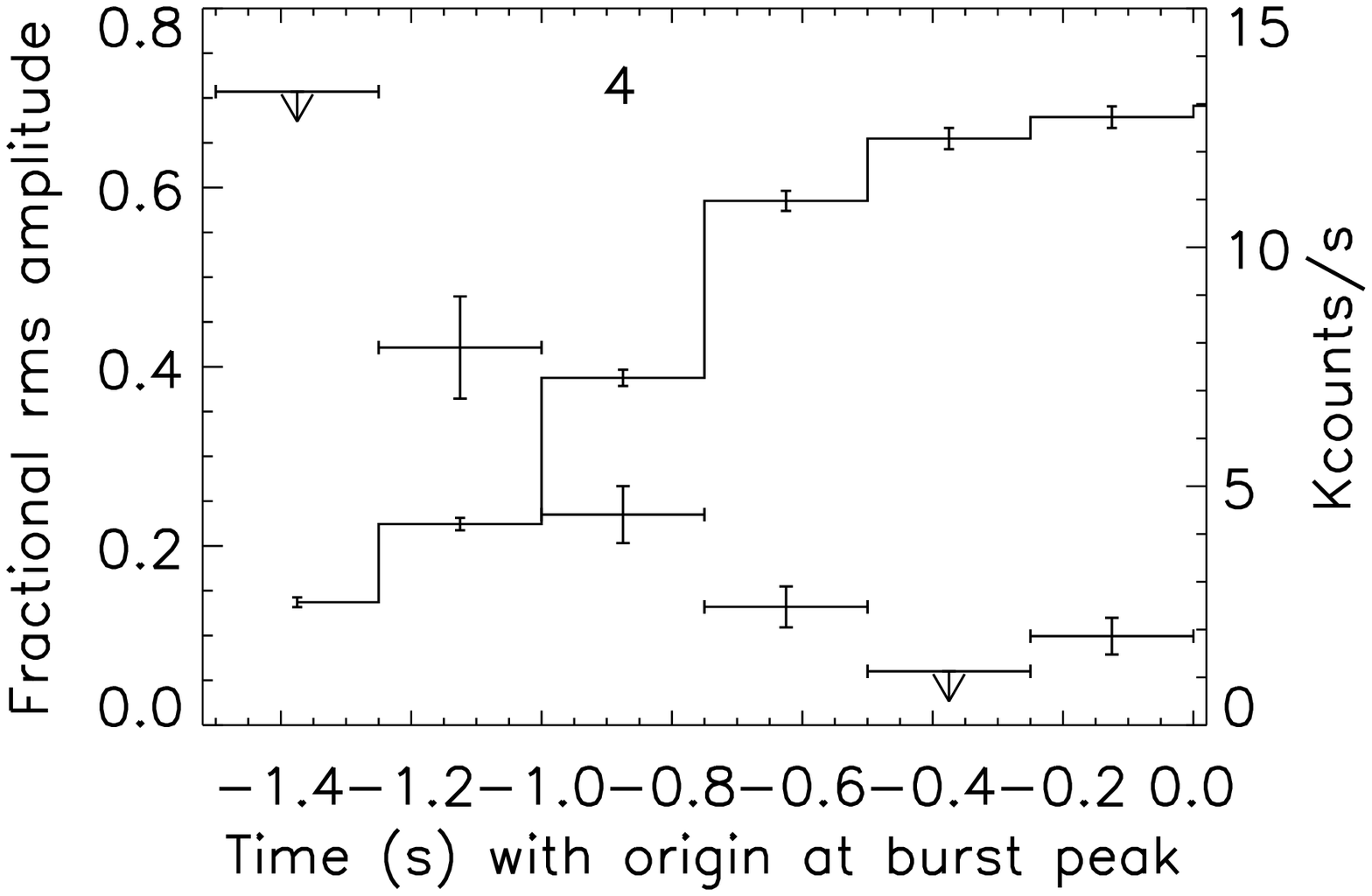} \\
\includegraphics[width=0.32\textheight]{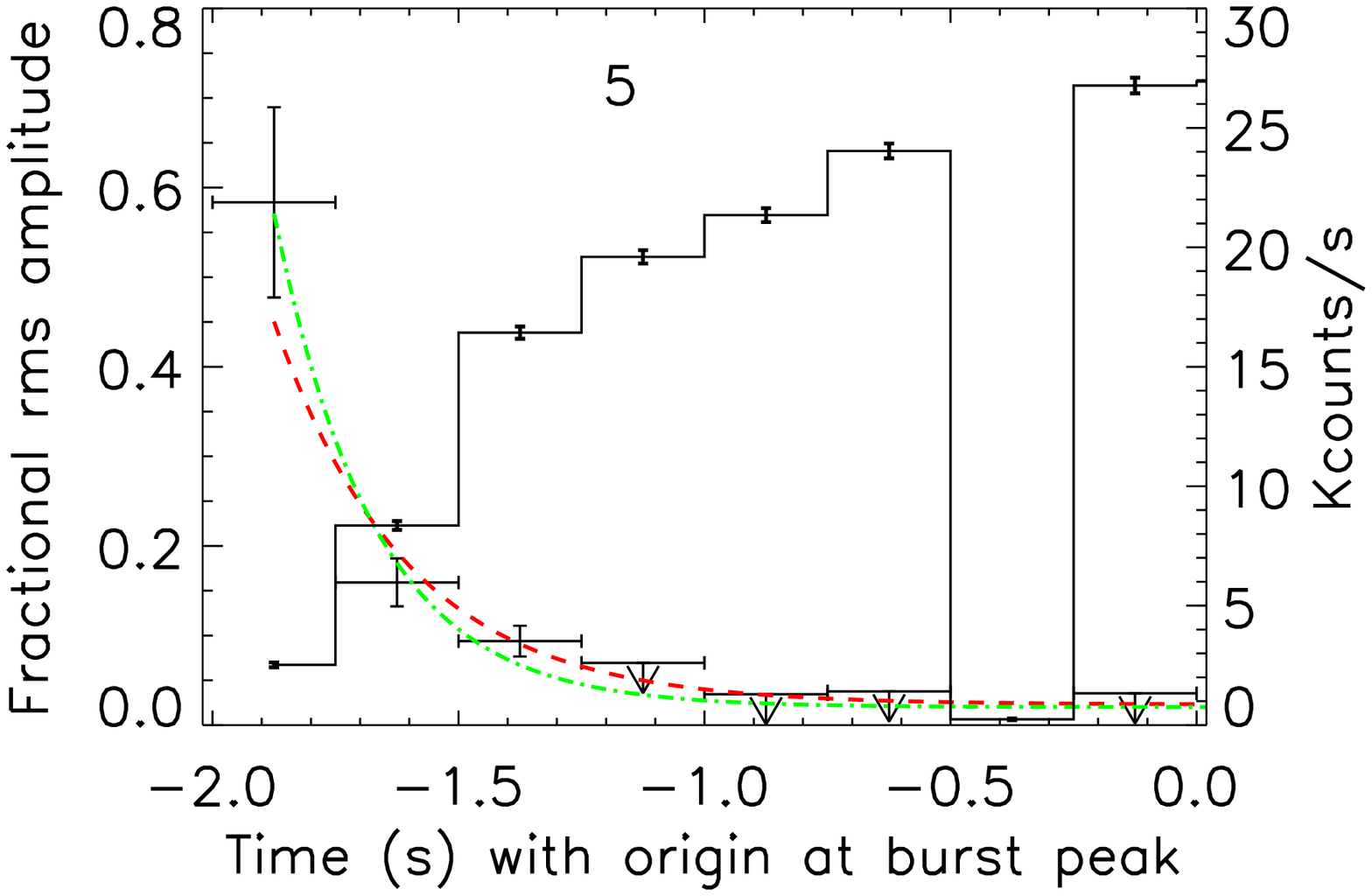} &
\includegraphics[width=0.32\textheight]{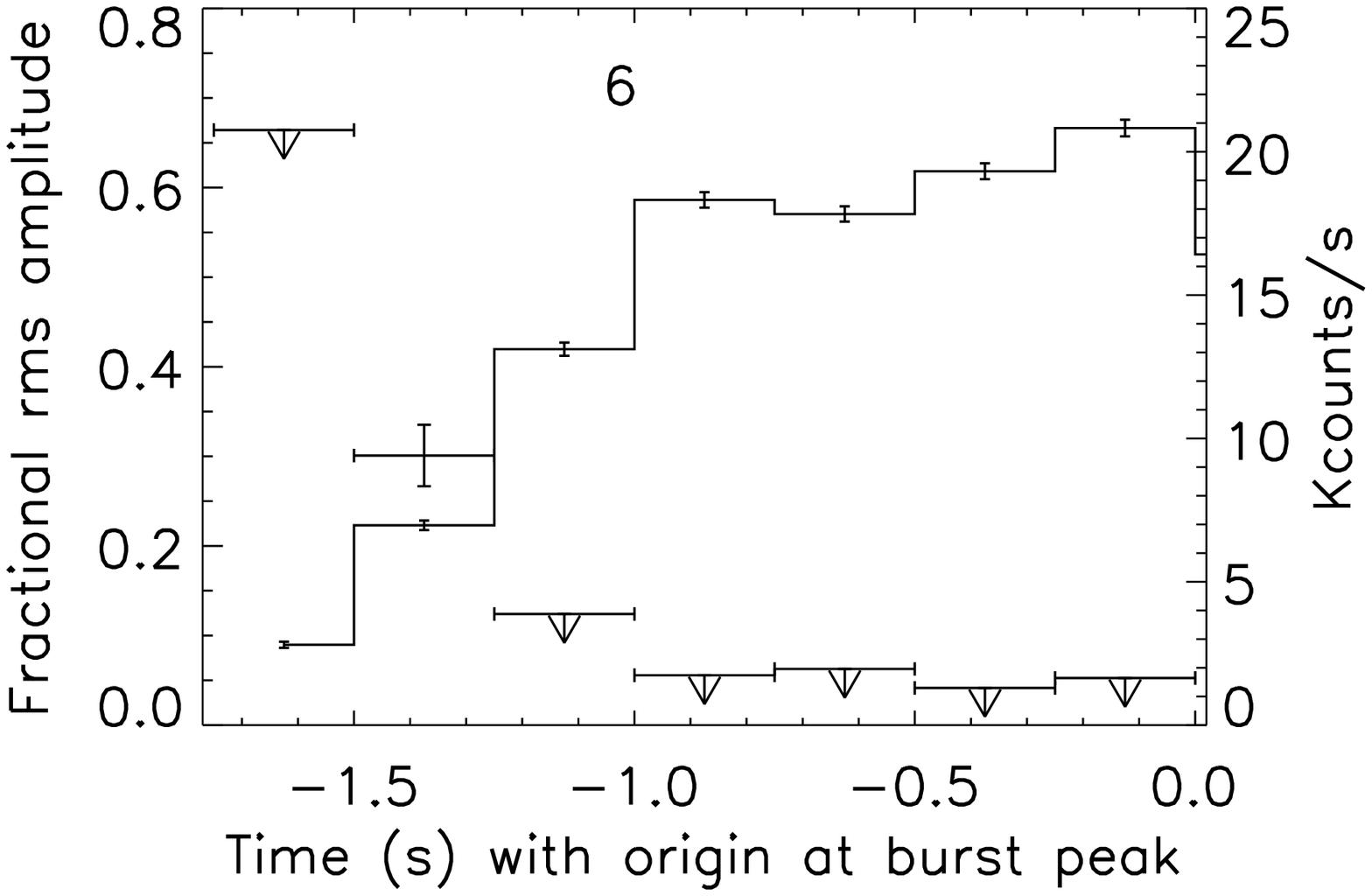} \\
\includegraphics[width=0.32\textheight]{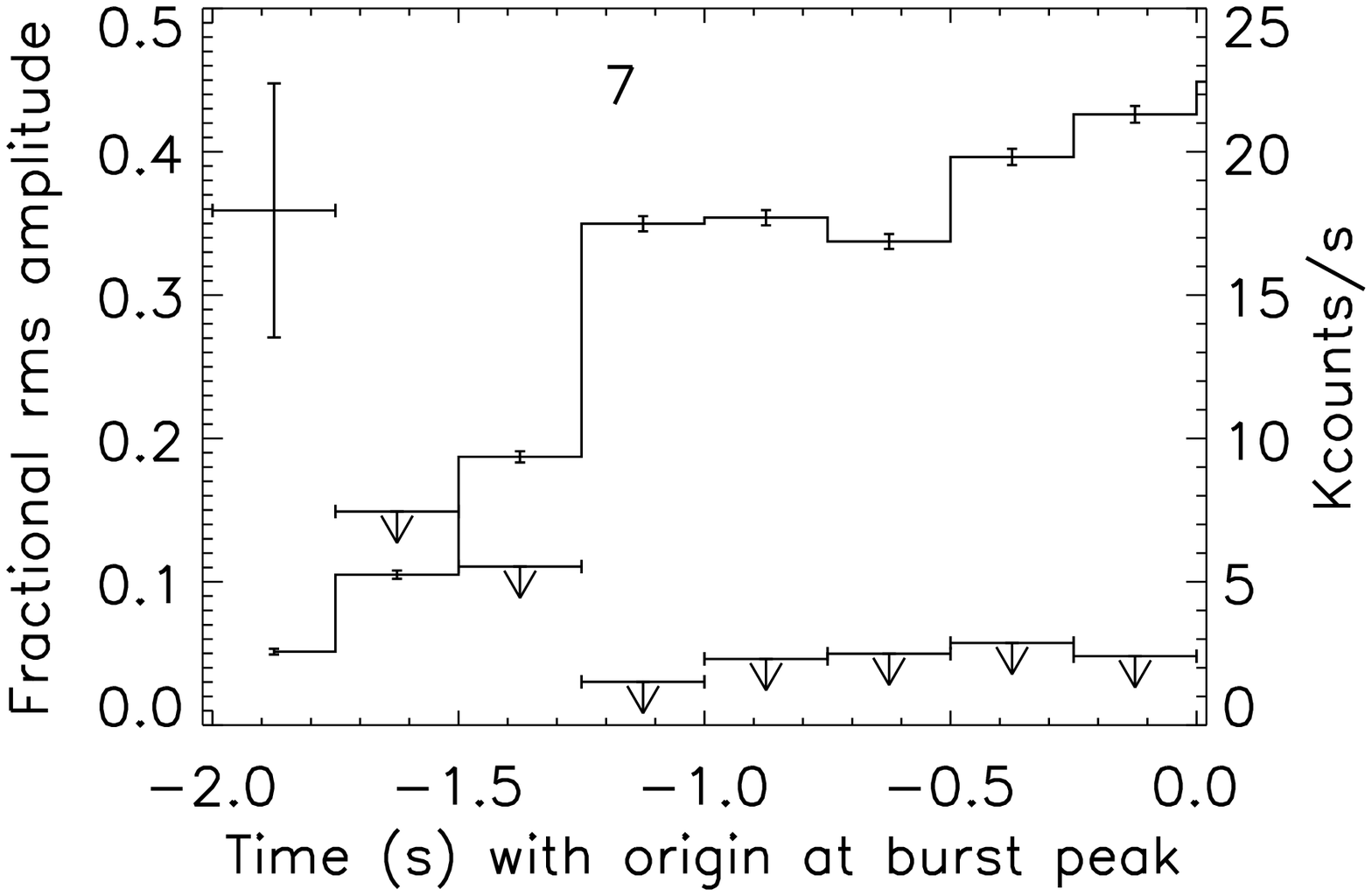} &
\includegraphics[width=0.32\textheight]{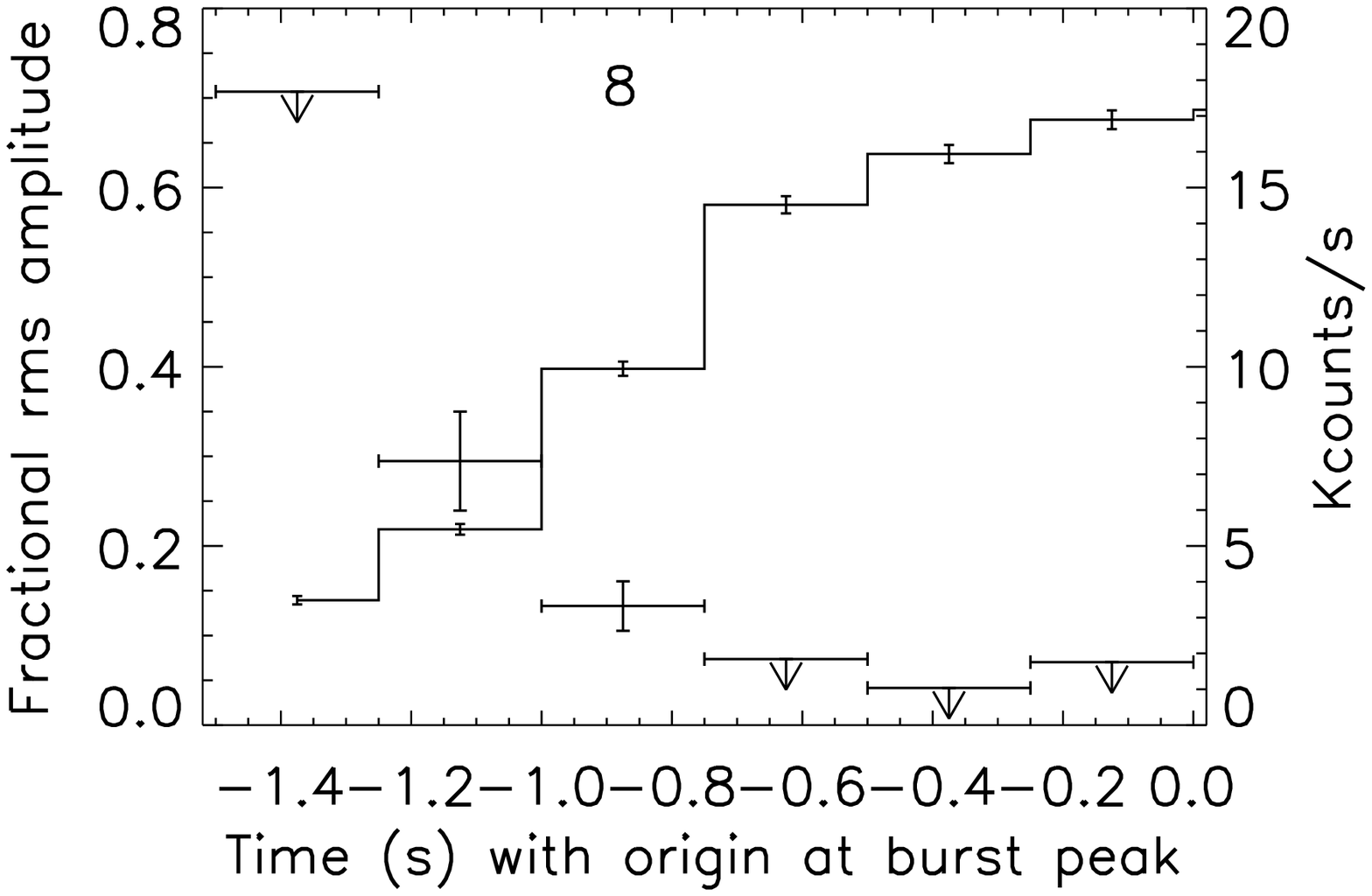} \\
\end{tabular}
\caption{Time evolution of fractional rms oscillation amplitude (left-hand y-axis) and PCA count rate
(histogram; right-hand y-axis) of thermonuclear bursts during rise. Each of 27 panels is for
a burst from 4U 1636--536 with rise oscillations detected with {\it RXTE} PCA 
(see \S~\ref{amplitude_calculation}).
The y-error bars give the $1\sigma$ errors and the horizontal bars show the time bin size of 0.25 s.
If oscillation is not detected (see \S~\ref{amplitude_calculation} for the criterion), an upper limit,
denoted by an arrow, is given. A panel number gives the burst index quoted in the first 
column of Table~\ref{Log}. Model fits to the fractional rms amplitude evolutions are also 
shown for two representative bursts with small (burst no. 5) and large (Burst no. 16) number of 
time bins with detected oscillations. The red dashed curves and the green dash-dot curves 
show the best-fit empirical model $a-bc(1-e^{-t/c})$ ($t$: time variable, $a, b, c$: parameters) 
for the weighted least square minimization method and the 
likelihood maximization technique respectively (see \S~\ref{Rms}).
This figure shows that the fractional rms oscillation amplitude usually decreases with time 
during burst rise (\S~\ref{Rms}).
\label{rmsampevol}} 
\end{figure*}

\clearpage
\addtocounter{figure}{ -1}
\begin{figure*}
\centering
\begin{tabular}{lr} 
\includegraphics[width=0.32\textheight]{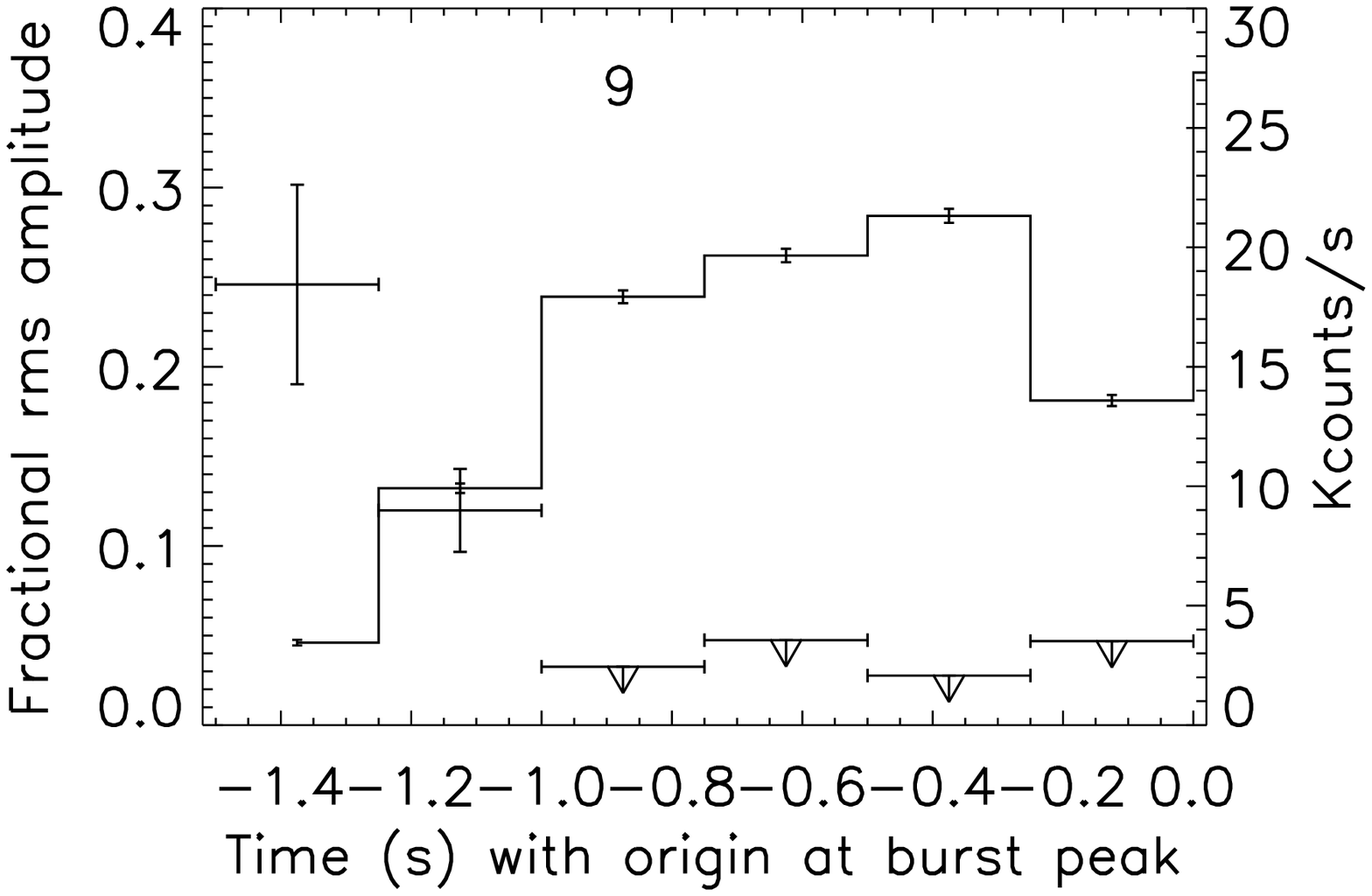} &
\includegraphics[width=0.32\textheight]{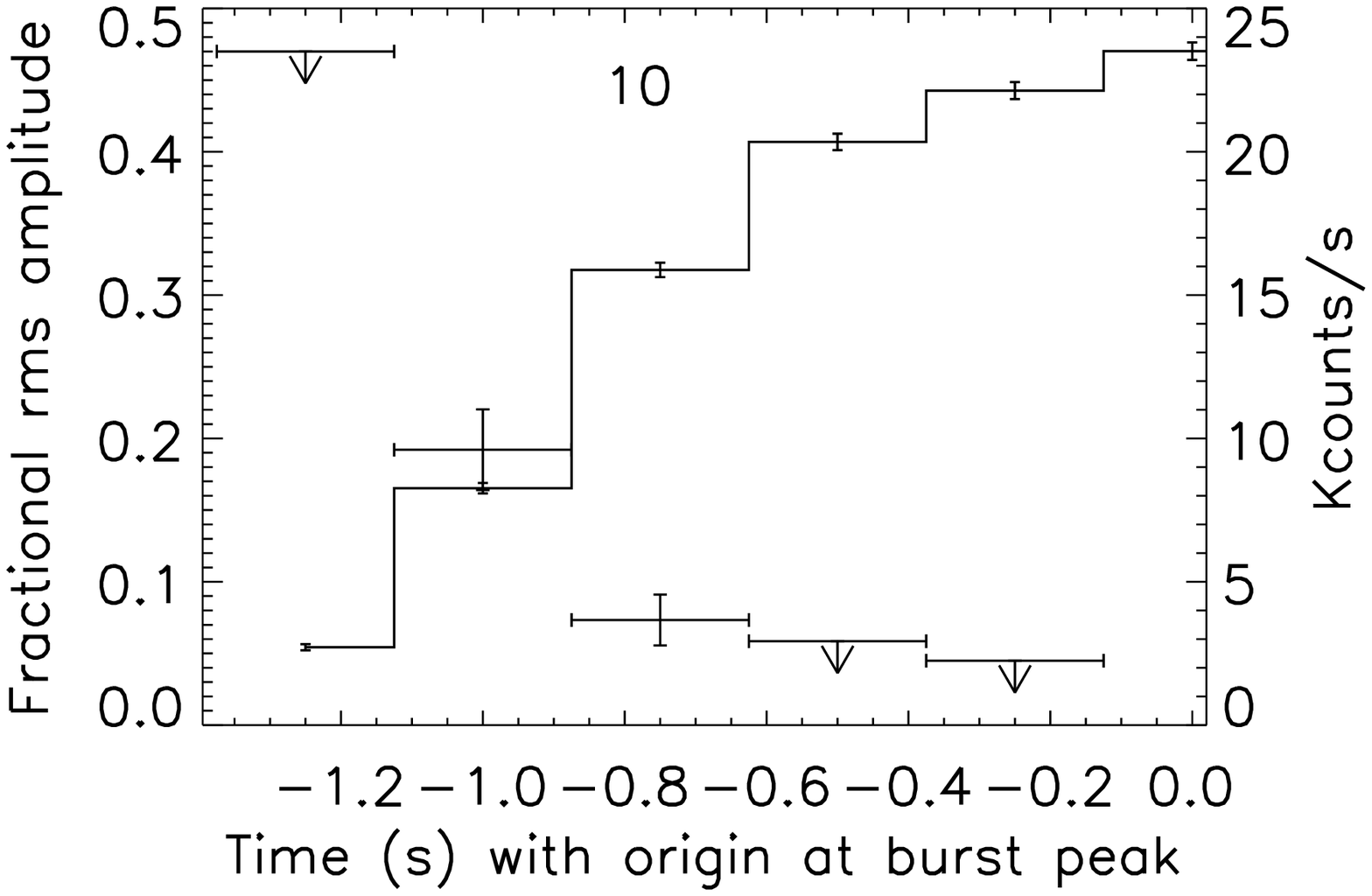} \\
\includegraphics[width=0.32\textheight]{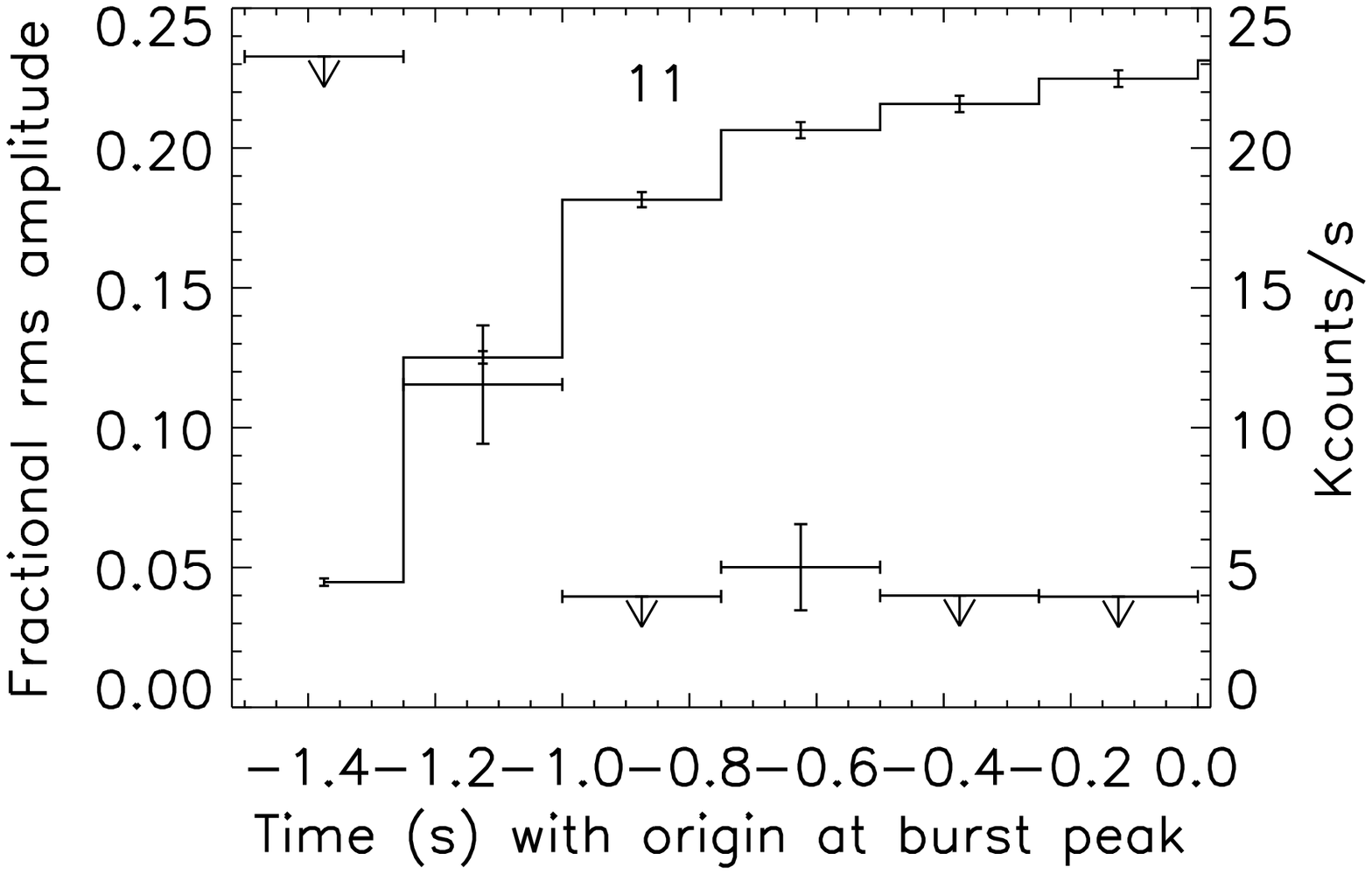} &
\includegraphics[width=0.32\textheight]{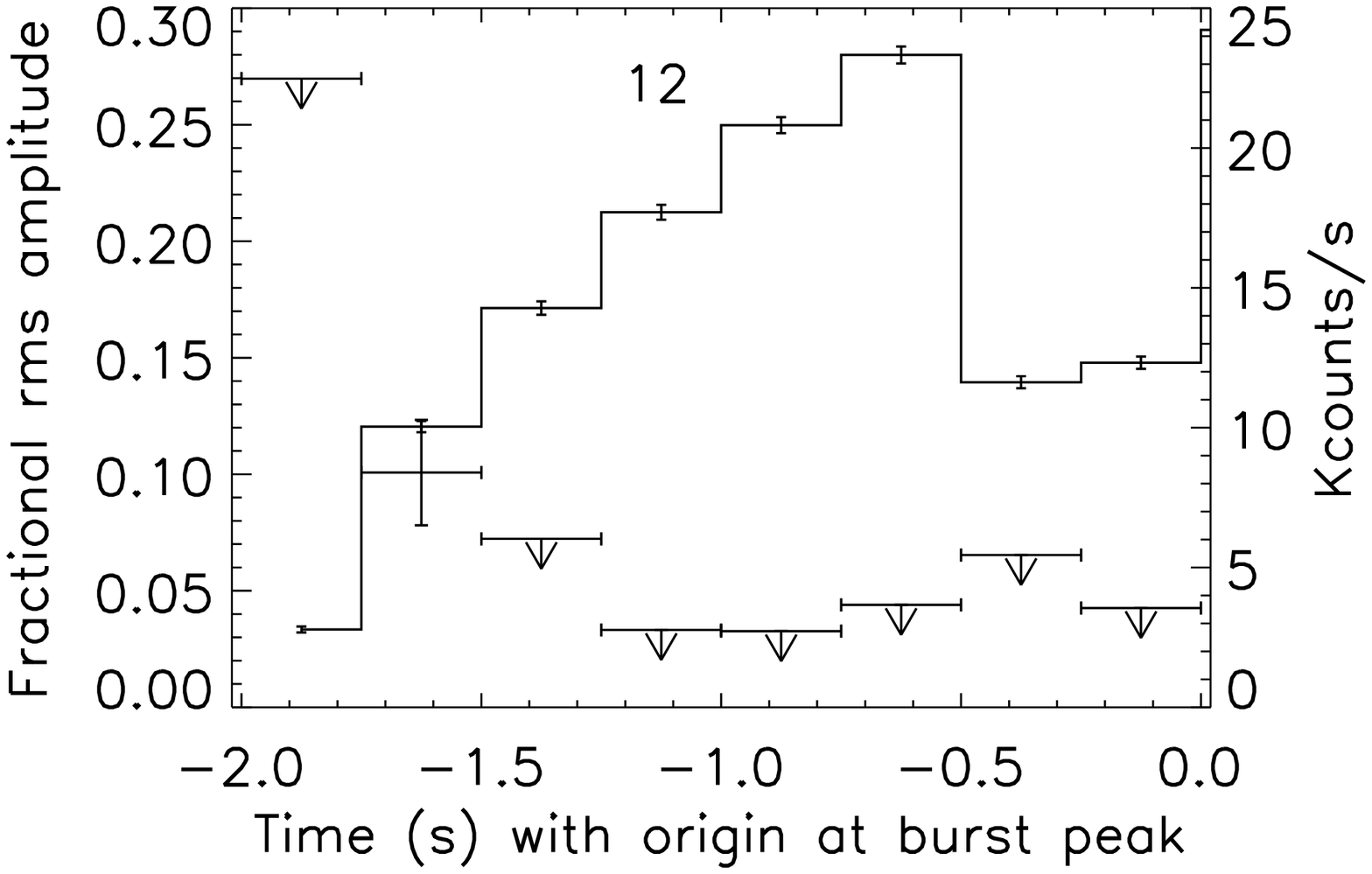} \\
\includegraphics[width=0.32\textheight]{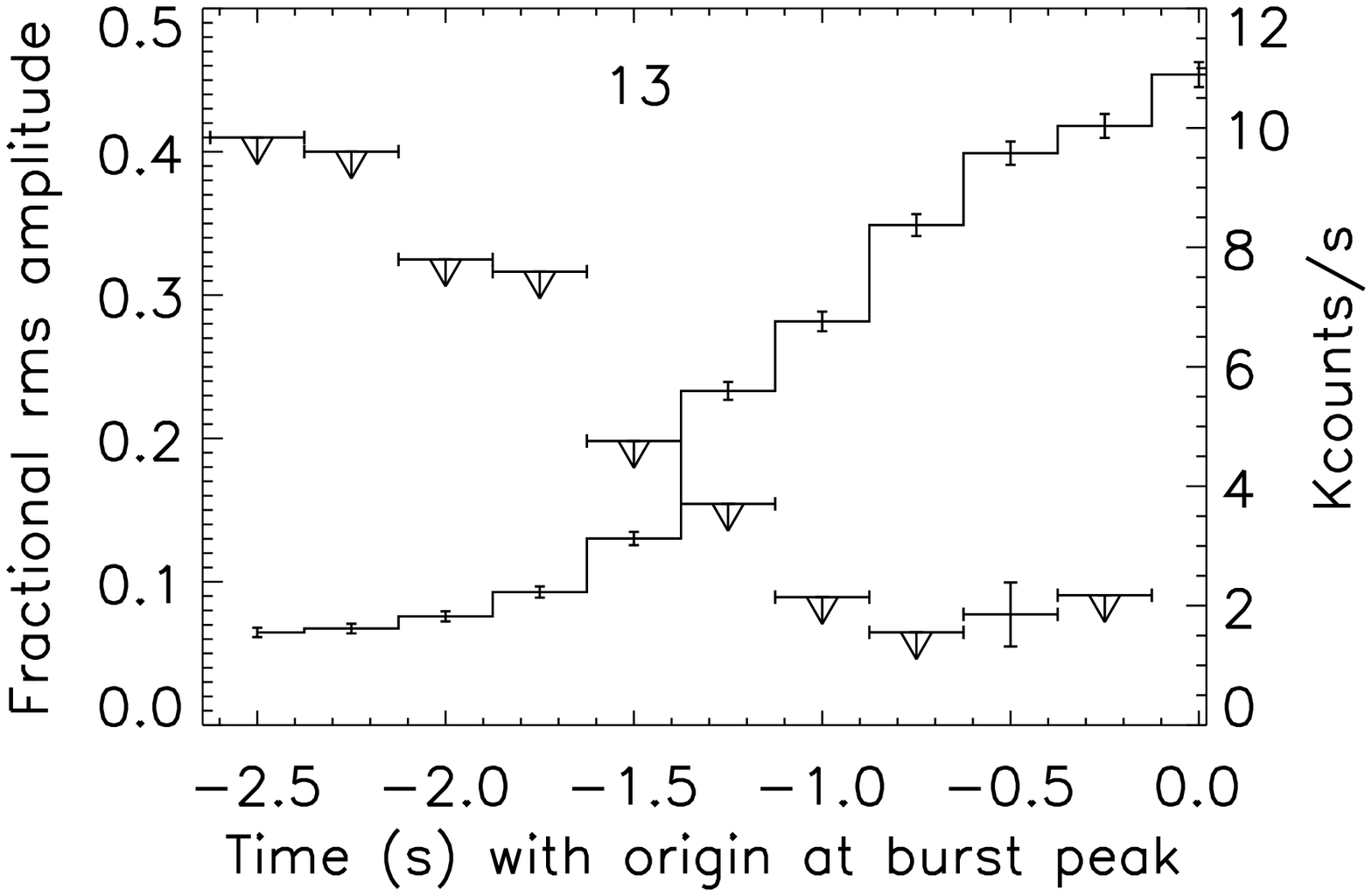} & 
\includegraphics[width=0.32\textheight]{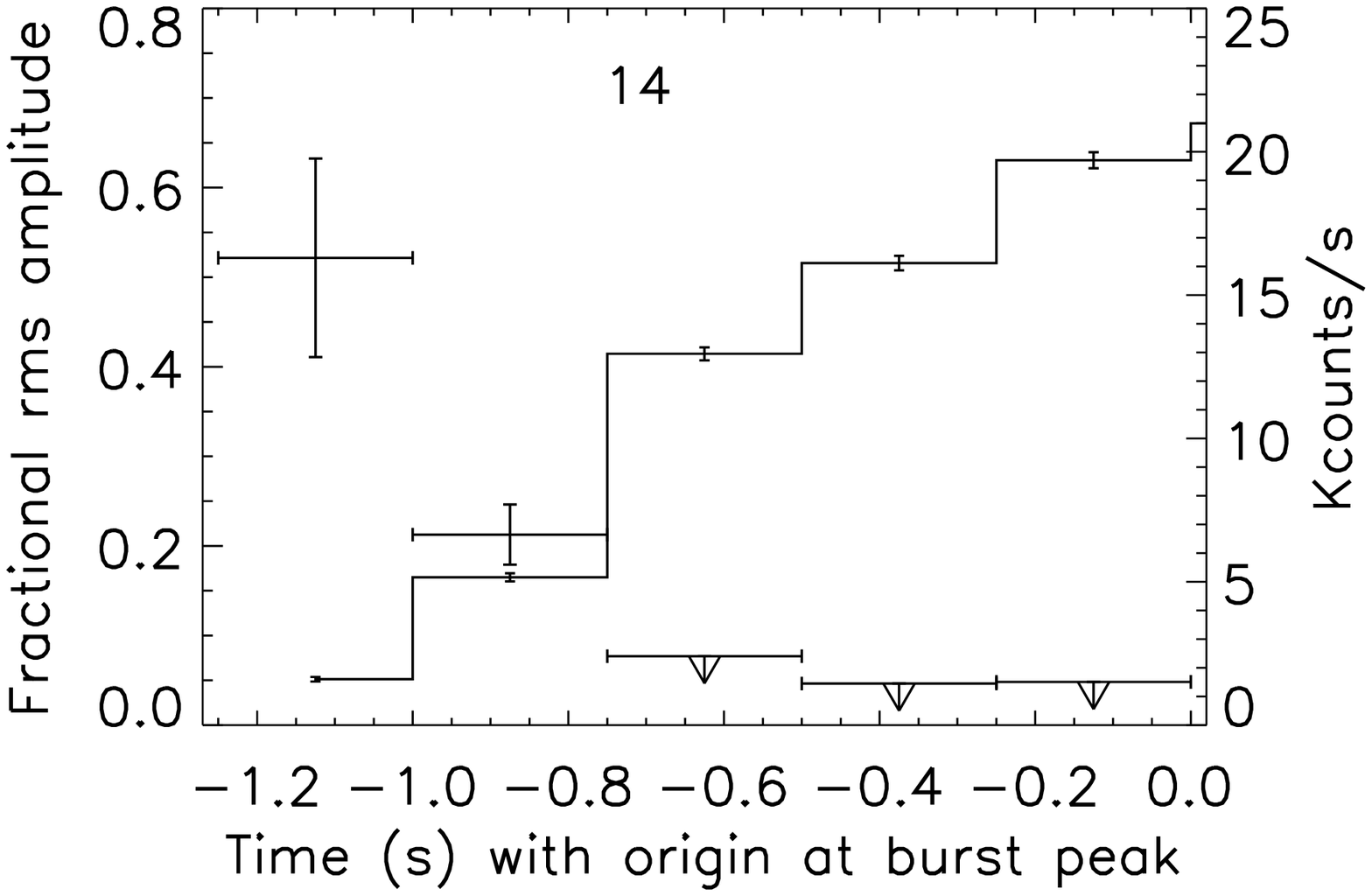} \\
\includegraphics[width=0.32\textheight]{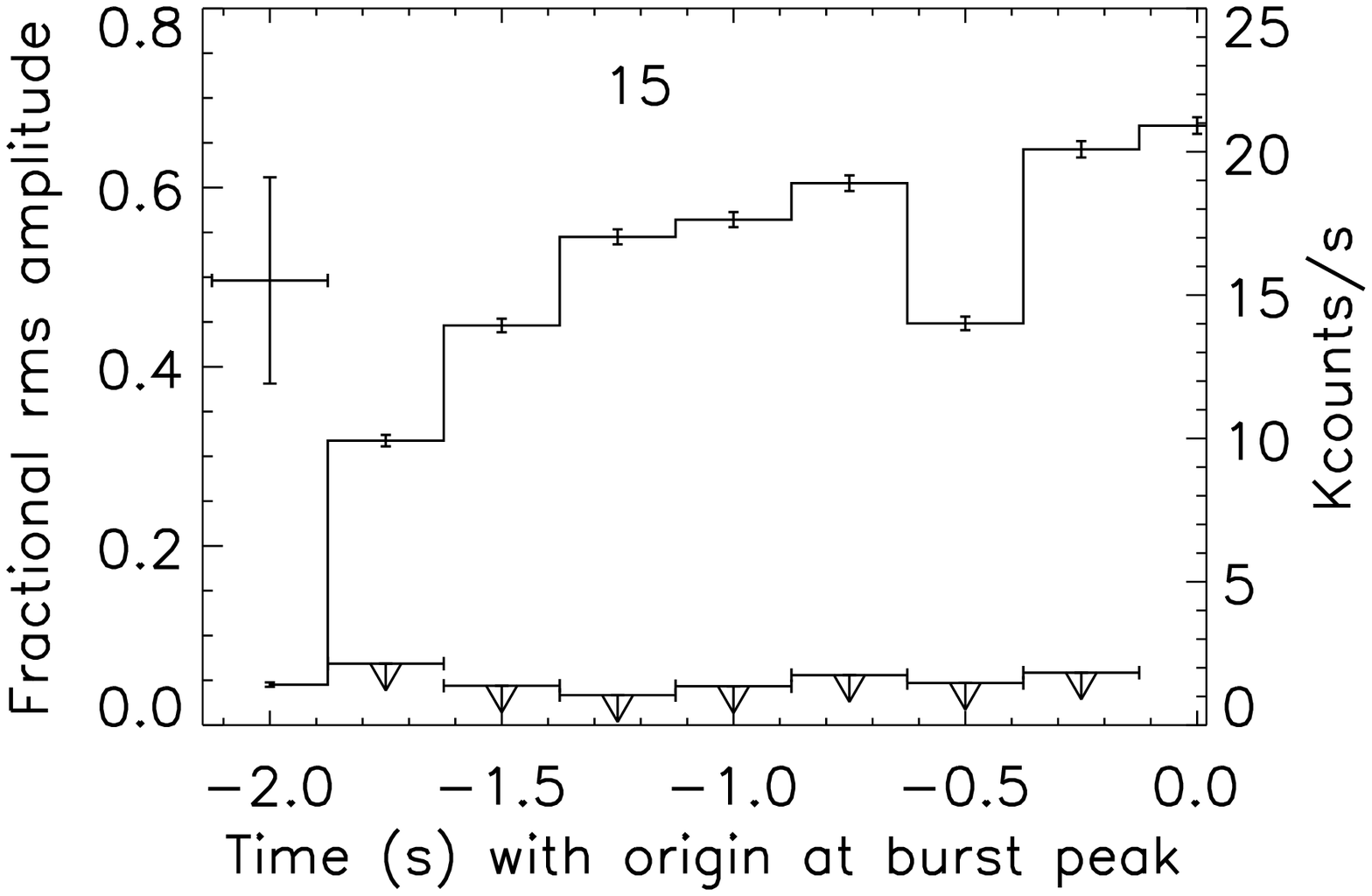} &
\includegraphics[width=0.32\textheight]{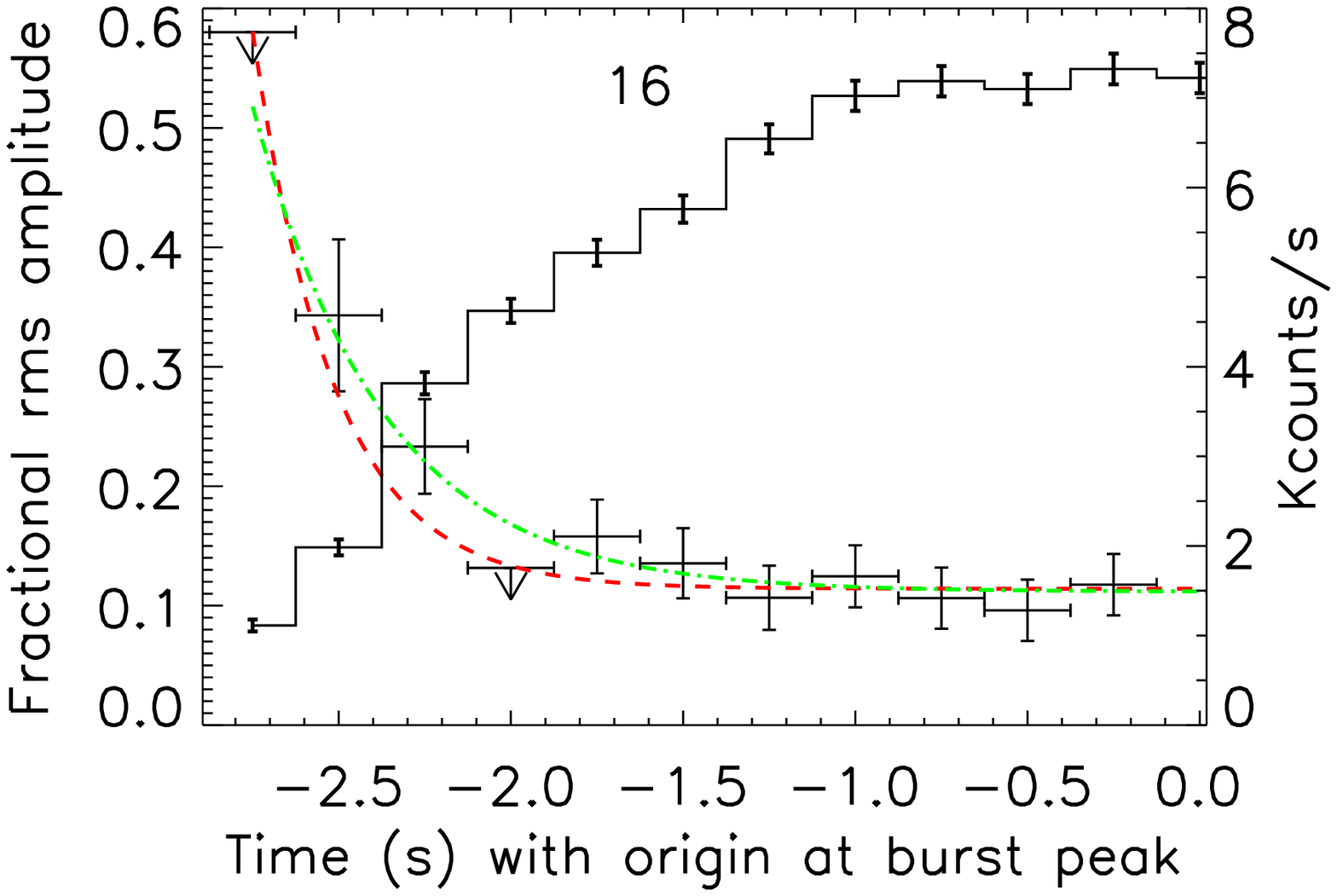} \\
\end{tabular}
\caption{Continued.}
\end{figure*}

\clearpage
\addtocounter{figure}{ -1}
\begin{figure*}
\centering
\begin{tabular}{lr}
\includegraphics[width=0.32\textheight]{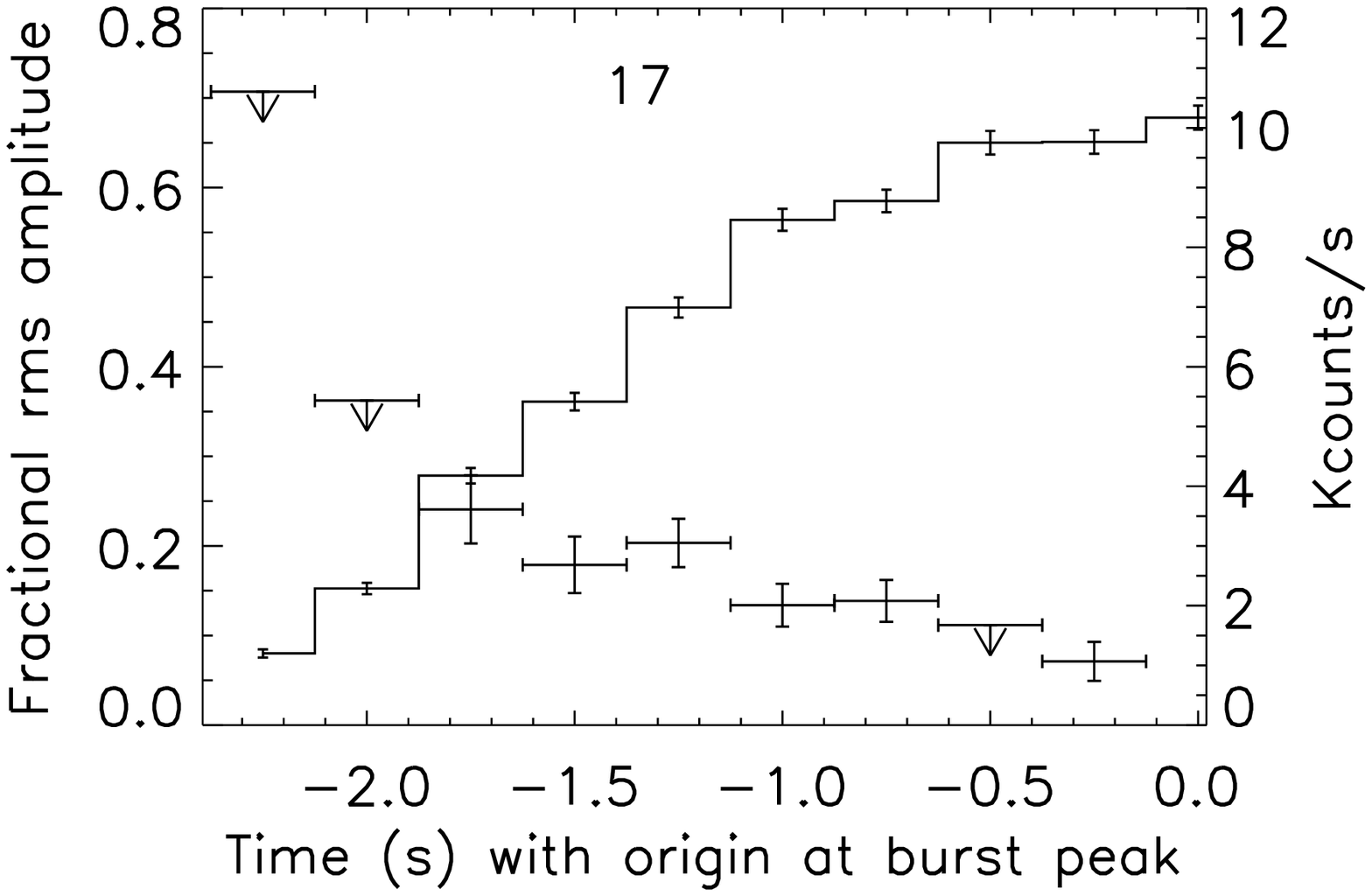} & 
\includegraphics[width=0.32\textheight]{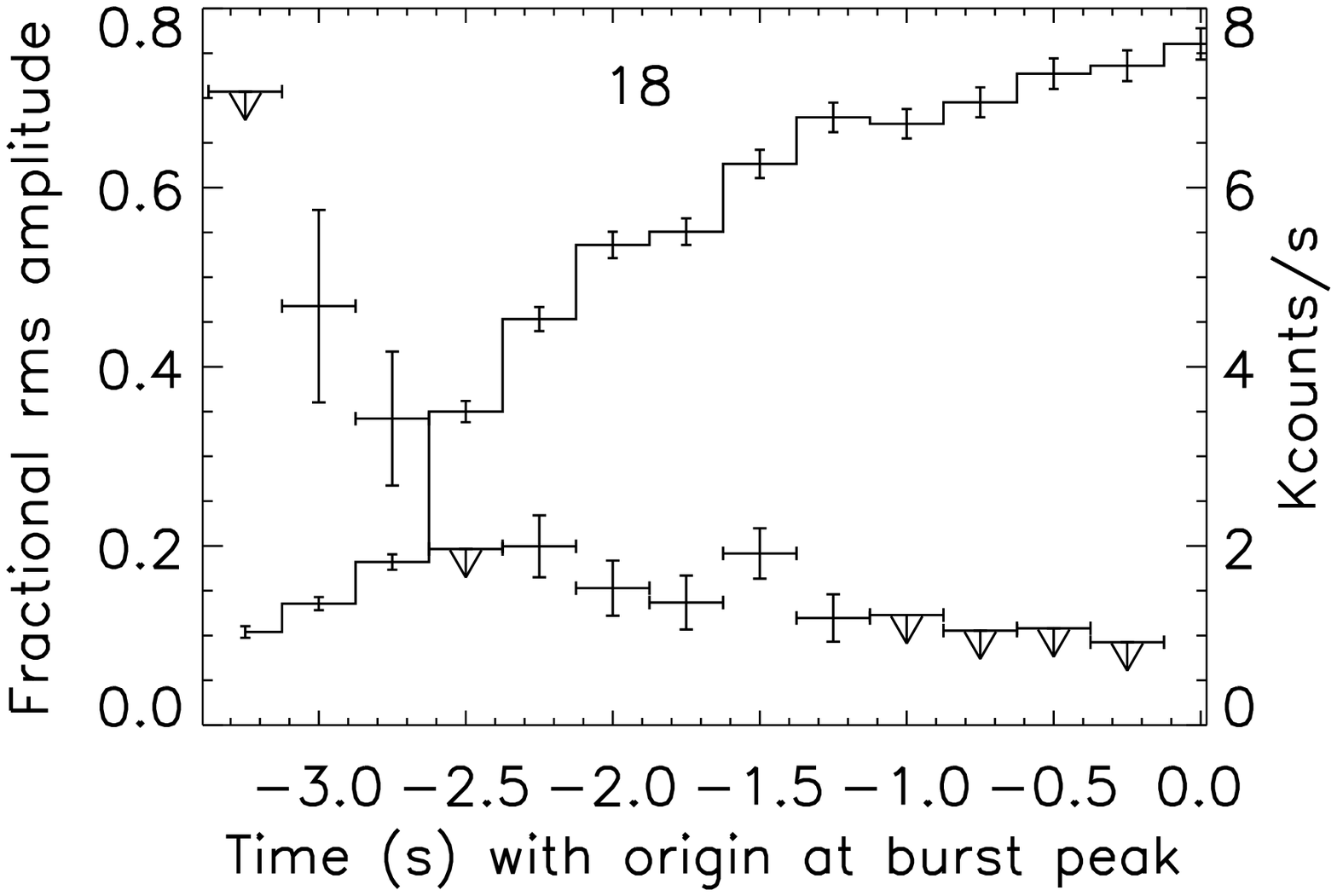} \\
\includegraphics[width=0.32\textheight]{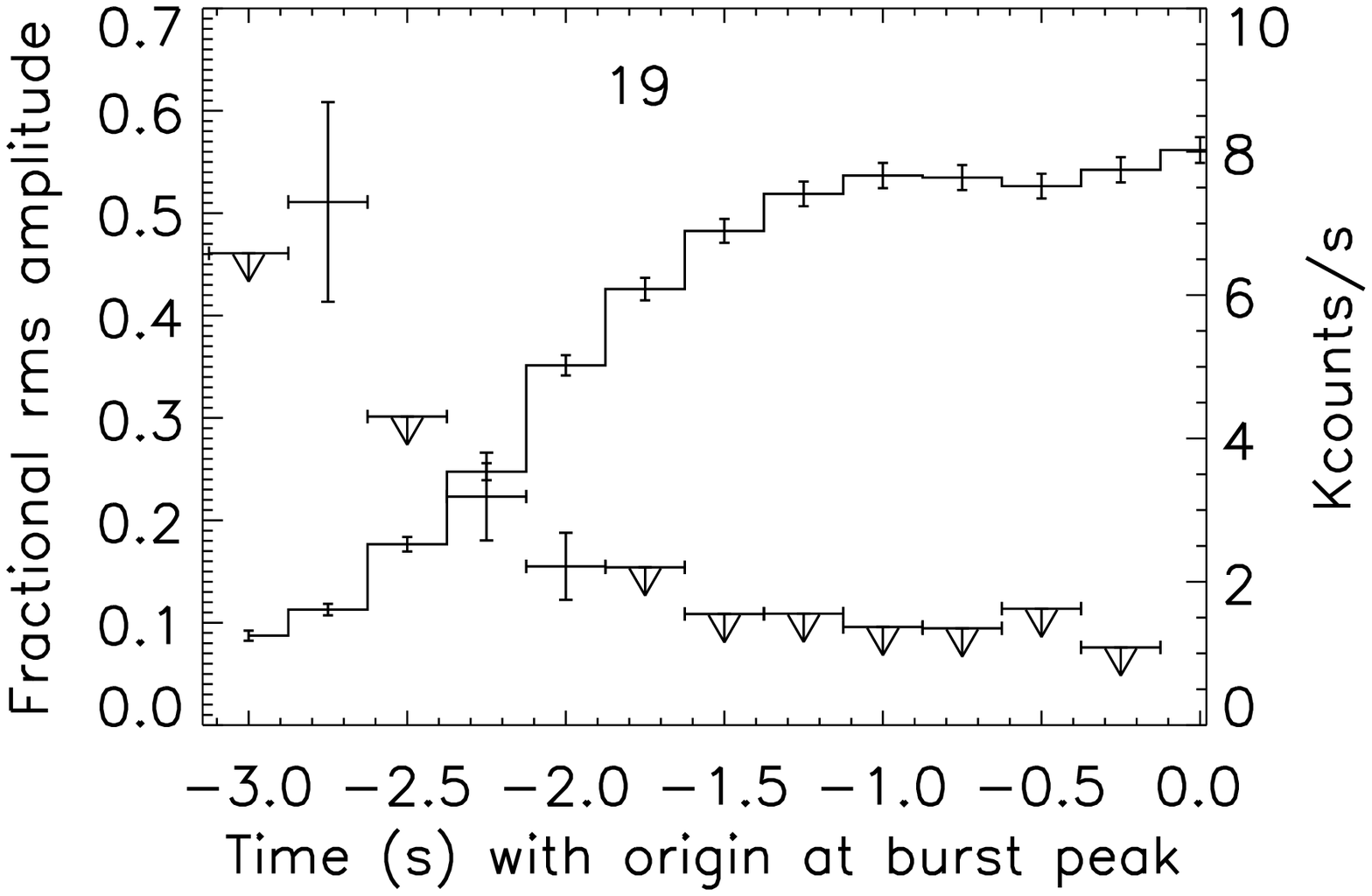} &
\includegraphics[width=0.32\textheight]{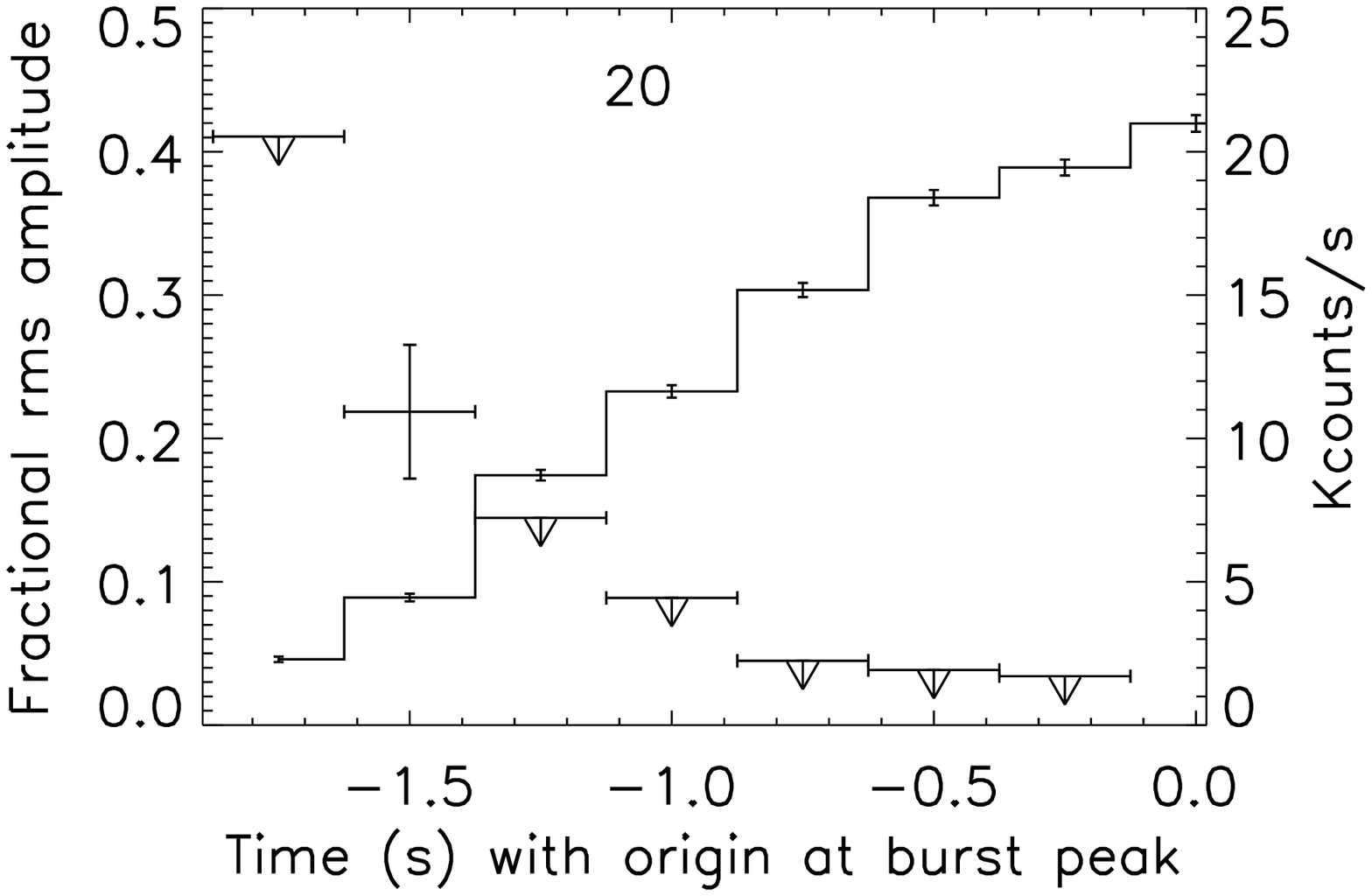} \\
\includegraphics[width=0.32\textheight]{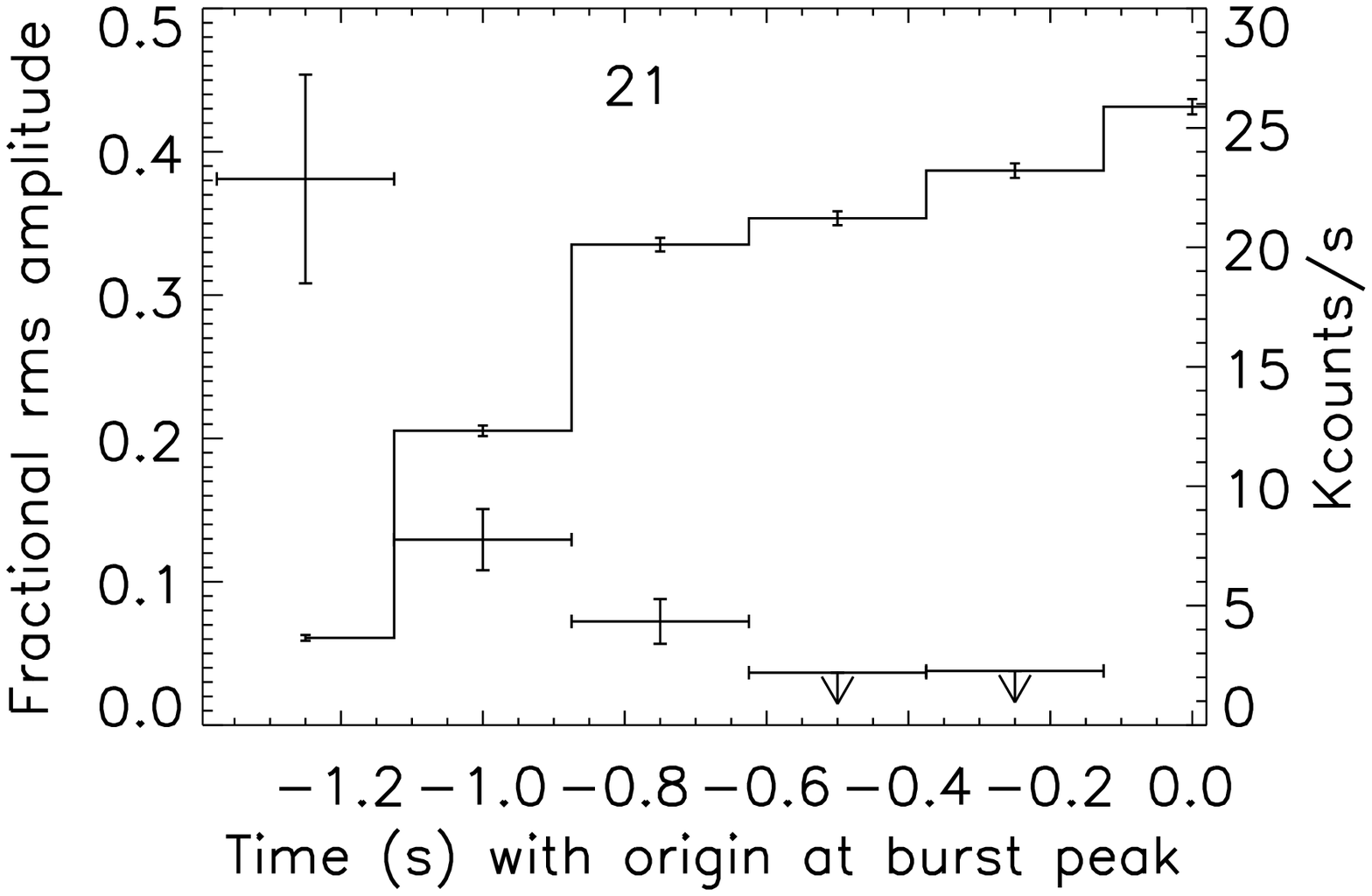} &
\includegraphics[width=0.32\textheight]{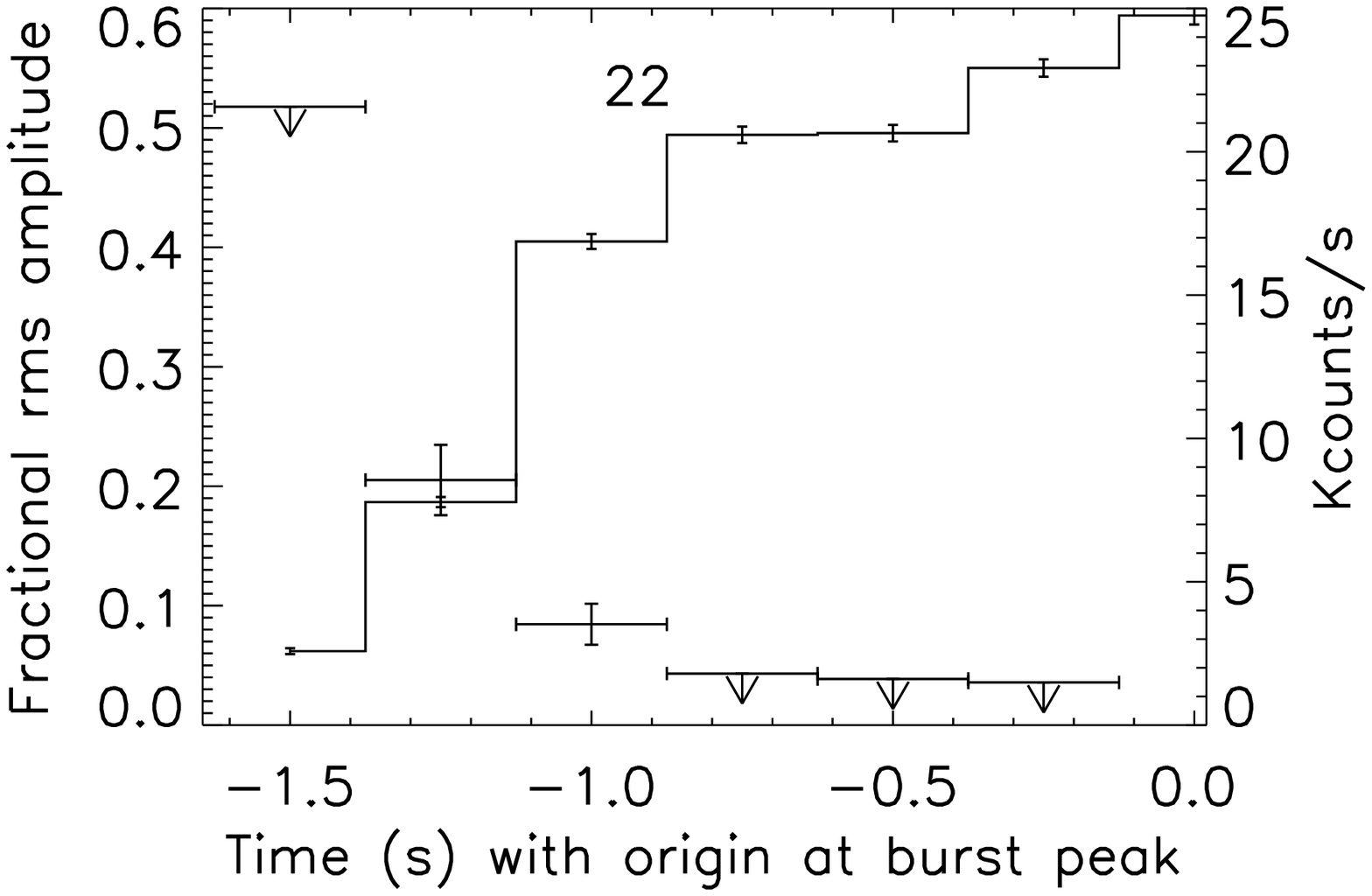} \\
\includegraphics[width=0.32\textheight]{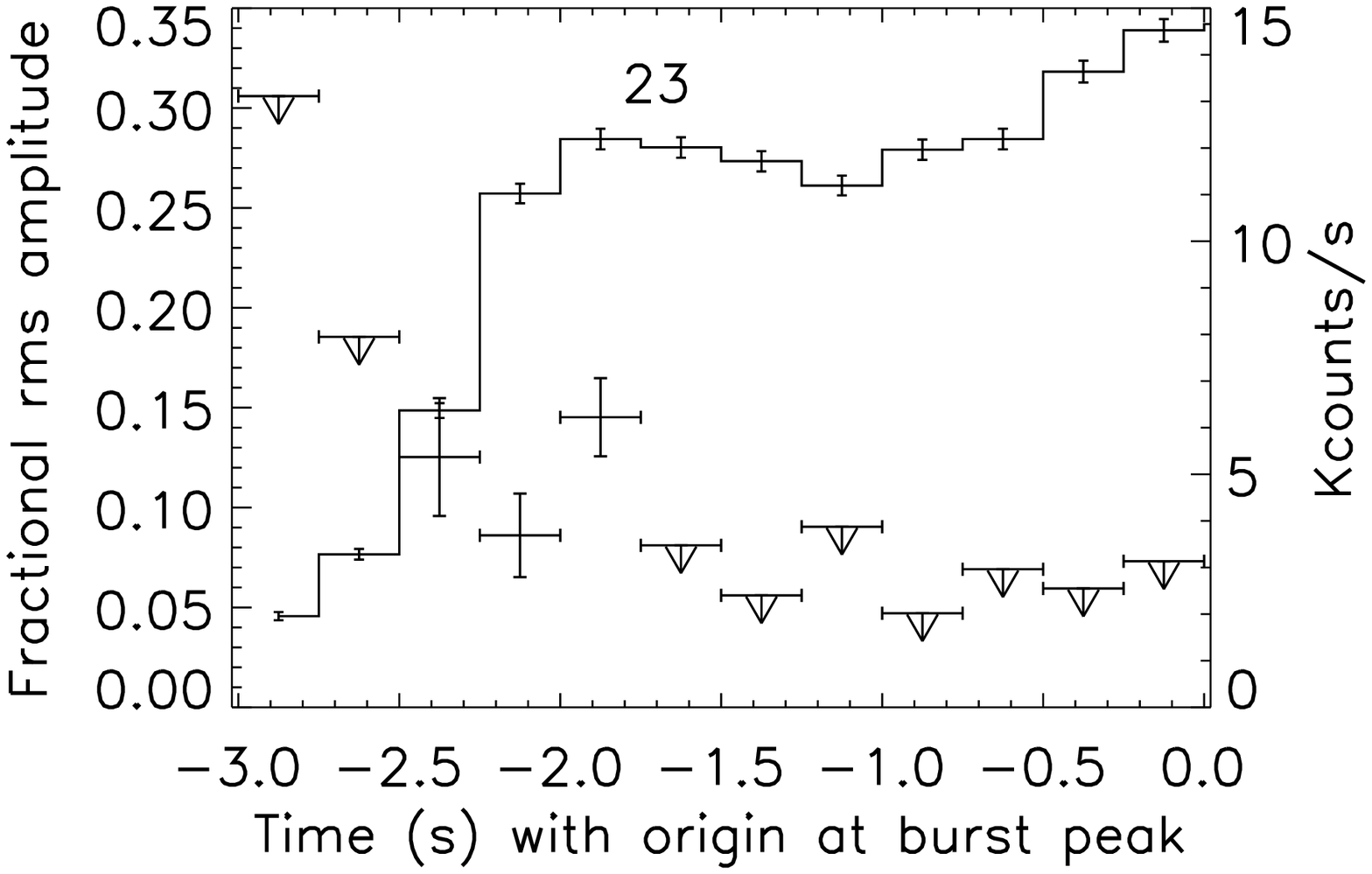} & 
\includegraphics[width=0.32\textheight]{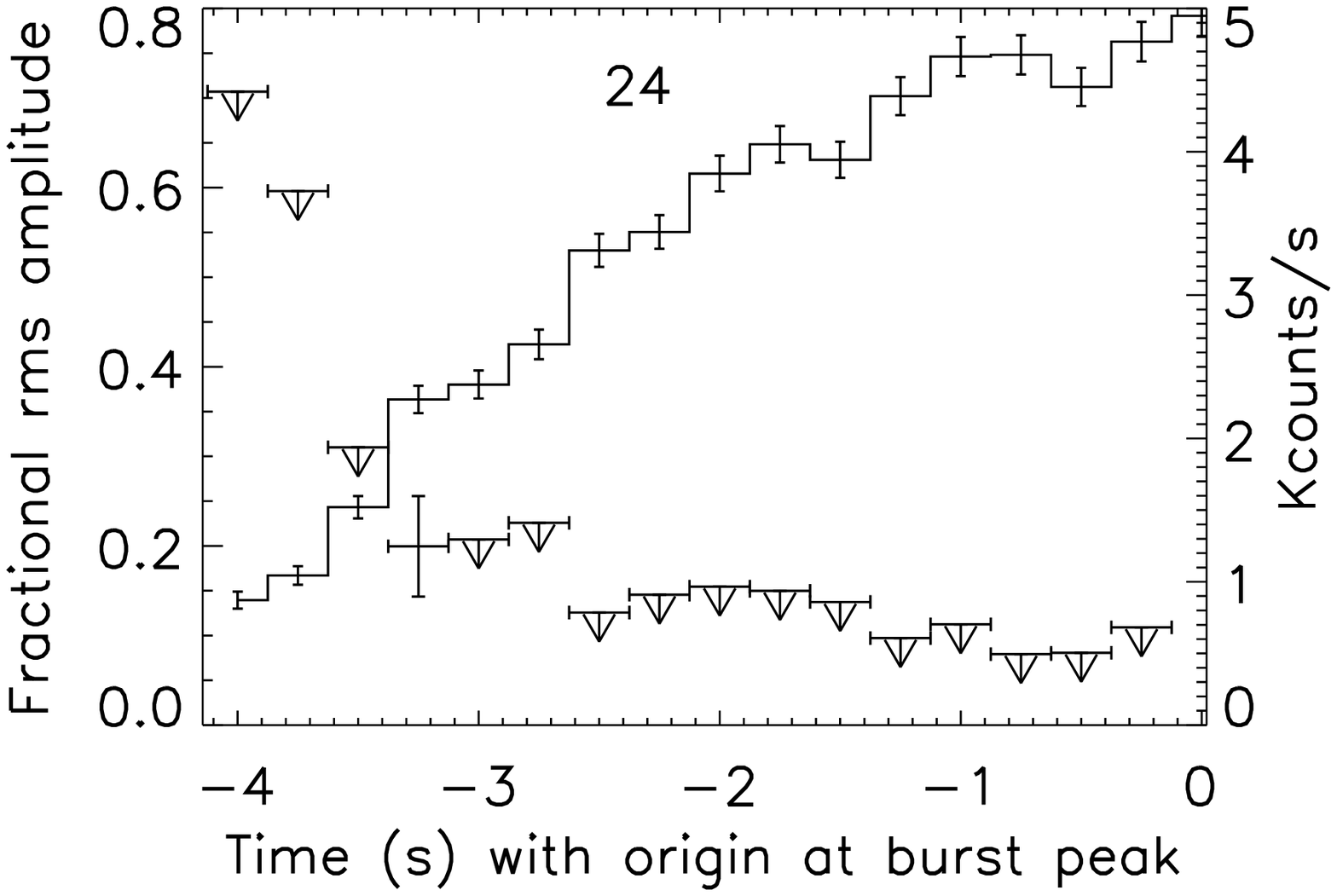} \\
\end{tabular}
\caption{Continued.}
\end{figure*}

\clearpage
\addtocounter{figure}{ -1}
\begin{figure*}
\centering
\begin{tabular}{lr}
\includegraphics[width=0.32\textheight]{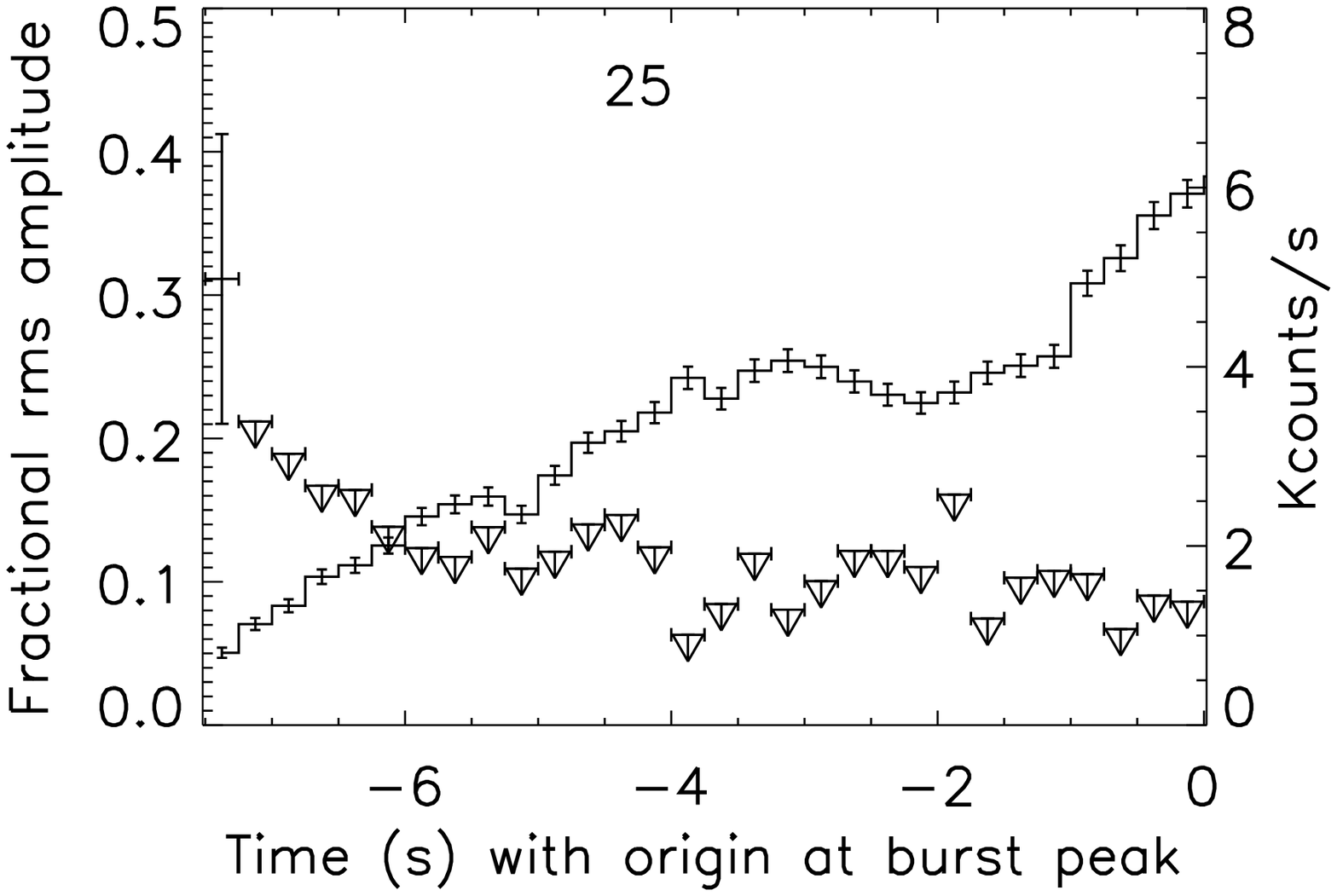} &
\includegraphics[width=0.32\textheight]{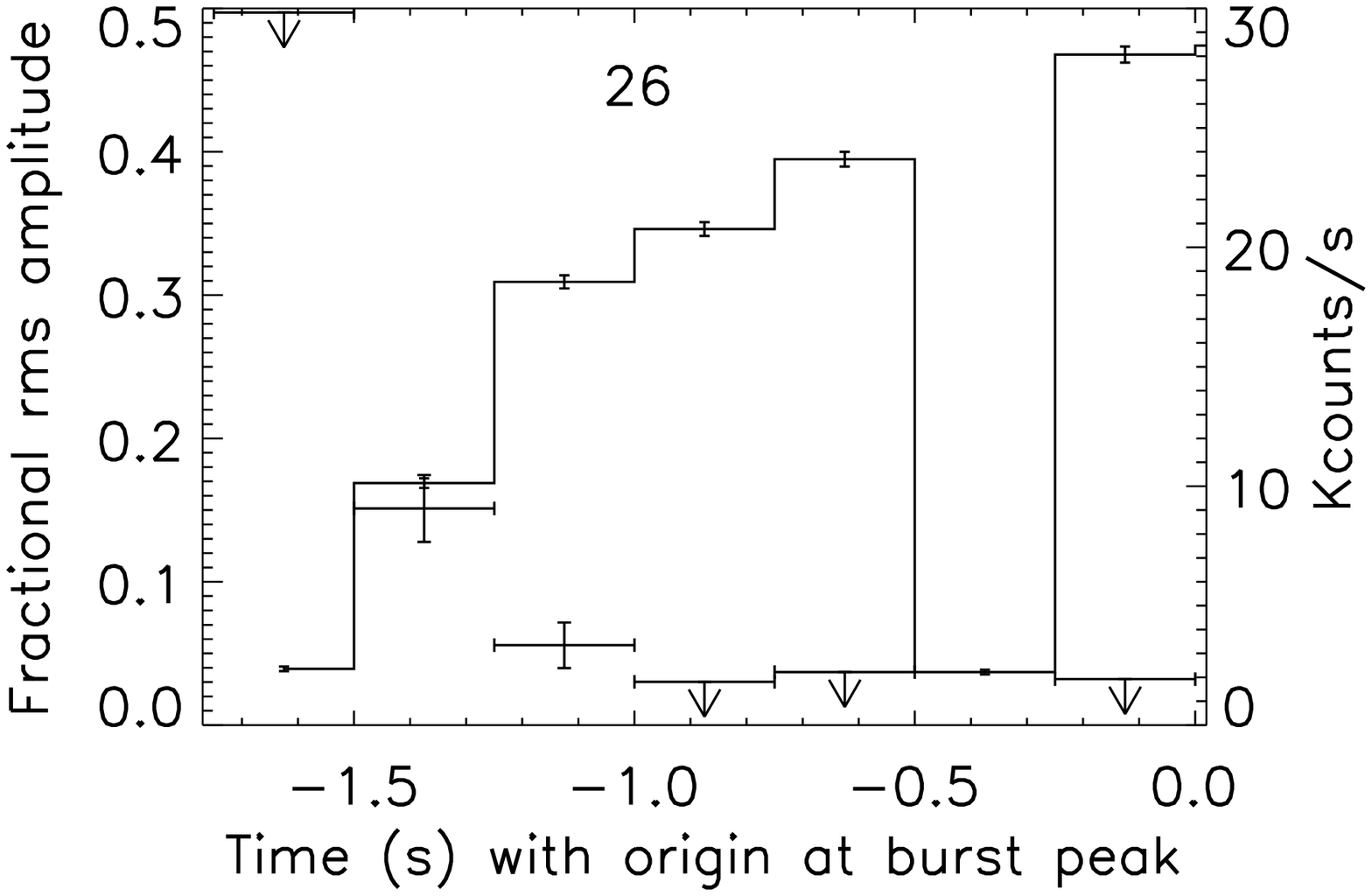} \\
\includegraphics[width=0.32\textheight]{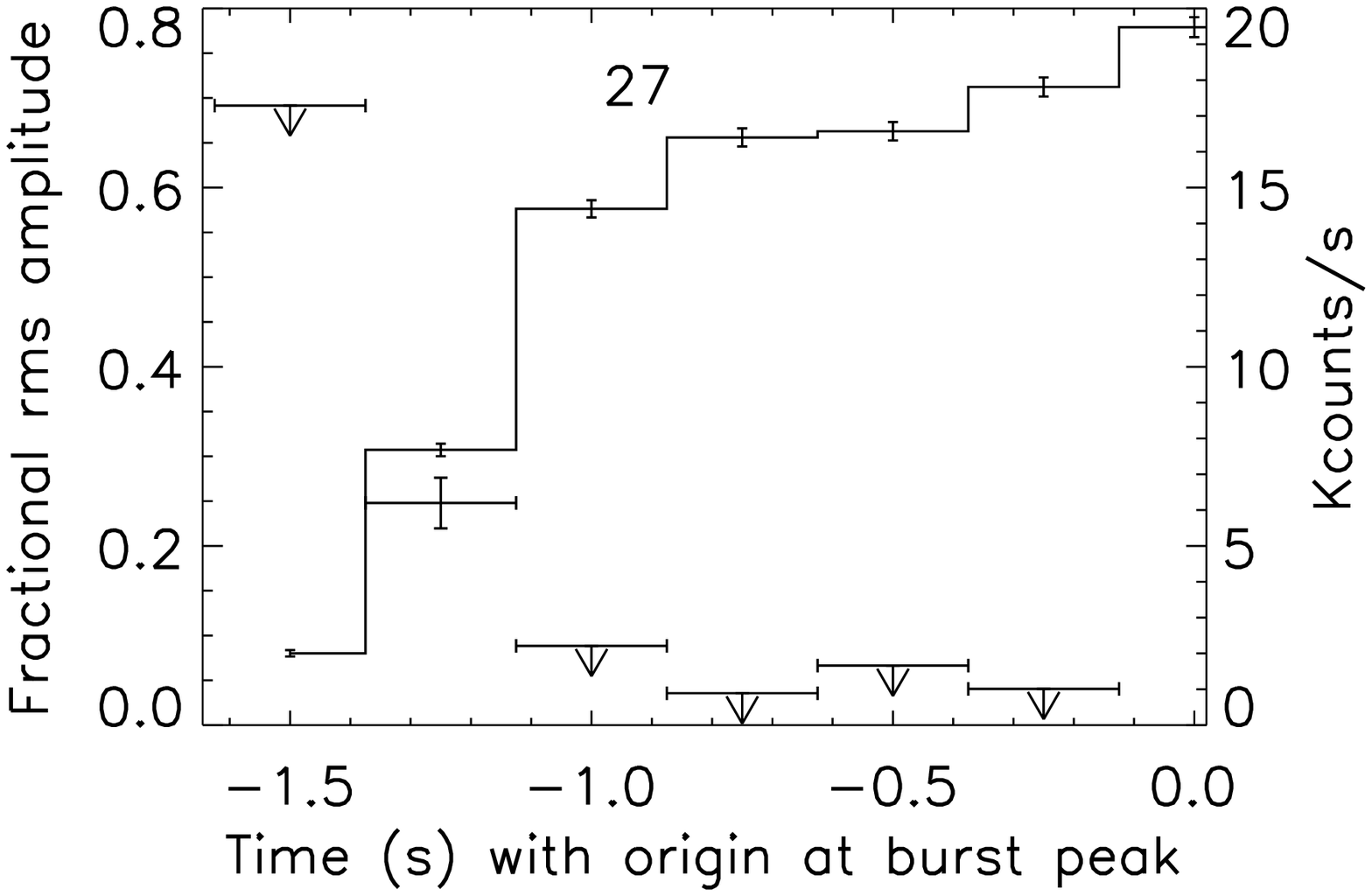}
\end{tabular}
\caption{Continued.}
\end{figure*}

\begin{figure*}
\centering
\begin{tabular}{lr}
\includegraphics[width=0.35\textheight]{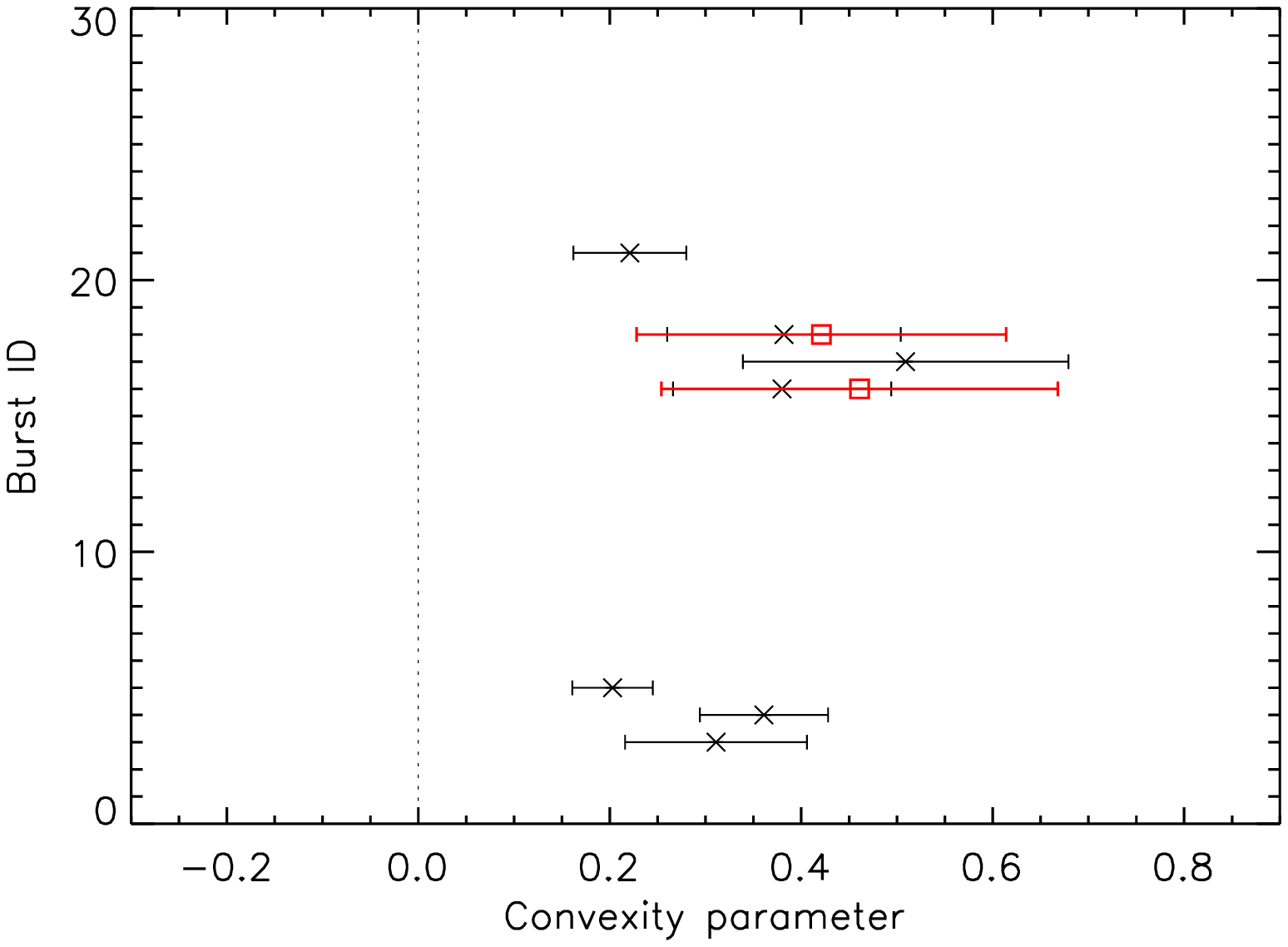} &
\includegraphics[width=0.35\textheight]{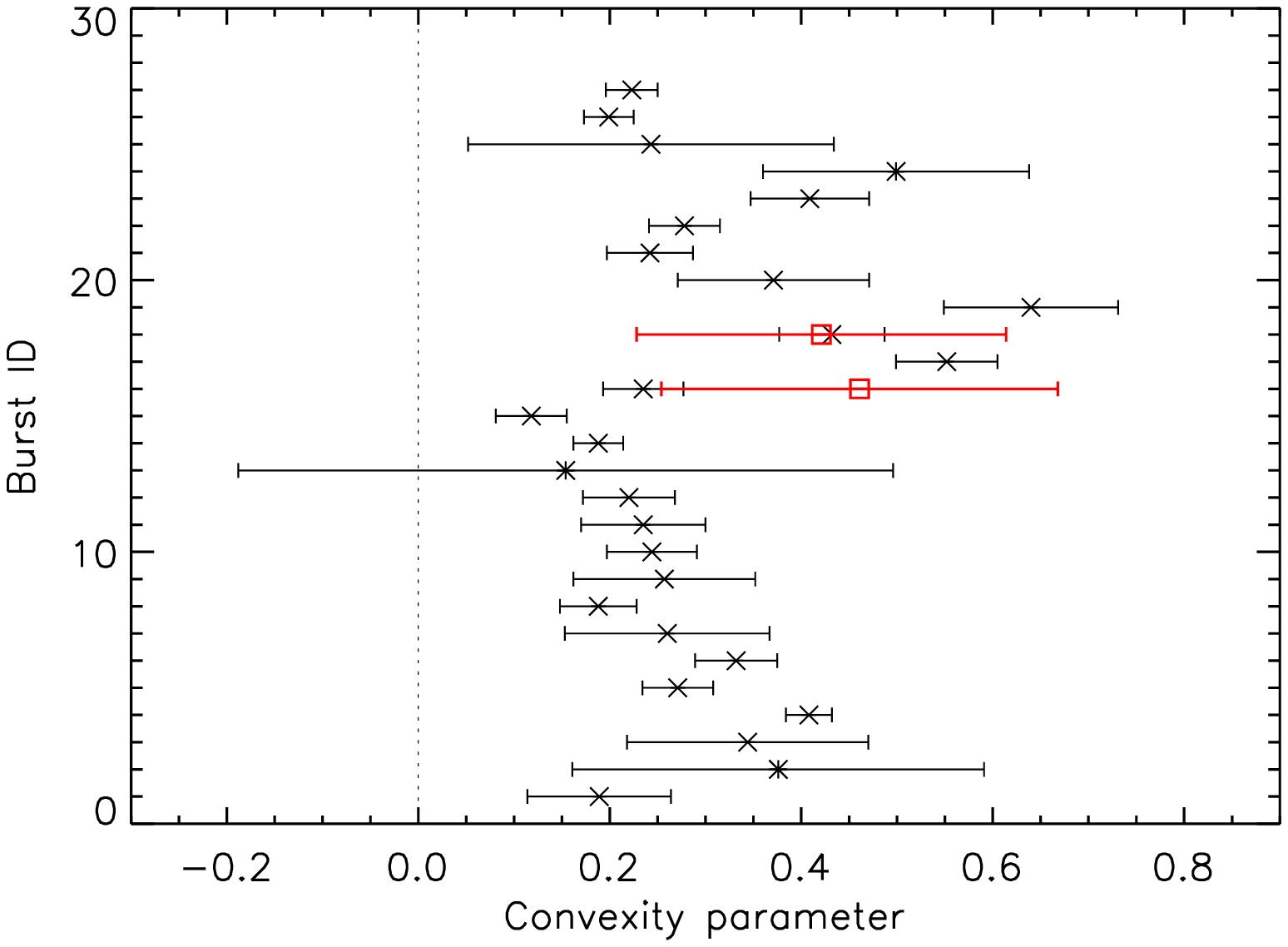} \\
\end{tabular}
\caption{{\it Left panel}: the best-fit convexity parameter $c$ (with $1\sigma$ error) for 
4U 1636--536 bursts having oscillations during rise detected with {\it RXTE} PCA
(see \S~\ref{amplitude_calculation}; Table~\ref{Log}). The parameter $c$ appears in the model $a-bc(1-e^{-t/c})$,
which is used to fit the time evolution of the fractional rms amplitude (see \S~\ref{Rms}).
The black crosses are for fitting (including the upper limit points) using the
likelihood maximization technique (see \S~\ref{Rms}). The red squares
are for fitting excluding the upper limit points for two bursts (16 and 18) with
detected oscillations in minimum five time bins, including at least one of the
first two bins (see \S~\ref{Rms}).
The dotted vertical line corresponds to $c = 0$.
{\it Right panel}: similar to the left panel, but the 
black crosses and stars are for fitting (including the upper limit points) using the
weighted least square minimization method, where 
the stars are for bursts with no detected oscillations in the first three time
bins (see \S~\ref{Rms}). 
This figure shows that all the best-fit values of $c$ cluster to the right of
the dotted vertical line, implying latitude-dependent flame speeds, possibly
due to the effects of the Coriolis force on thermonuclear
flame spreading (see \S~\ref{Discussions}). 
\label{convexpar}}
\end{figure*}

\begin{figure*}
\centering
\begin{tabular}{lr}
\includegraphics[width=0.35\textheight]{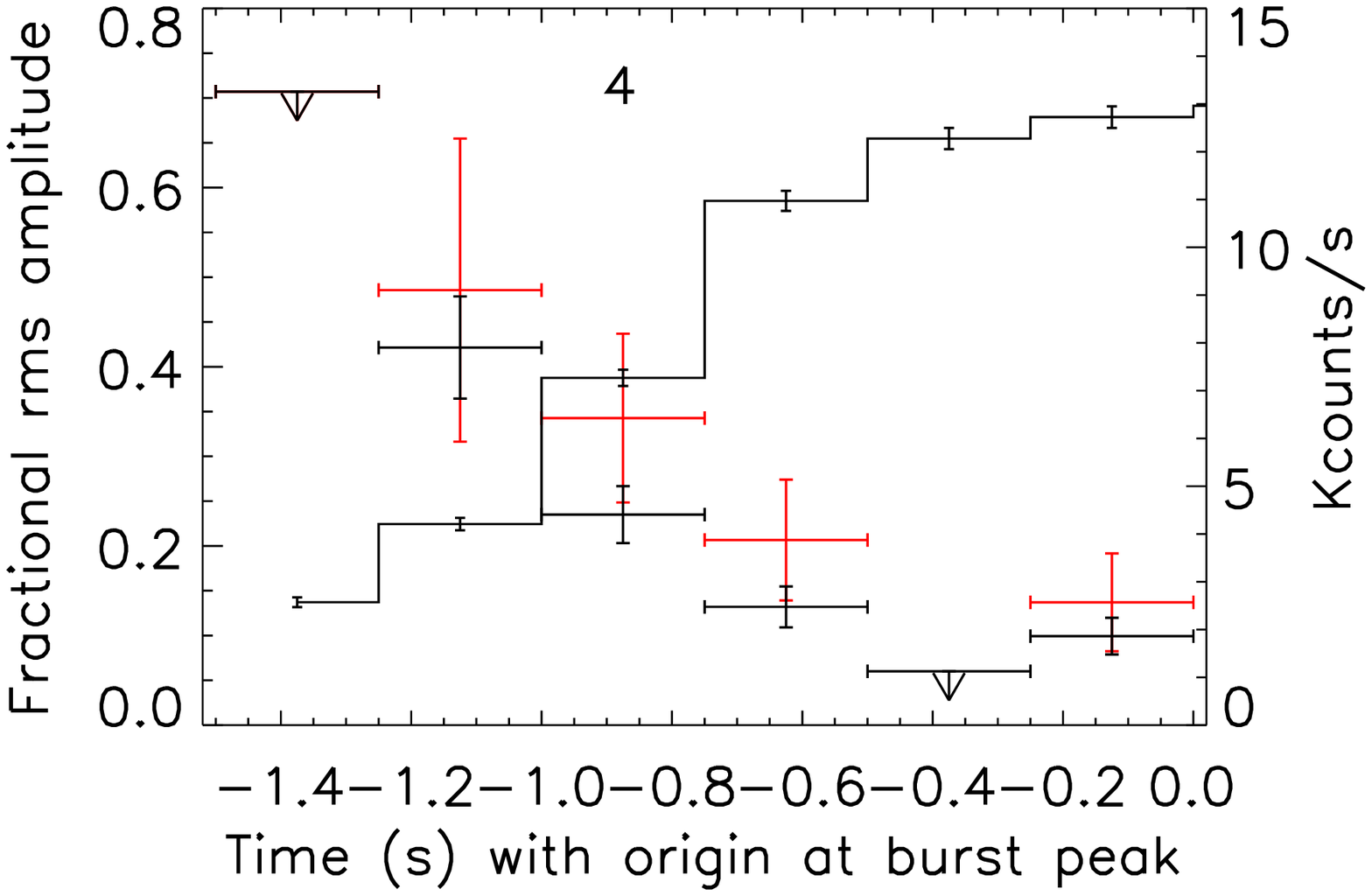} &
\includegraphics[width=0.35\textheight]{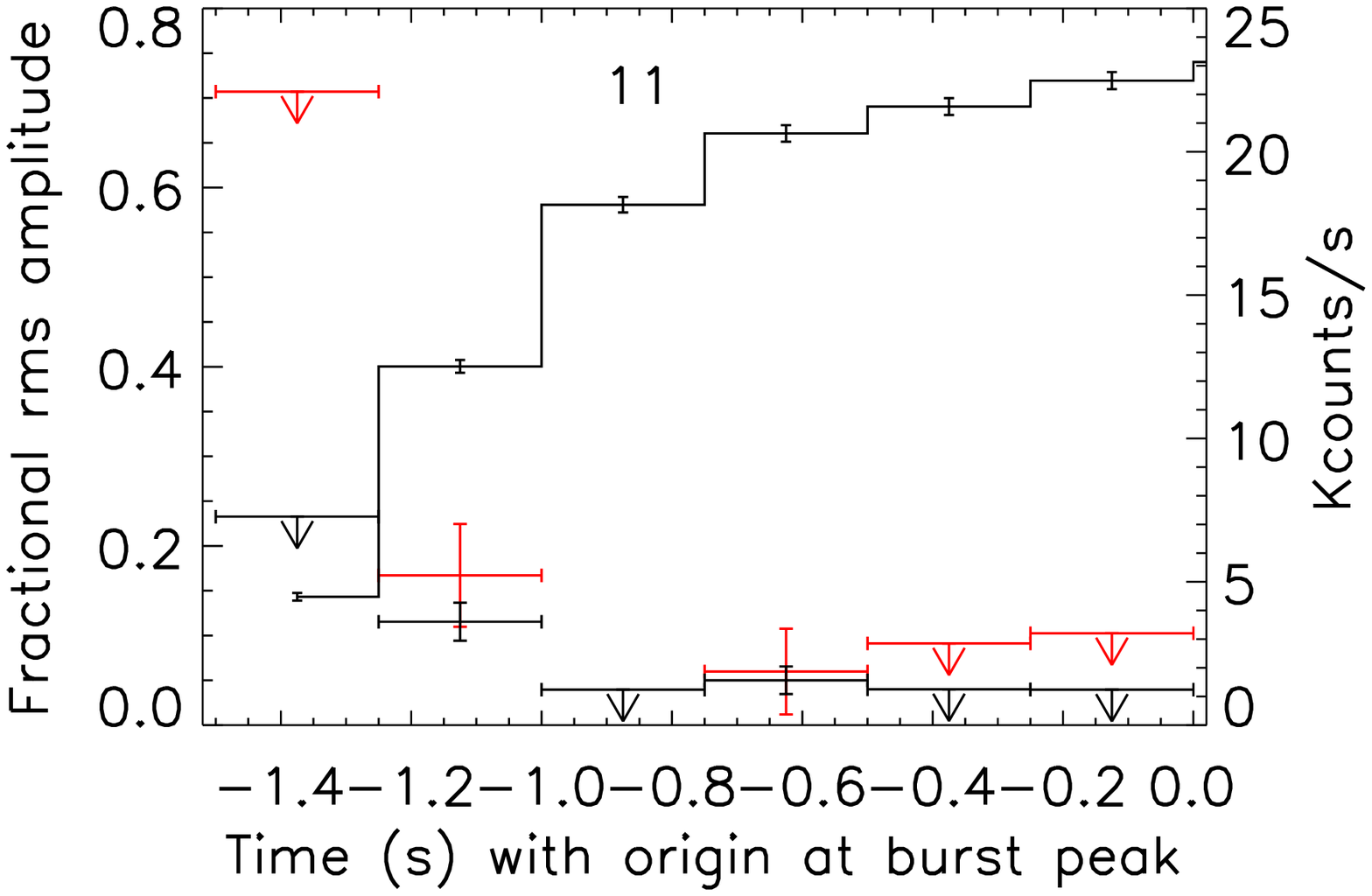} \\
\end{tabular}
\caption{Similar to panels 4 and 11 of Fig.~\ref{rmsampevol}, but in addition to the
points without considering $f_a$ (shown in black), the points considering $f_a$ are
also shown (in red).
This figure shows that the fractional rms oscillation amplitude usually decreases with time
during burst rise, even when the effect of the changing persistent emission is considered 
(\S~\ref{Rms}).
\label{rmsampevolfa}} 
\end{figure*}

\clearpage
\begin{figure*}
\centering
\begin{tabular}{lr}
\includegraphics[width=0.32\textheight]{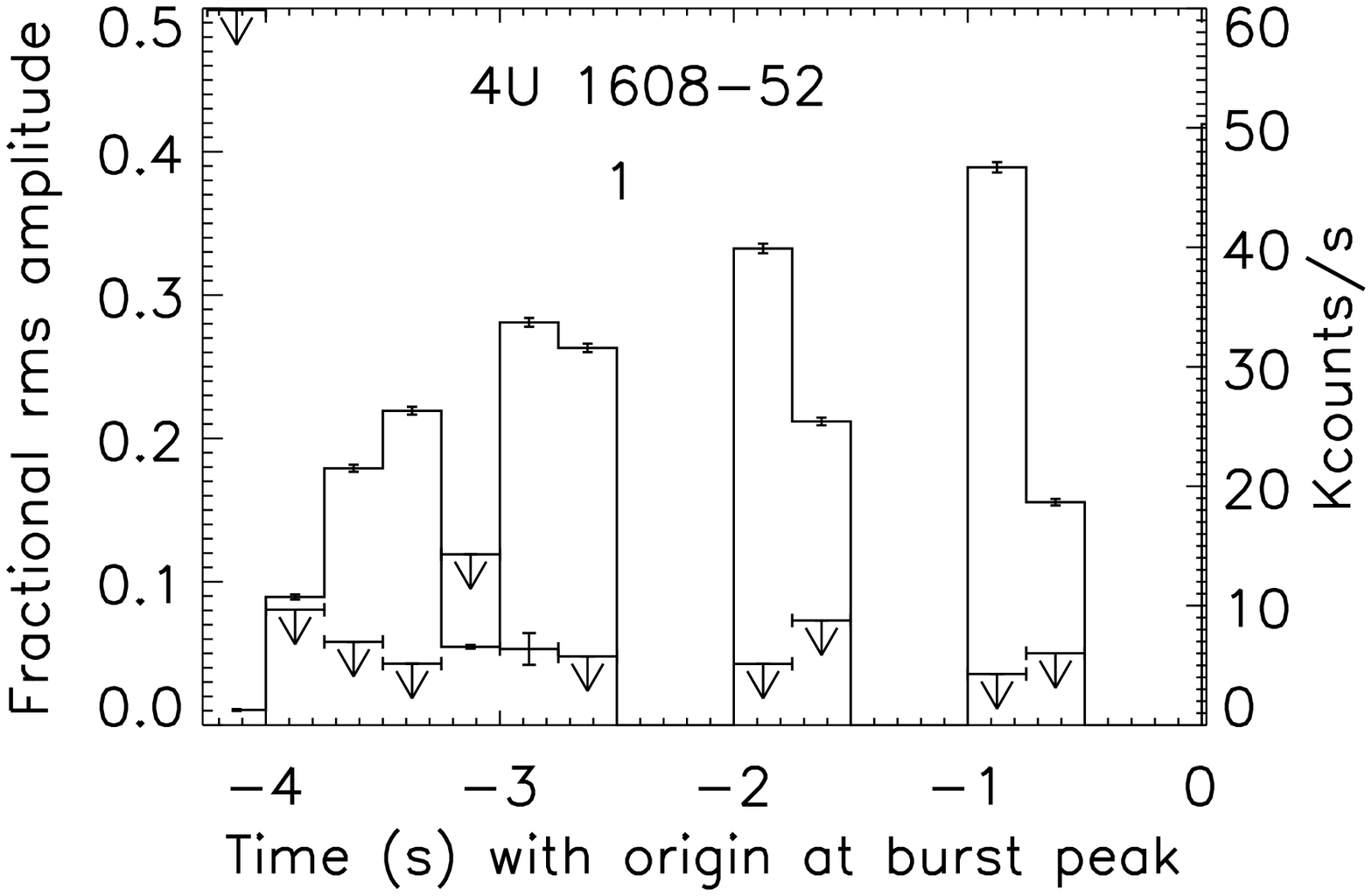} &
\includegraphics[width=0.32\textheight]{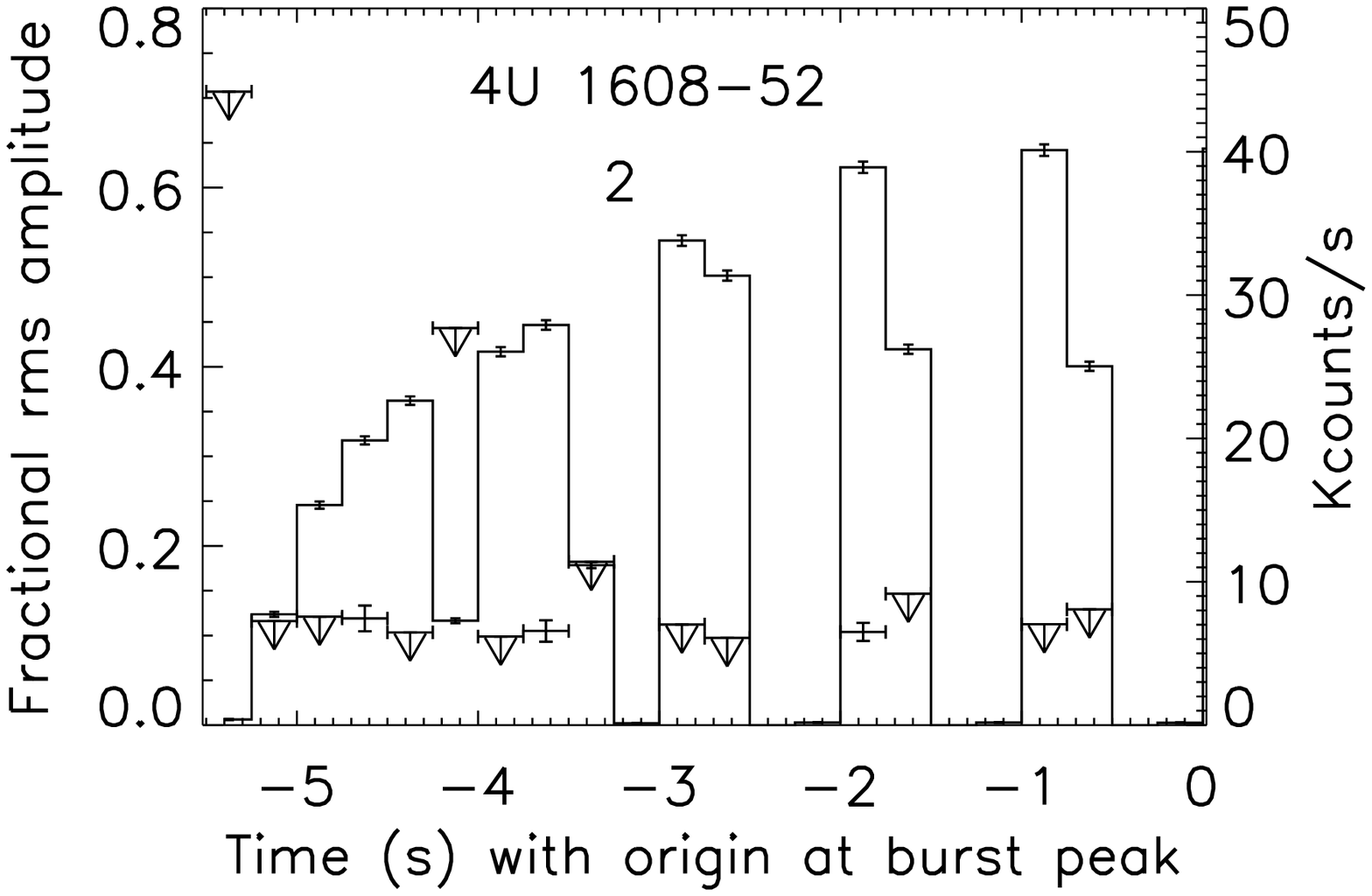} \\ 
\includegraphics[width=0.32\textheight]{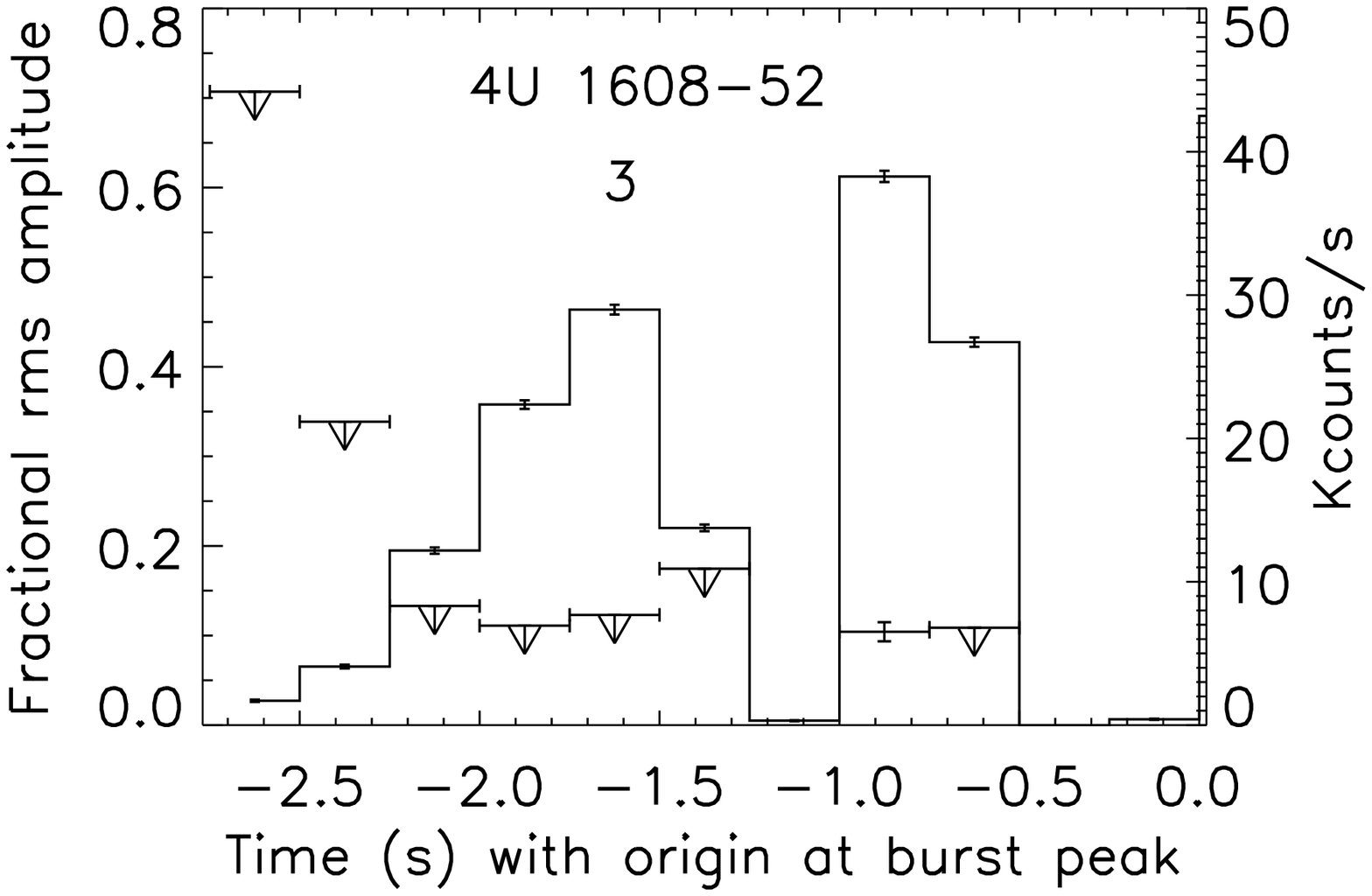} &
\includegraphics[width=0.32\textheight]{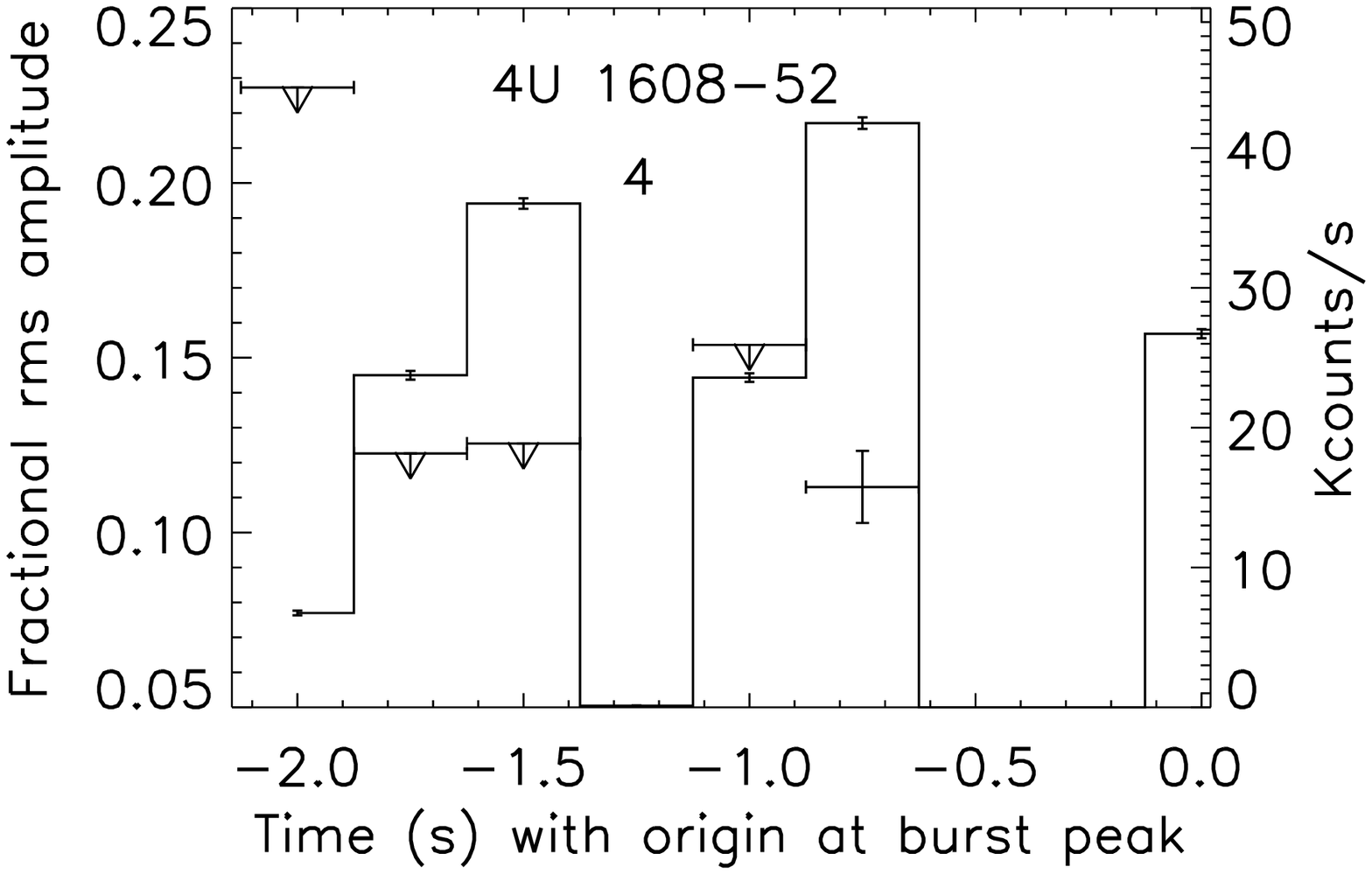} \\
\includegraphics[width=0.32\textheight]{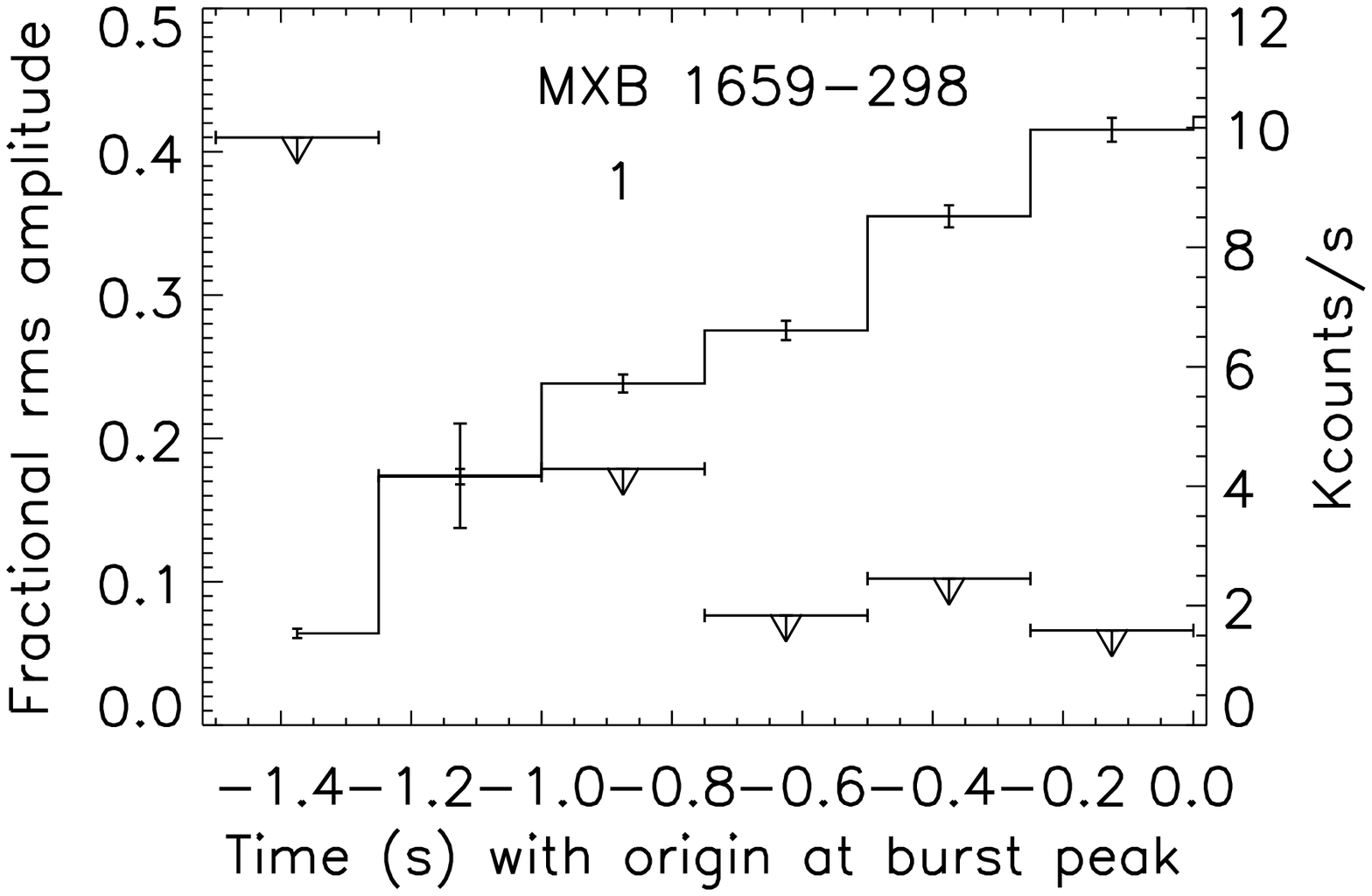} &
\includegraphics[width=0.32\textheight]{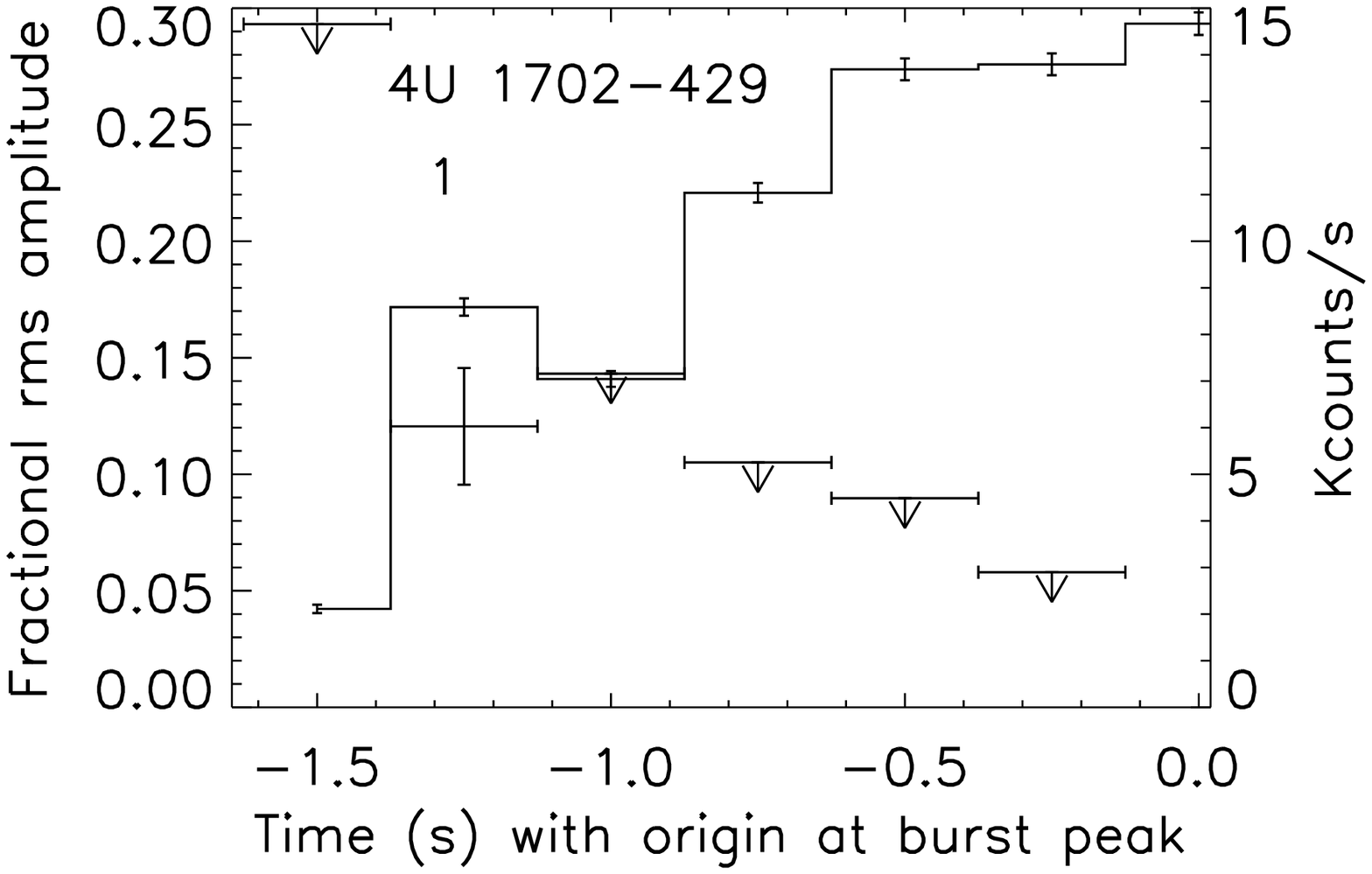} \\
\includegraphics[width=0.32\textheight]{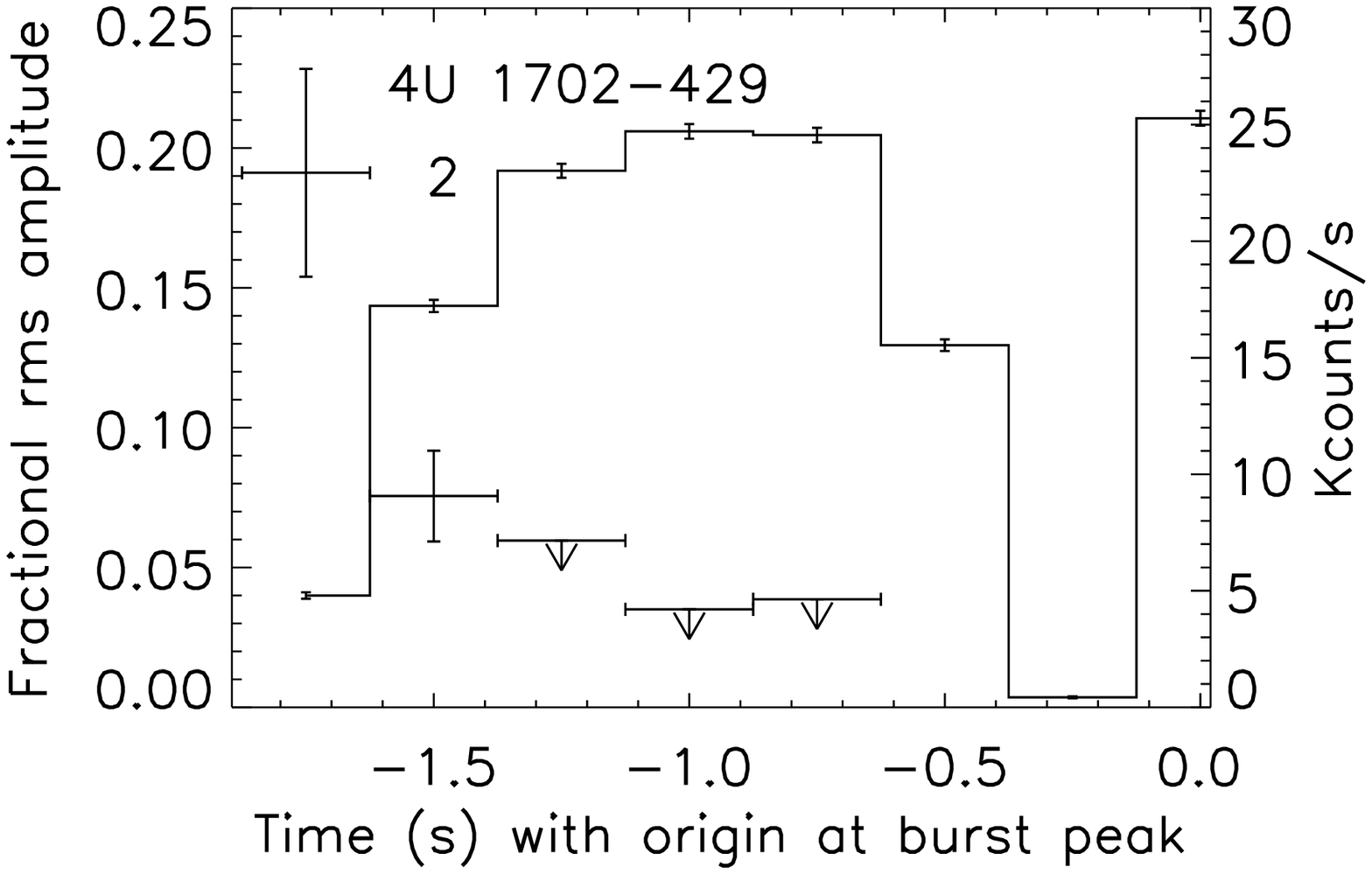} &
\includegraphics[width=0.32\textheight]{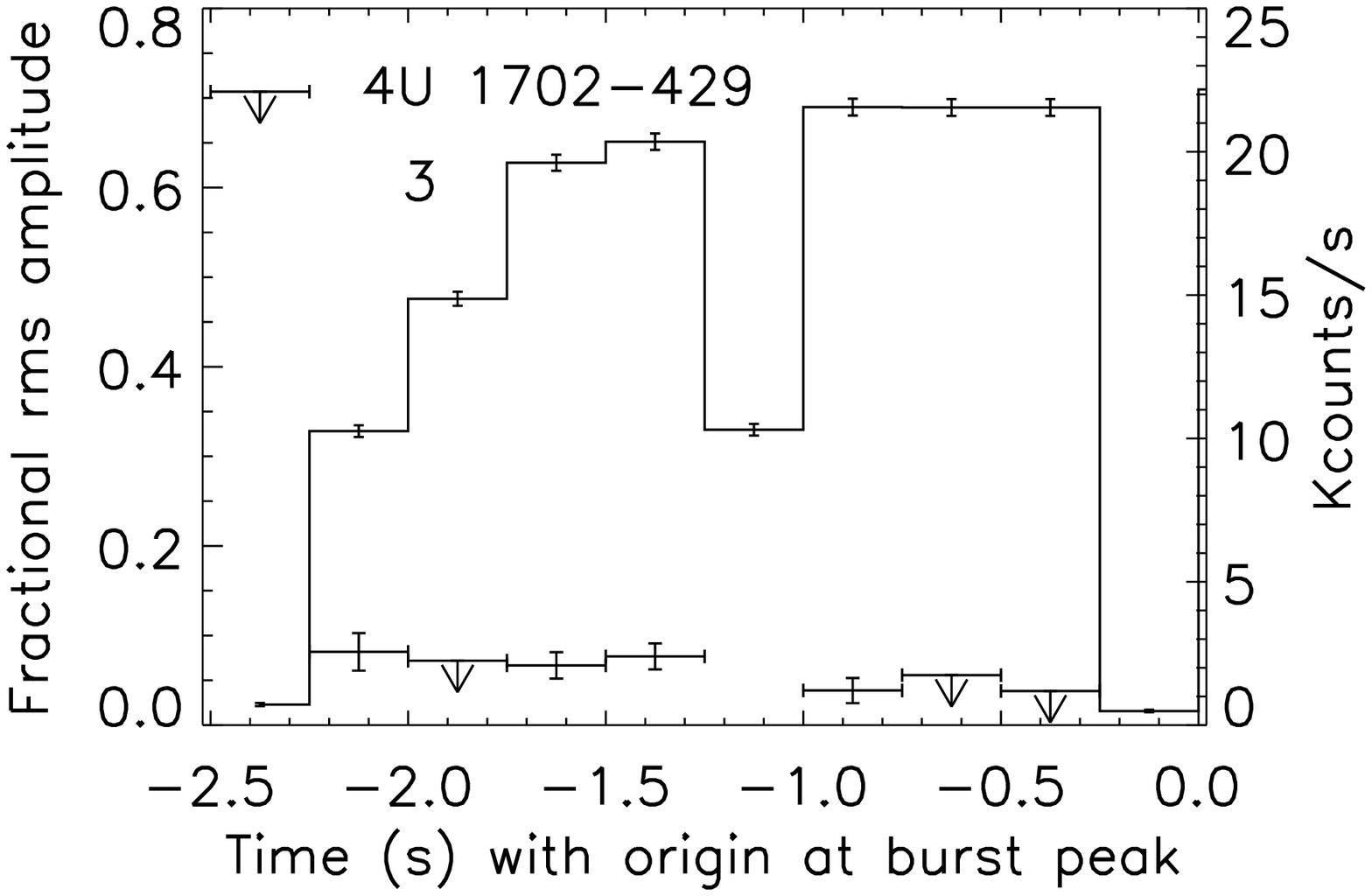} \\
\end{tabular}
\caption{Same as Fig.~\ref{rmsampevol}, but for 24 bursts with rise oscillations 
from nine sources
(mentioned in Table~\ref{Log2}). Some bursts from some sources, 
especially 4U 1608--52, have a number of data gaps because of high intensity. 
These data gaps are excluded for the fractional rms amplitude calculation 
when possible (\S~\ref{othersources}). 
\label{rmsampevol2}}
\end{figure*}

\clearpage
\addtocounter{figure}{ -1}
\begin{figure*}
\centering
\begin{tabular}{lr} 
\includegraphics[width=0.32\textheight]{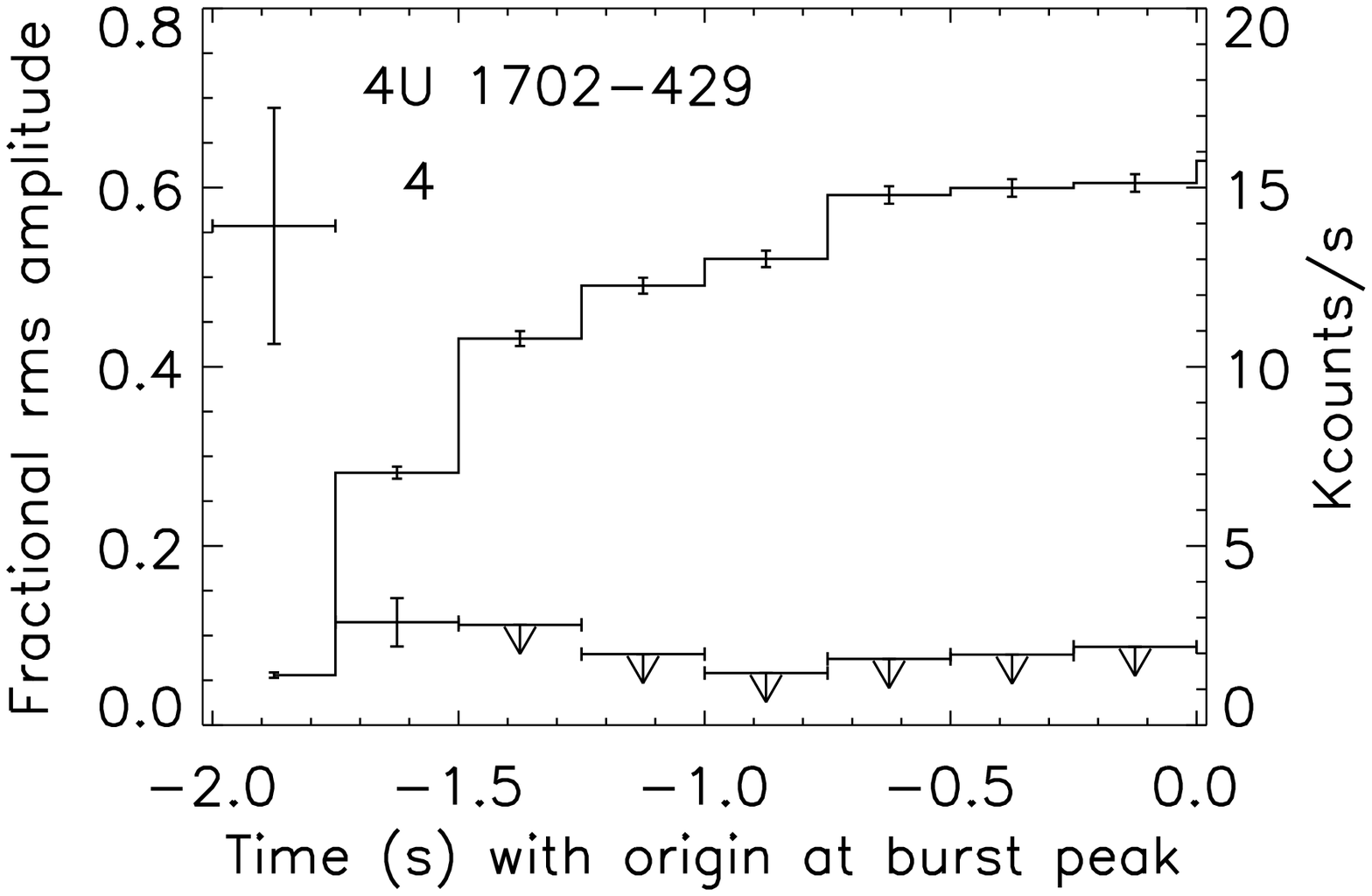} &
\includegraphics[width=0.32\textheight]{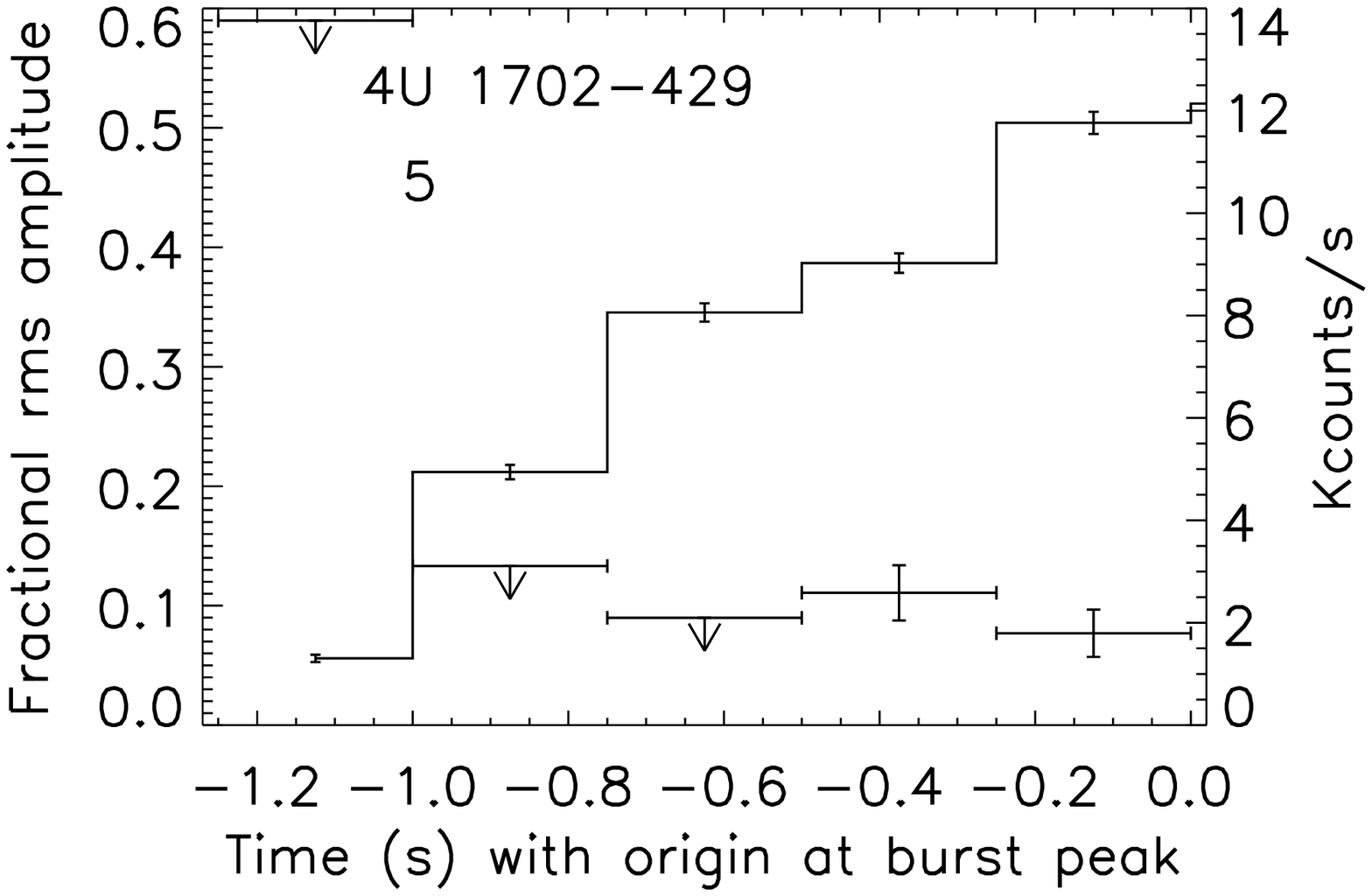} \\ 
\includegraphics[width=0.32\textheight]{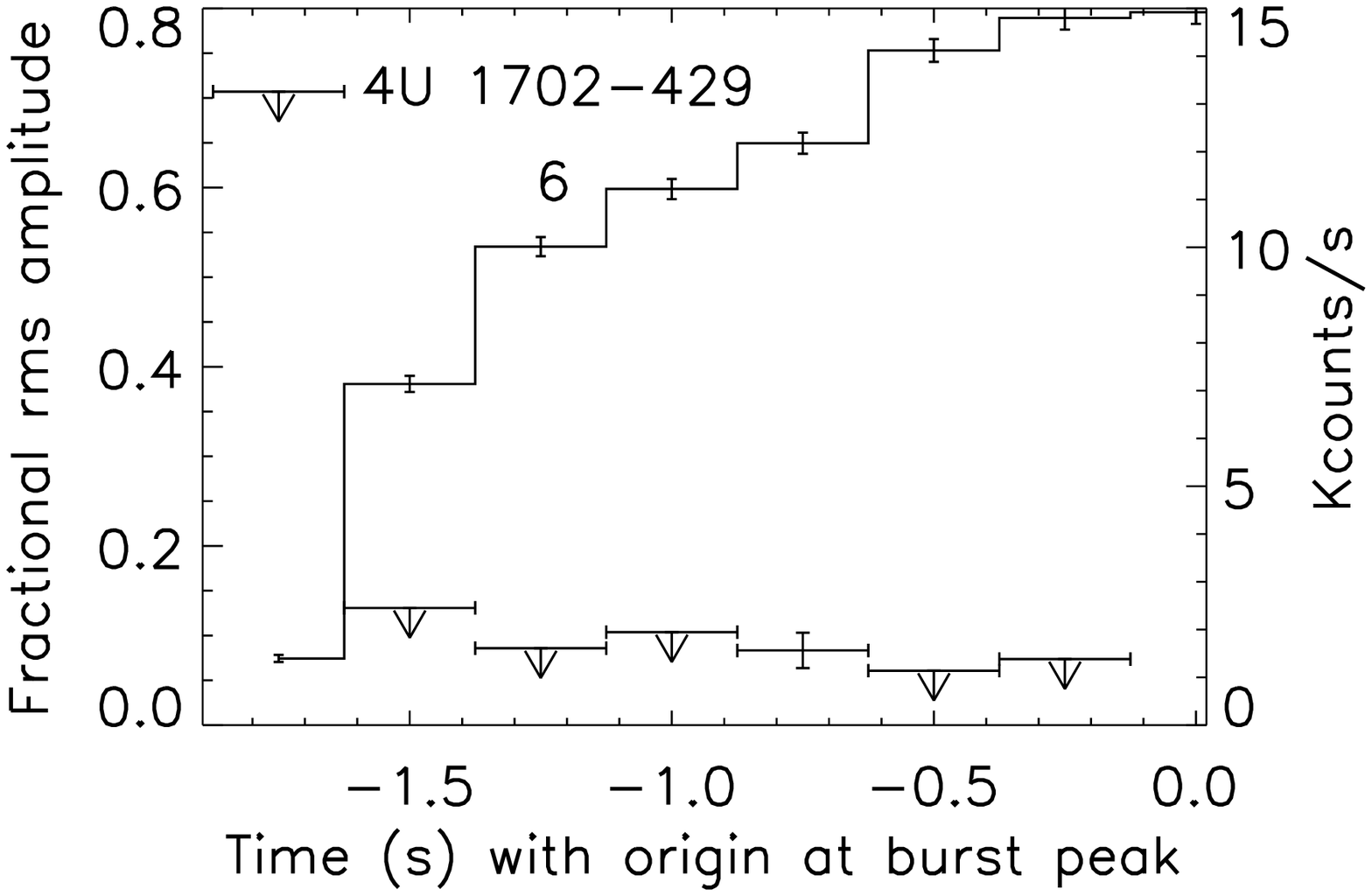} &
\includegraphics[width=0.32\textheight]{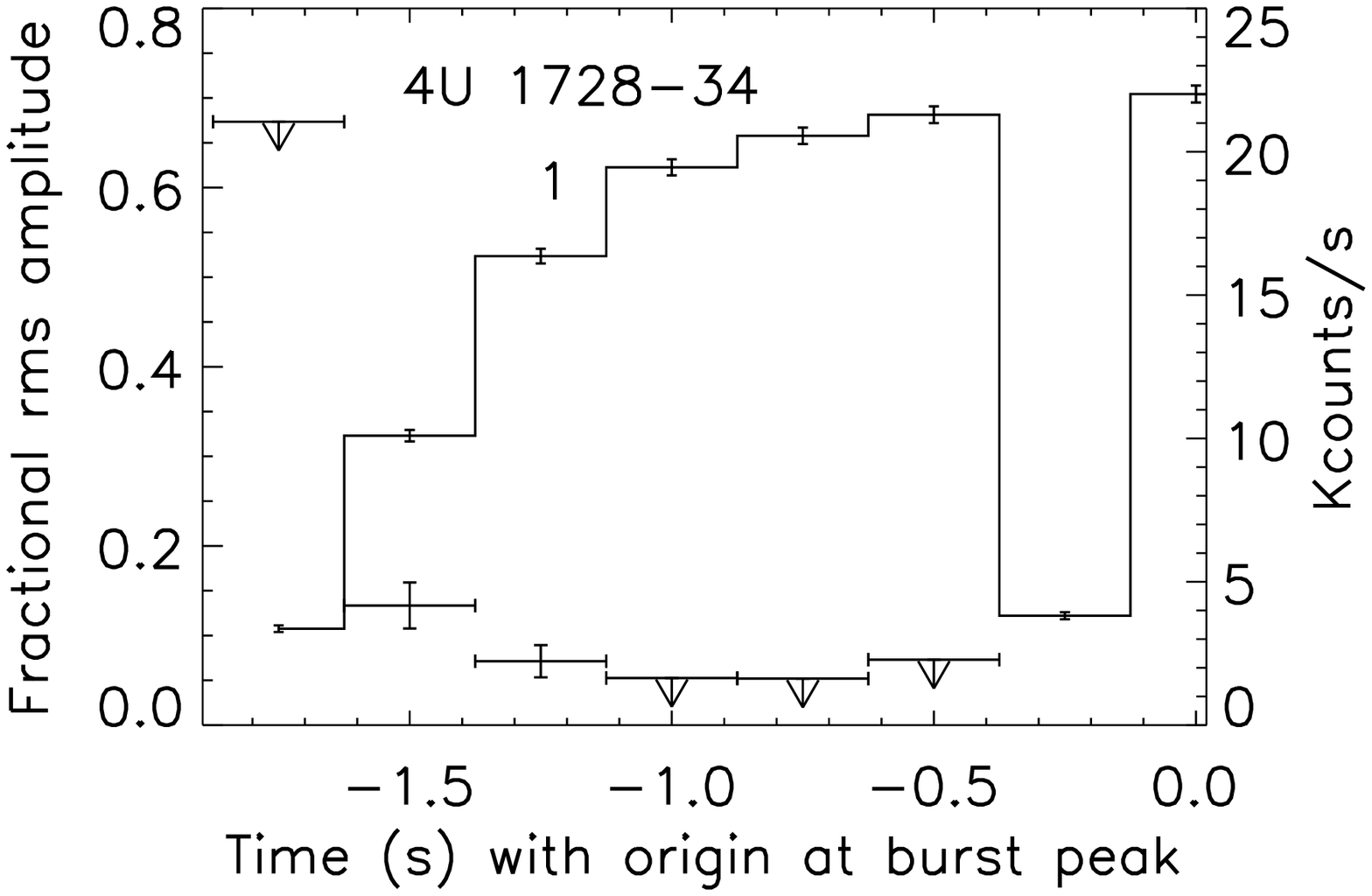} \\ 
\includegraphics[width=0.32\textheight]{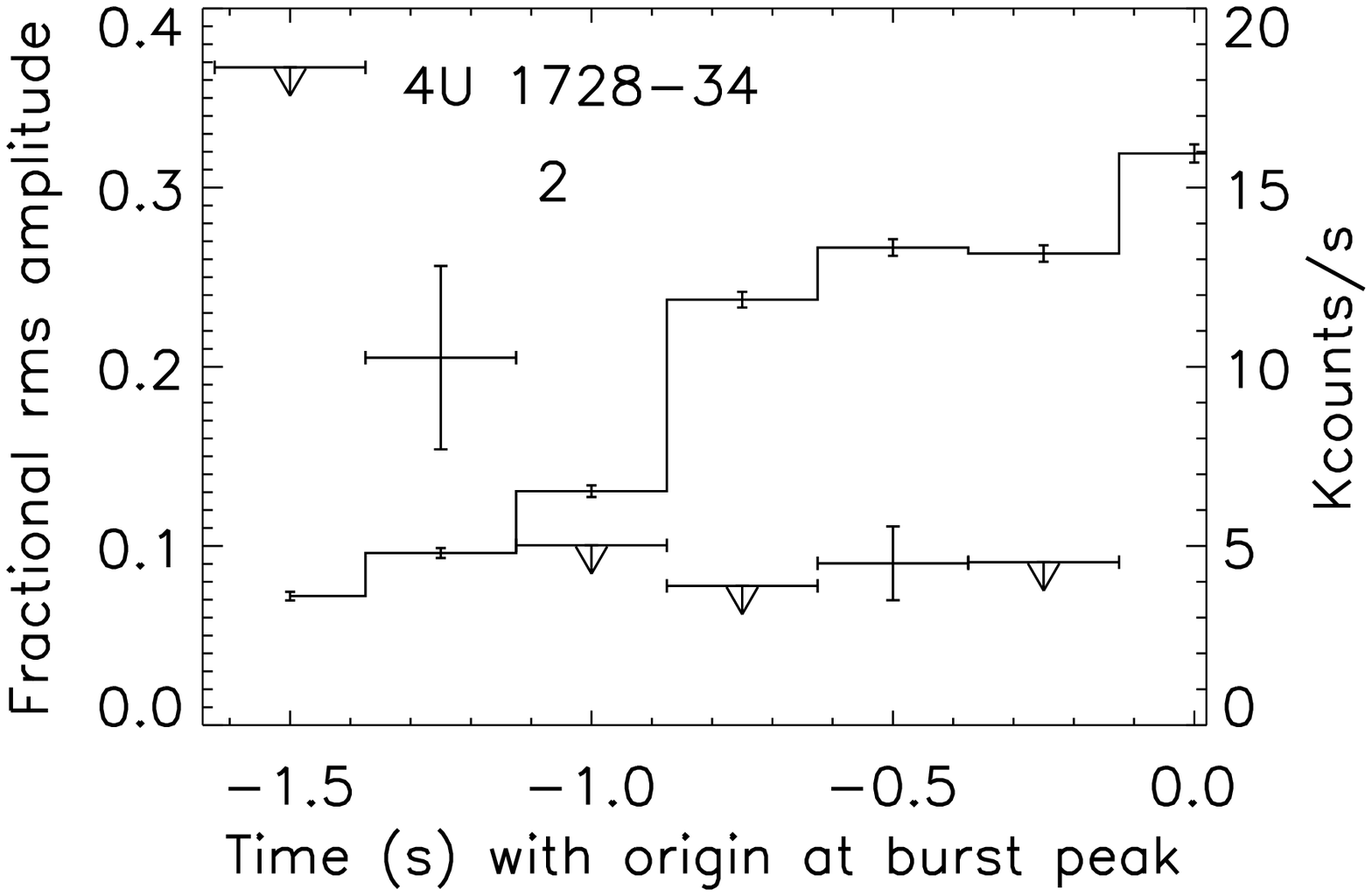} &
\includegraphics[width=0.32\textheight]{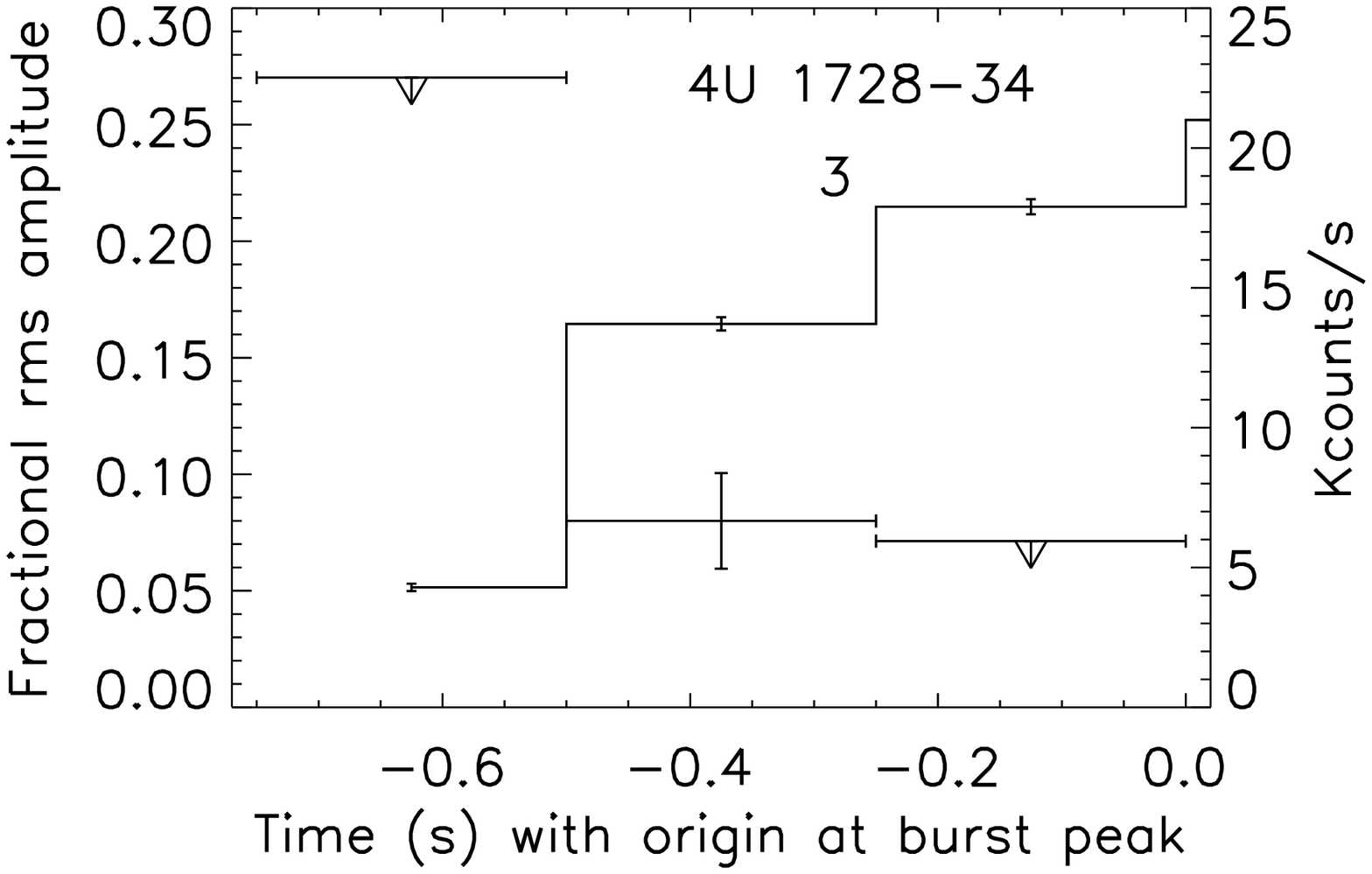} \\
\includegraphics[width=0.32\textheight]{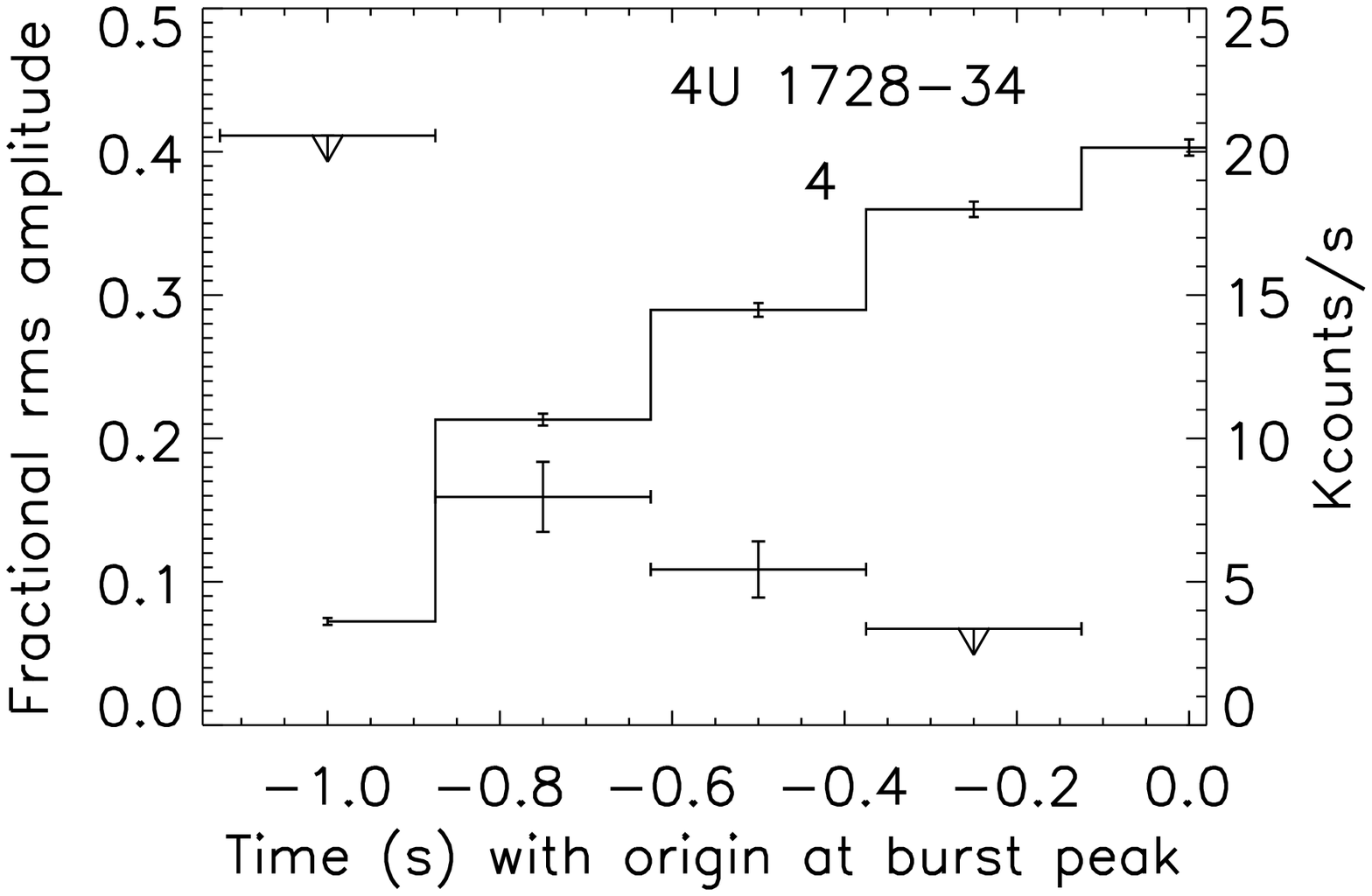} &
\includegraphics[width=0.32\textheight]{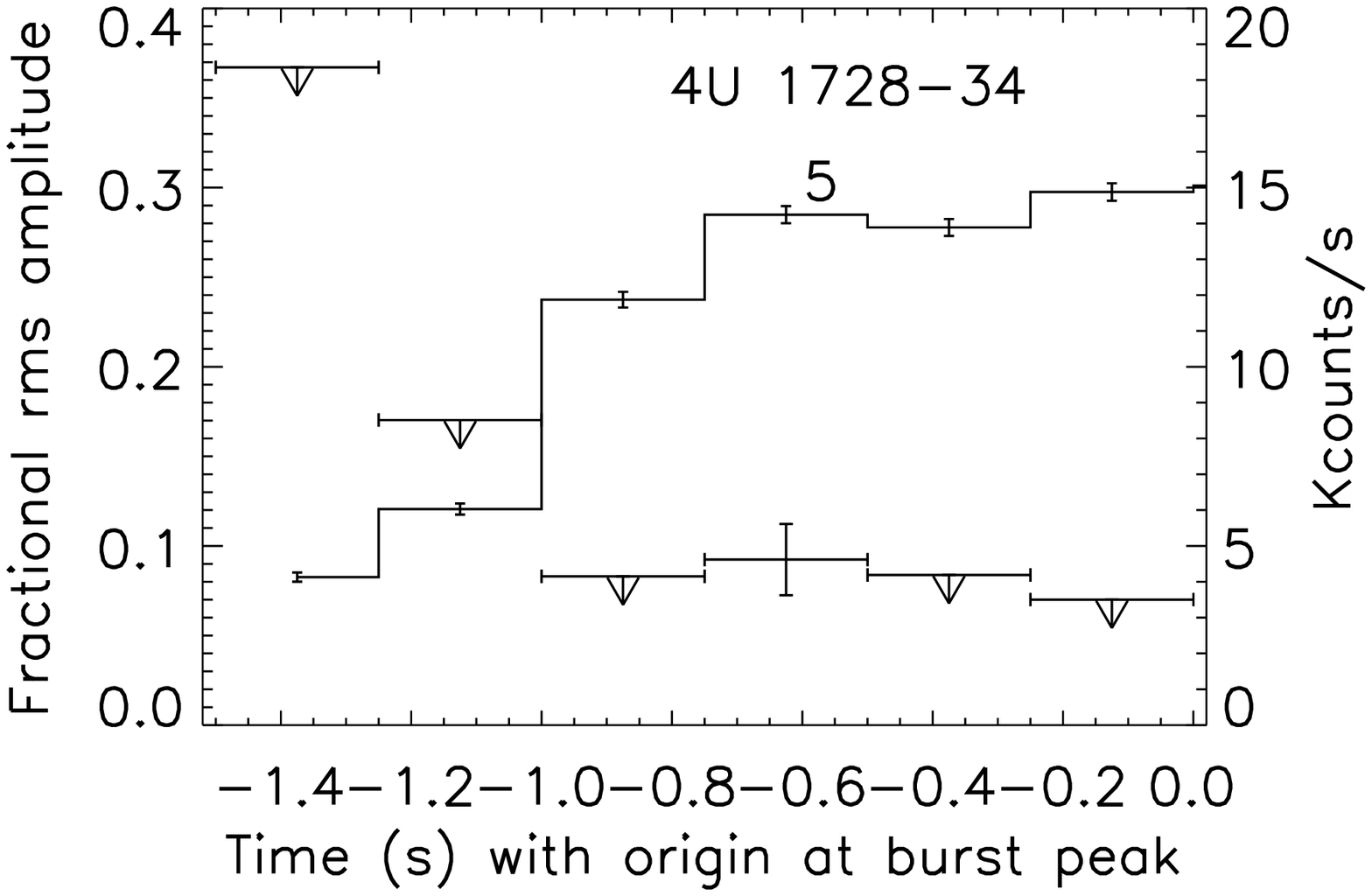} \\
\end{tabular}
\caption{Continued.}
\end{figure*}

\clearpage
\addtocounter{figure}{ -1}
\begin{figure*}
\centering
\begin{tabular}{lr}

\includegraphics[width=0.32\textheight]{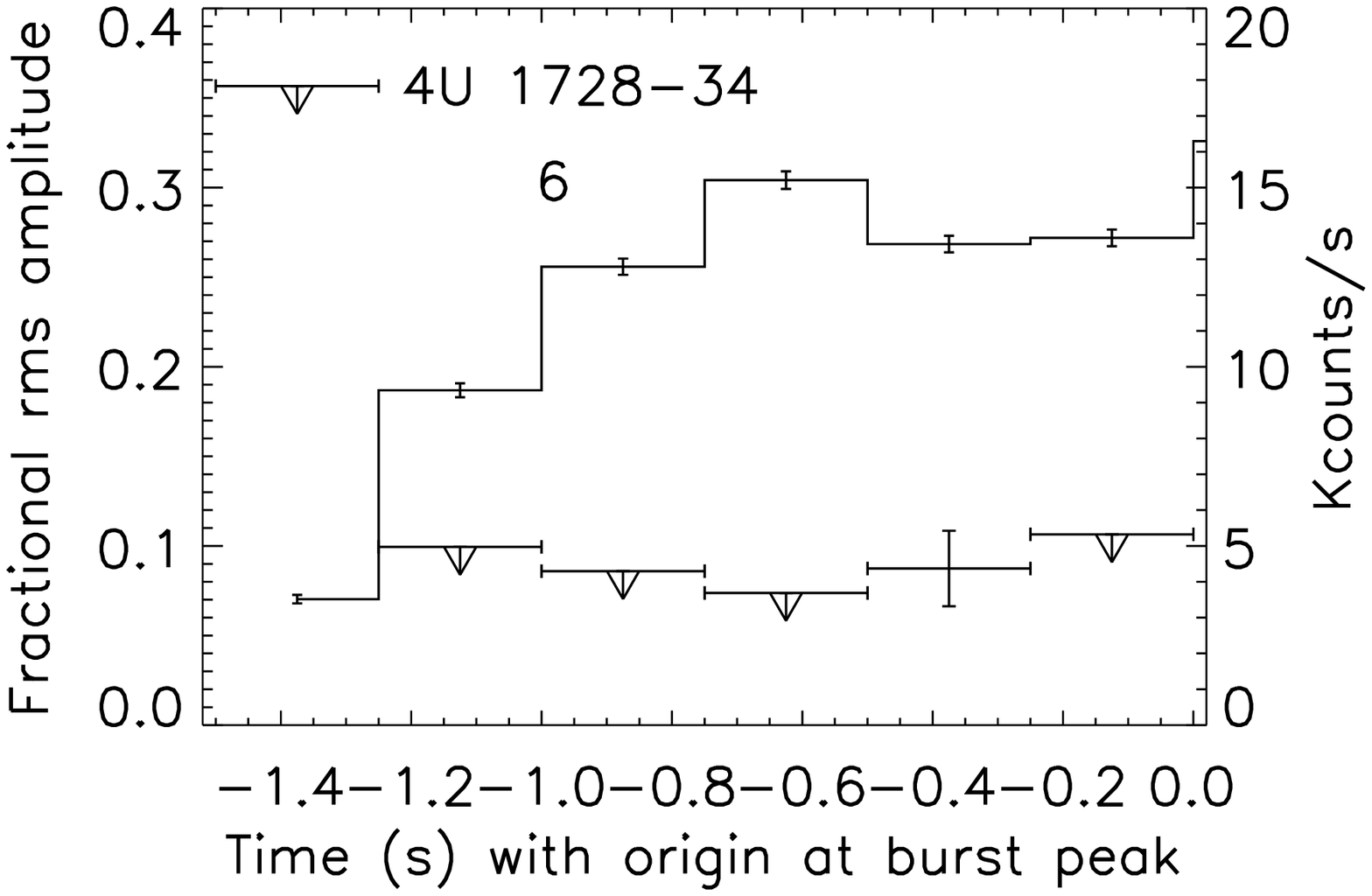} &
\includegraphics[width=0.32\textheight]{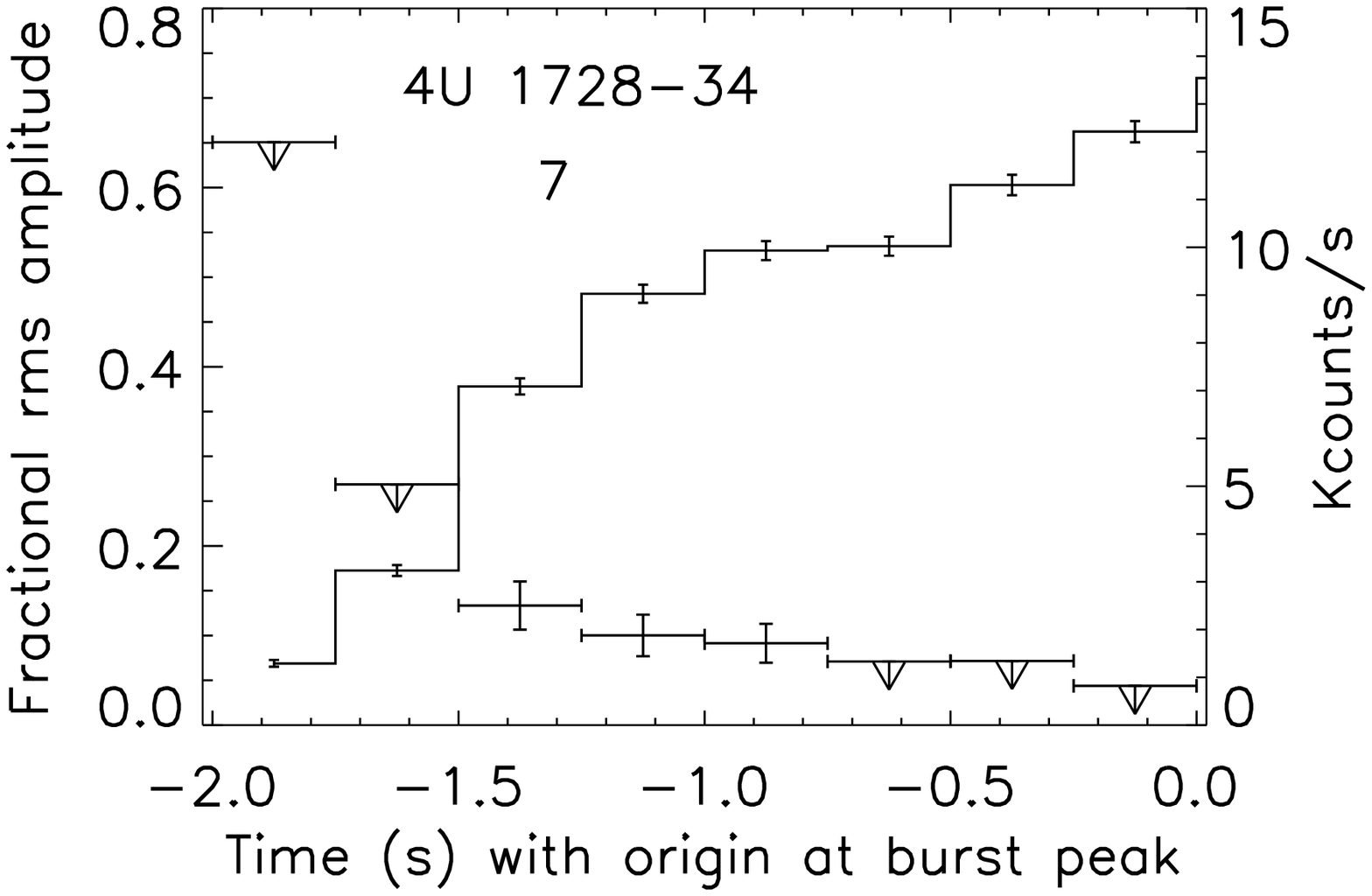} \\
\includegraphics[width=0.32\textheight]{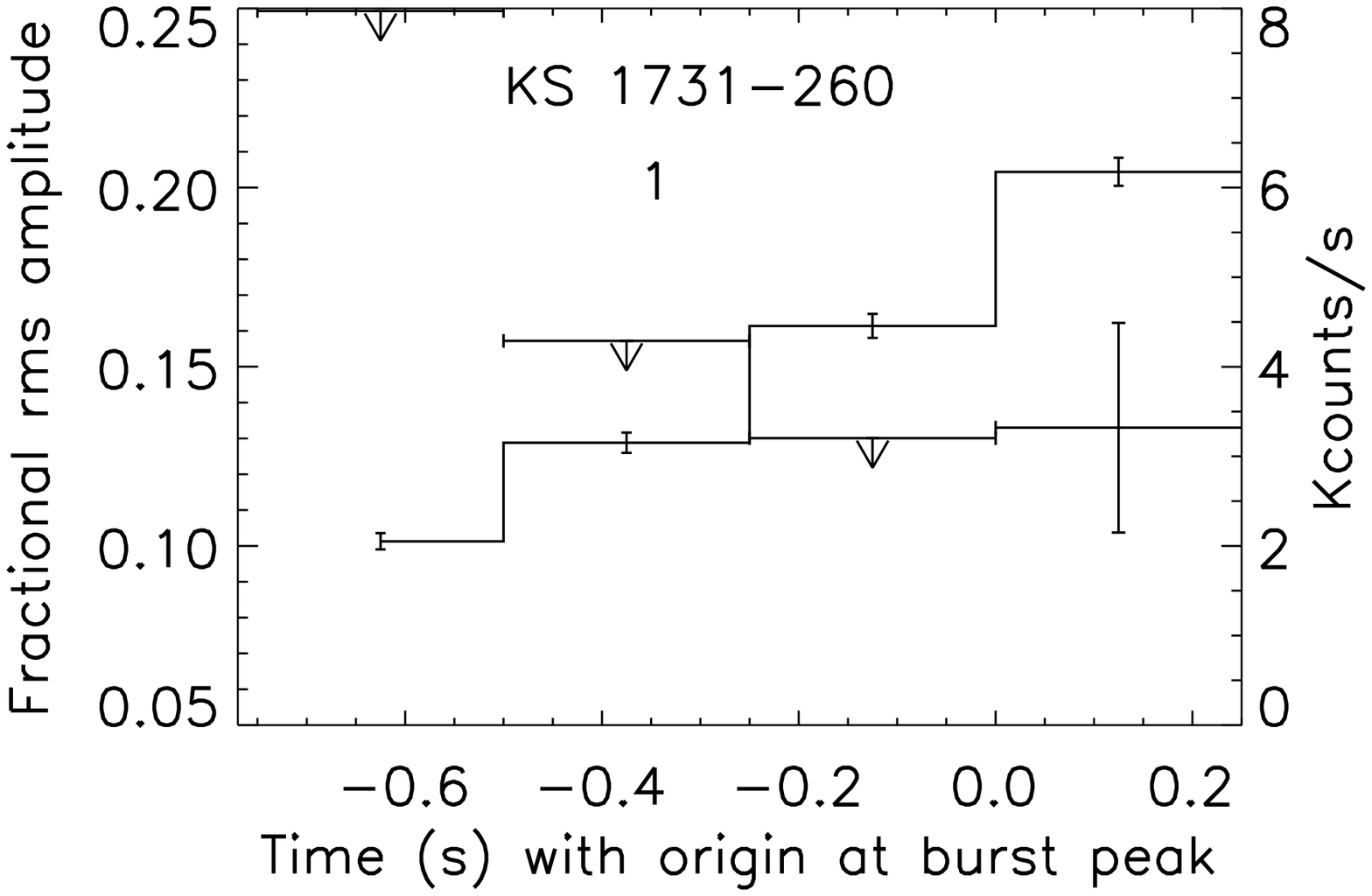} &
\includegraphics[width=0.32\textheight]{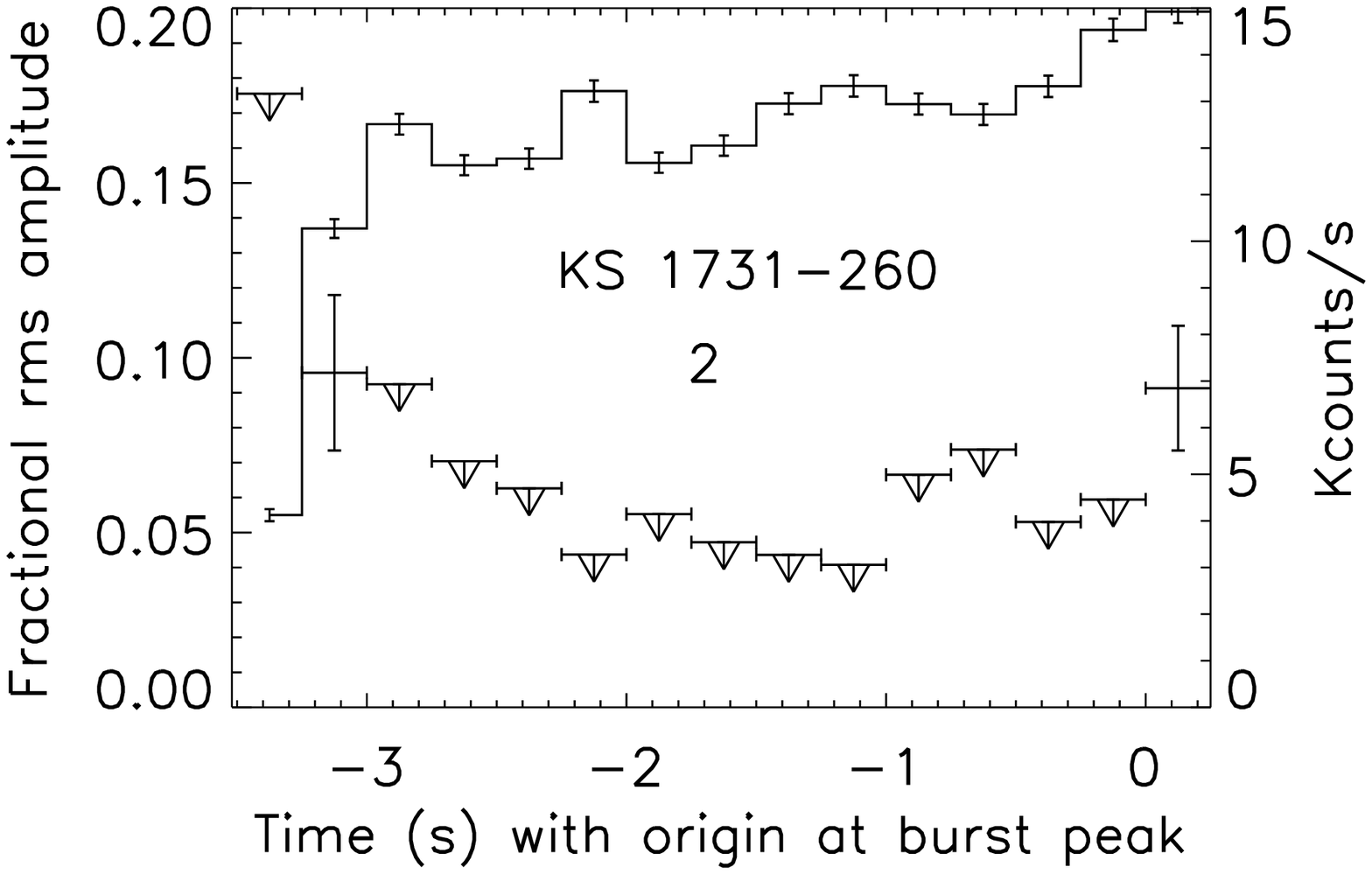} \\
\includegraphics[width=0.32\textheight]{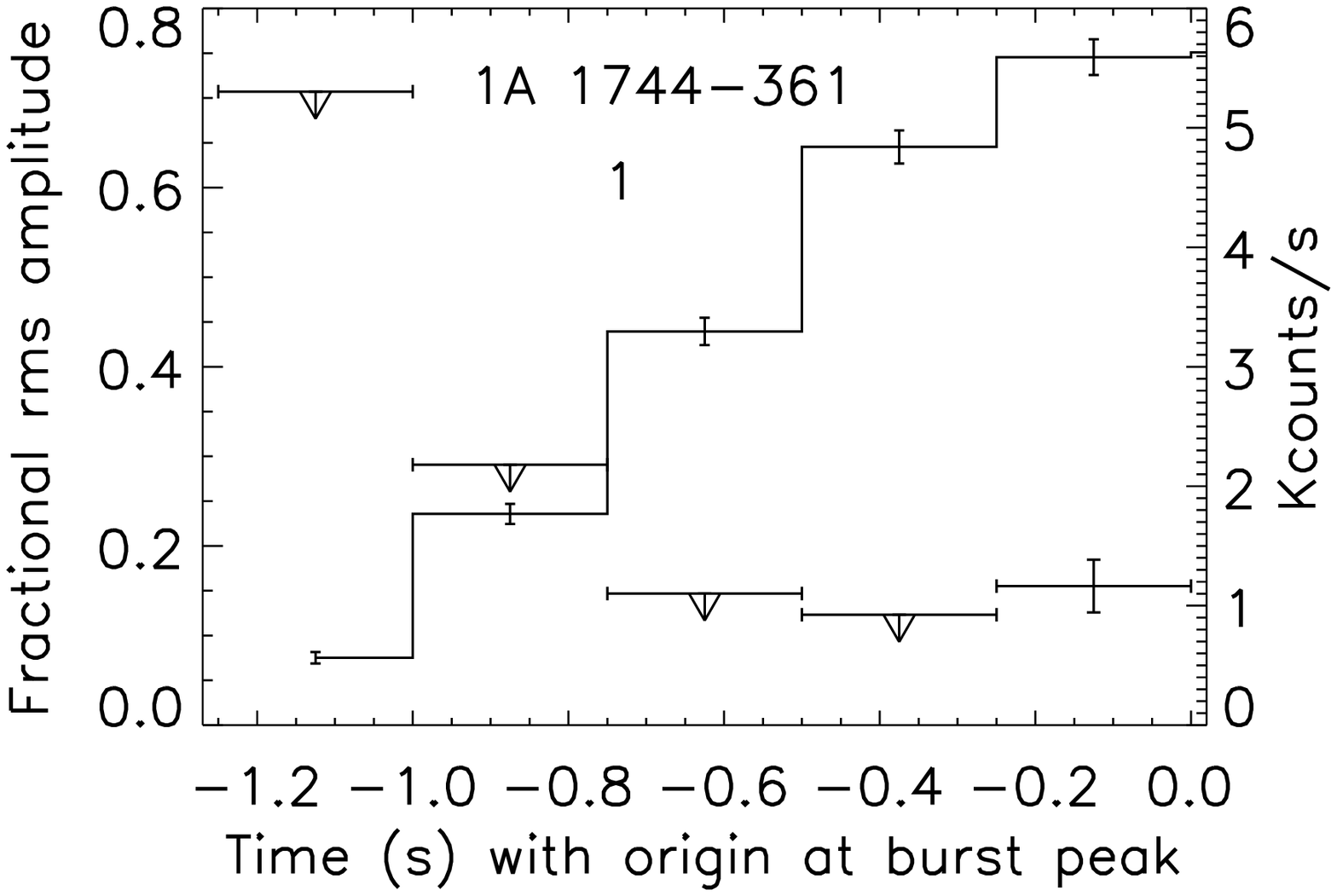}	&
\includegraphics[width=0.32\textheight]{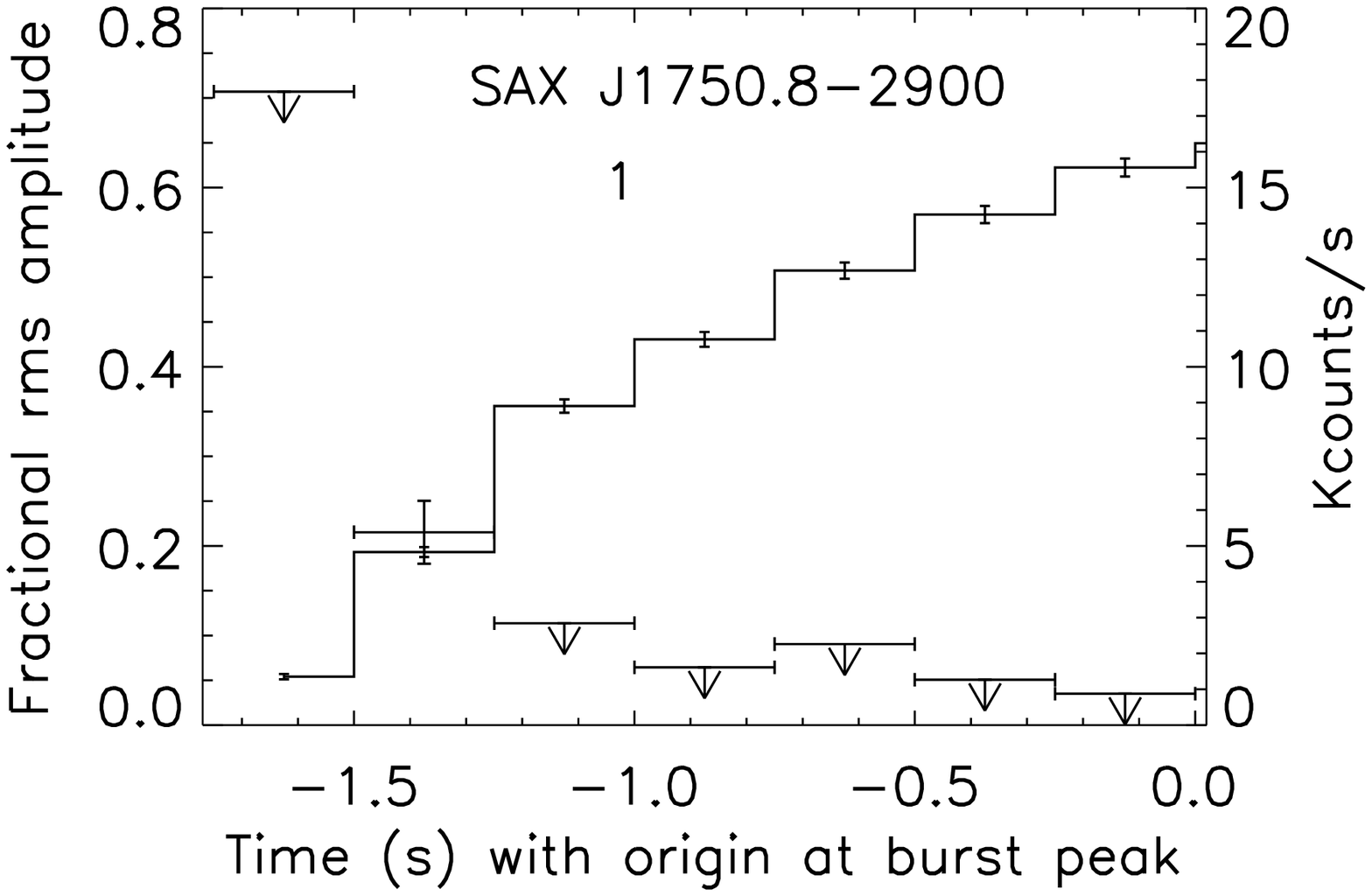} \\
\includegraphics[width=0.32\textheight]{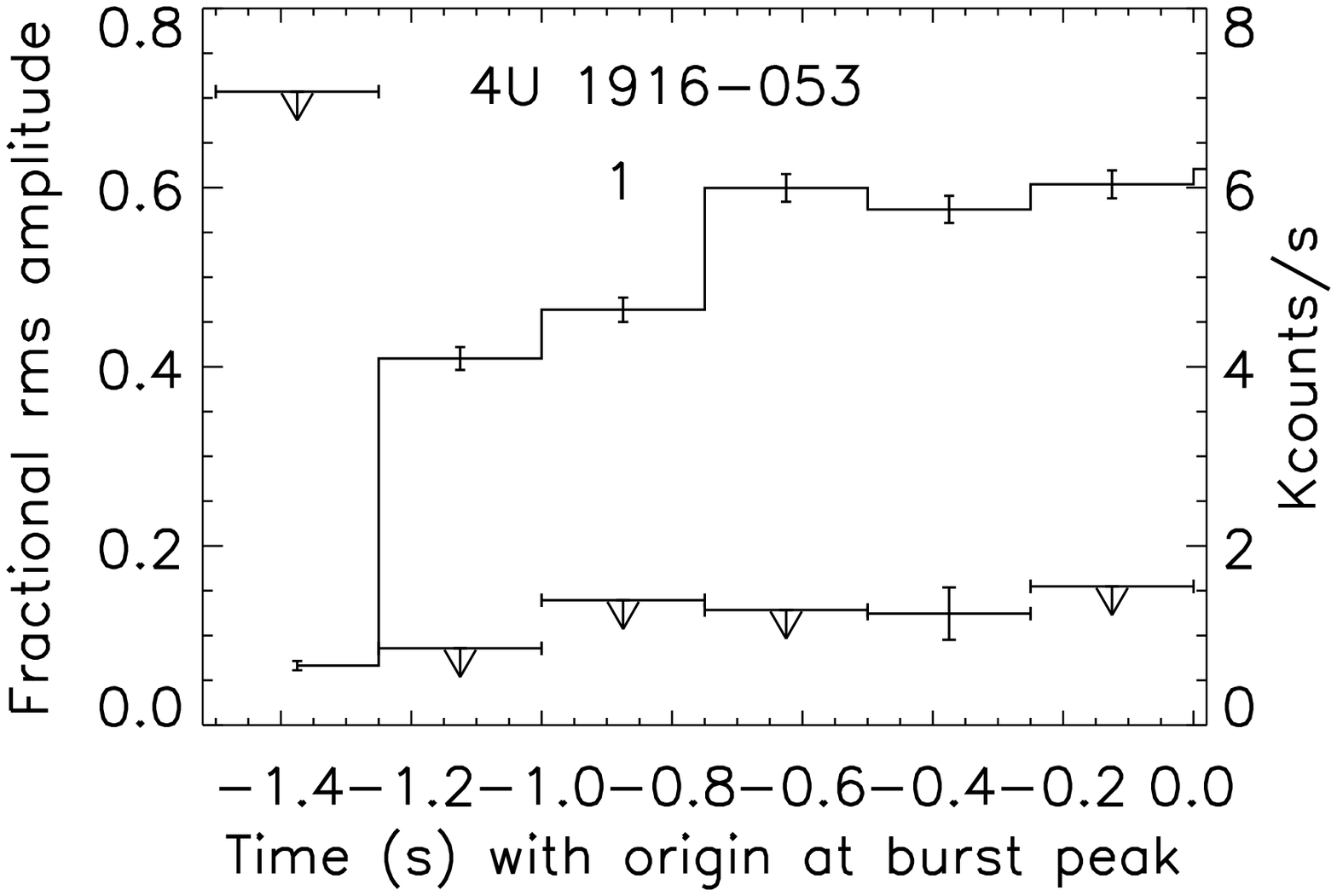} &
\includegraphics[width=0.32\textheight]{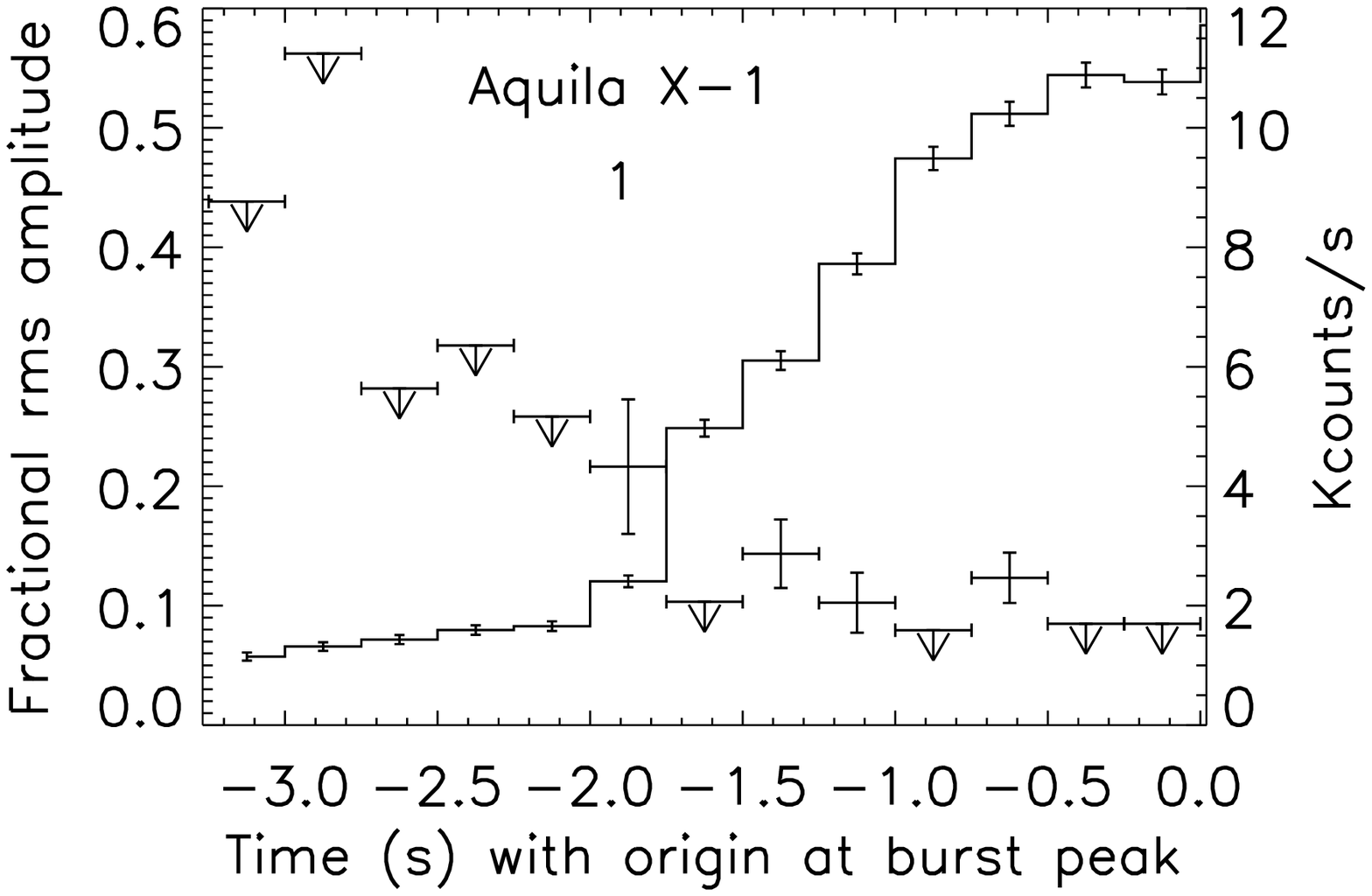} \\
\end{tabular}
\caption{Continued.}
\end{figure*}

\clearpage
\begin{figure*}
\centering
\includegraphics[width=0.66\textheight]{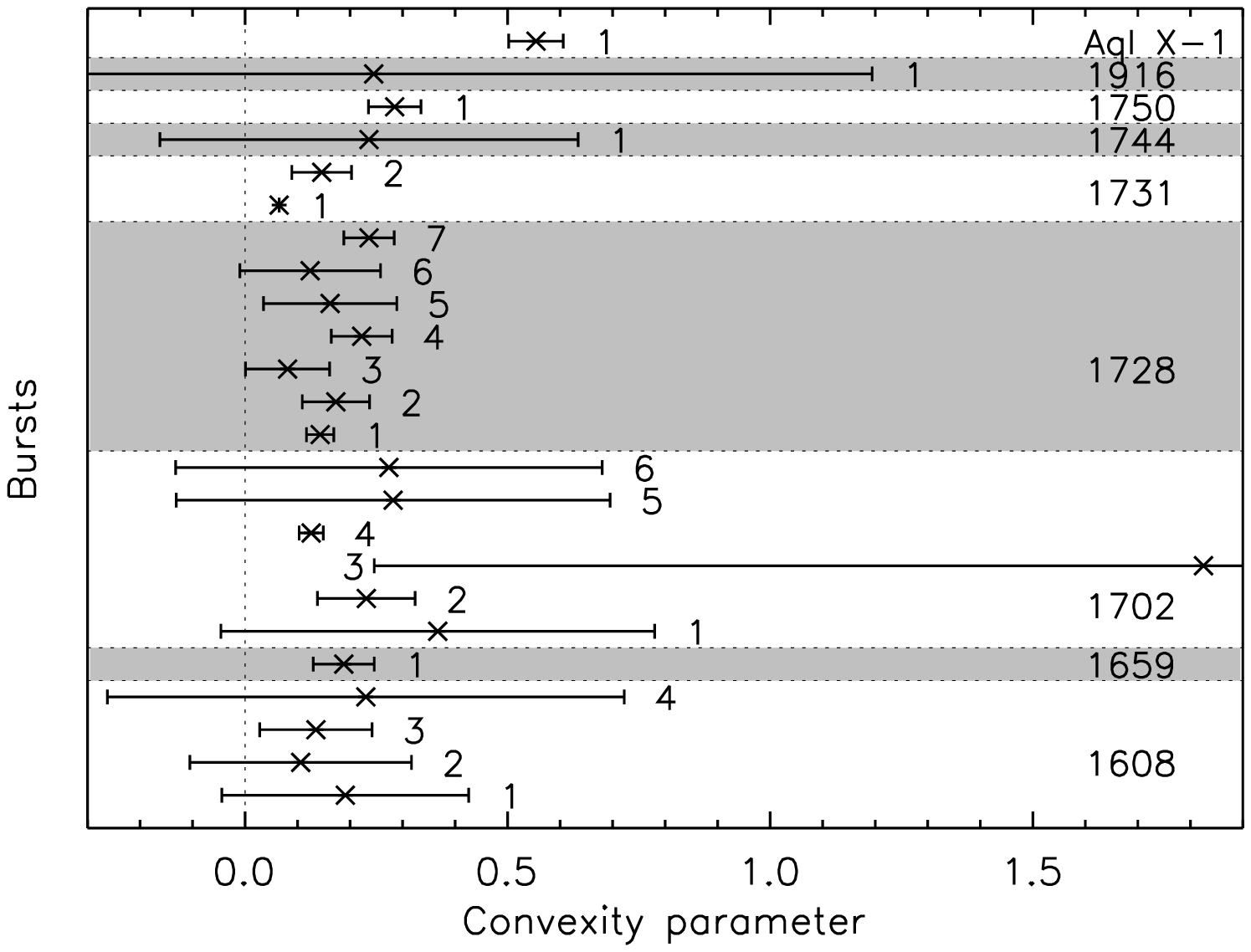} 
\caption{Similar to the right panel of Fig.~\ref{convexpar}, but 
for 24 bursts with rise oscillations from nine sources (mentioned in Table~\ref{Log2}).
In this case the y-axis represents the bursts from different sources (see Table~\ref{Log2}
for source names and the corresponding burst numbers). The dotted horizontal lines separate the 
bursts from different sources (\S~\ref{othersources}).
\label{convexpar2}}
\end{figure*}

\begin{figure*}
\centering
\includegraphics[width=0.7\textheight]{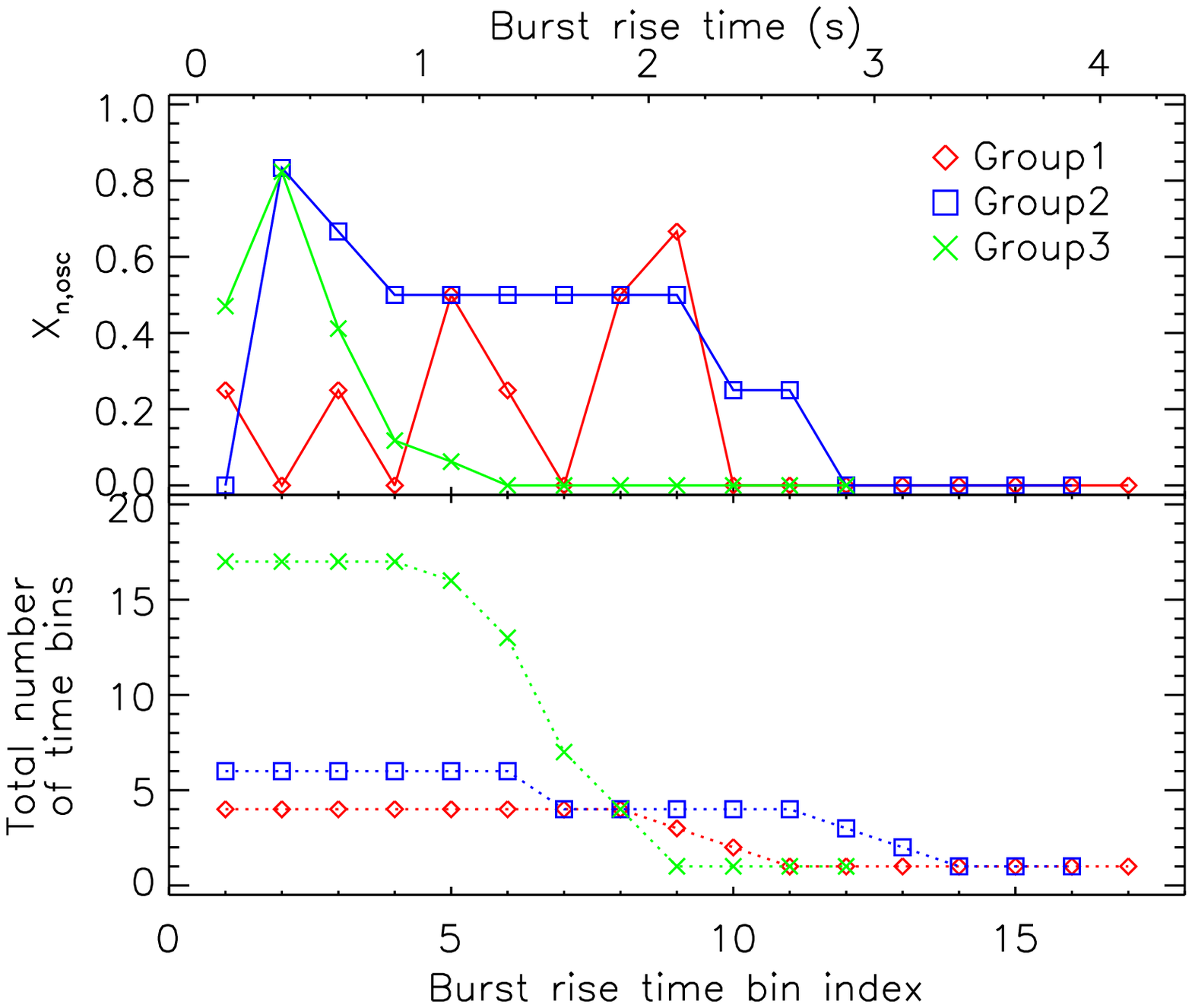}
\vspace{1.0cm}
\caption{{\it Top panel}: fraction ($X_{\rm n,osc}$) of thermonuclear burst rise time bins 
(each 0.25 s) with detected oscillations versus time bin index
for each morphological group as discussed in \S~\ref{morphology}. 
{\it Bottom panel}: for each morphological group, this panel shows the total number 
of burst rise time bins available for a given time bin index. 
In this figure, 27 4U 1636--536 bursts with oscillations during rise detected with 
{\it RXTE} PCA are considered (see \S~\ref{amplitude_calculation}).
This figure shows that rise times of Group 3 bursts are usually shorter than those of Group 1 and 2
bursts. Moreover, burst rise oscillations for Group 3 disappear typically much faster than
those for other two groups (see \S~\ref{morphology} and \S~\ref{Discussions}).
\label{detectiowithgroup}}
\end{figure*}

\begin{figure*}
\centering
\hspace{-1.5cm}
\includegraphics[width=0.7\textheight]{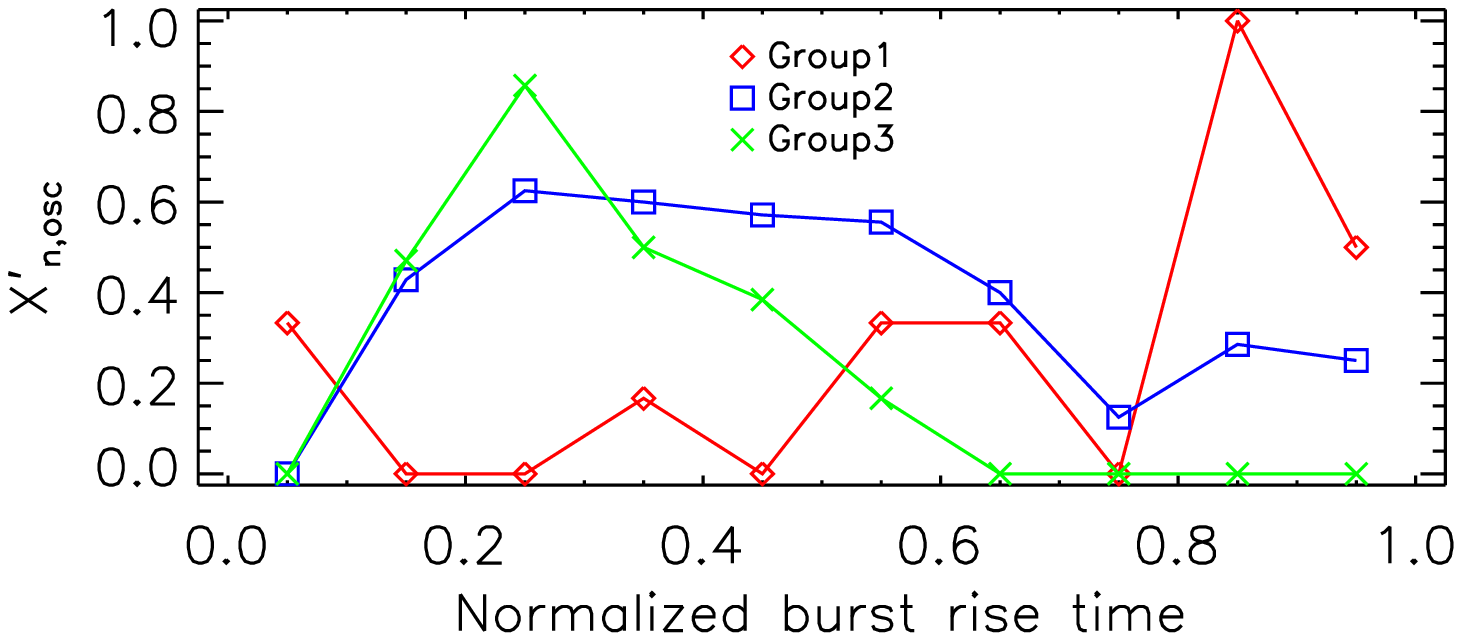} 
\caption{Similar to the upper panel of Fig.~\ref{detectiowithgroup}, 
but the x-axis is normalized as follows. The rise time of each burst is normalized
to 1, and subsequently divided into 10 equal bins. Then considering all the bursts
for a morphological group, the fraction ($X'_{\rm n,osc}$) of the $n^{\rm th}$ bins with
detected burst rise oscillations is calculated ($n = 1, 2, ..., 10$).
Finally, $X'_{\rm n,osc}$ is plotted with the normalized burst rise time (defined in the
range $0-1$), which is calculated
from $n$. This figure shows that typically oscillations last for a shorter fraction of burst rise
for Group 3 than for other two groups (see \S~\ref{morphology} and \S~\ref{Discussions}).
\label{detectiowithgroupscaled}}
\end{figure*}

\end{document}